\documentclass[useAMS,usenatbib,usegraphicx]{mn2e}

\newcommand{\cl}{C_{\ell}}

\title[Discriminating between unresolved point sources and
``negative'' SZ clusters]{Discriminating between unresolved point sources and
``negative'' SZ clusters in CMB maps}
\author[J.A. Rubi\~no-Mart\'{\i}n and R.A.Sunyaev]{J.A. 
Rubi\~no-Mart\'{\i}n$^{1,2}$\thanks{E-mail:jalberto@mpa-garching.mpg.de, 
jalberto@ll.iac.es} 
and R.A. Sunyaev$^{1,3}$\\
$^{1}$Max-Planck Institut f\"ur Astrophysik, Karl-Schwarzschild-Str. 1,
D-85740 Garching, Germany\\
$^{2}$Instituto de Astrofisica de Canarias, C/ Via Lactea, s/n, 38200
La Laguna, Tenerife, Spain\\
$^{3}$Space Research Institute (IKI), Russian Academy of Sciences,
Moscow, Russia}
\begin{document}

\date{}

\pagerange{\pageref{firstpage}--\pageref{lastpage}} \pubyear{2003}

\maketitle

\label{firstpage}

\begin{abstract}
Clusters of galaxies produce negative features at 
wavelengths $\lambda > 1.25$ mm in CMB maps,
by means of the thermal SZ effect, while point radio  
sources produce positive peaks. 
This fact implies that a distribution
of unresolved SZ clusters could be detected using the negative
asymmetry introduced in the odd-moments of the brightness 
map (skewness and higher), or in the probability distribution 
function (PDF) for the 
fluctuations, once the map has been filtered in order to
remove the contribution from primordial CMB fluctuations from
large scales.
This property provides a consistency check to the
recent detections from CBI and BIMA experiments 
of an excess of power at small angular scales, in order
to confirm that they are produced by a distribution of unresolved 
SZ clusters. 
However it will require at least 1.5 - 2 times more observing time than
detection of corresponding power signal. 
This approach could also be used with the data of
the planned SZ experiments (e.g. ACT, AMI, AMIBA, APEX, 8 m South Pole
telescope). 

\end{abstract}


\section{Introduction}

Fluctuations in the Cosmic Microwave Background (CMB) radiation 
can provide information about hot gas in galaxy clusters 
over a wide range of redshifts (Sunyaev \& Zeldovich 1972,1980 
(hereafter SZ), and \citealt{birkinshaw99,carlstrom02}).
On arcminute angular scales and
smaller, the thermal SZ contribution to the CMB anisotropy is
expected to dominate that of the primary anisotropies 
(\citealt{sunyaev70}, \citealt{springel01} (hereafter SWH),
\citealt{holder02}).
A new generation of experiments is  
measuring the CMB sky at these angular scales. In particular, 
two recent experiments, BIMA \citep{dawson02} and CBI \citep{mason02},
both observing at frequencies around 30 GHz, have
detected an excess of power in the multipole region 
$\ell \ga 2000$, where the SZ power is expected to be dominant over
the CMB signal.
Nevertheless, at these observing frequencies, radio point sources
are known to also produce a significant contribution 
to the power \citep{longair69,franceschini89,toffolatti98} 
if they are not subtracted properly from the CMB maps. 
The reported detections of power have argued that this point-source
contamination is not a problem, thus suggesting
that the signal could be due to the SZ effect 
\citep{dawson02,bond02,komatsu02}.
These arguments are based on analytical models or simulations of
what we would expect to measure. Thus, it would be interesting
to explore, in a model-independent way, the nature of these
contributions. The importance of this topic has been stressed 
recently by \cite{cooray02}, who  
suggested to use a cross-correlation of CMB maps with maps of
the large scale structure. This idea has been applied for this 
purpose to other datasets with larger angular resolutions 
(e.g. \citealt{banday96}, to the COBE data, or 
\citealt{rubino00}, to the Tenerife data). 

\begin{figure}
\includegraphics[width=\columnwidth]{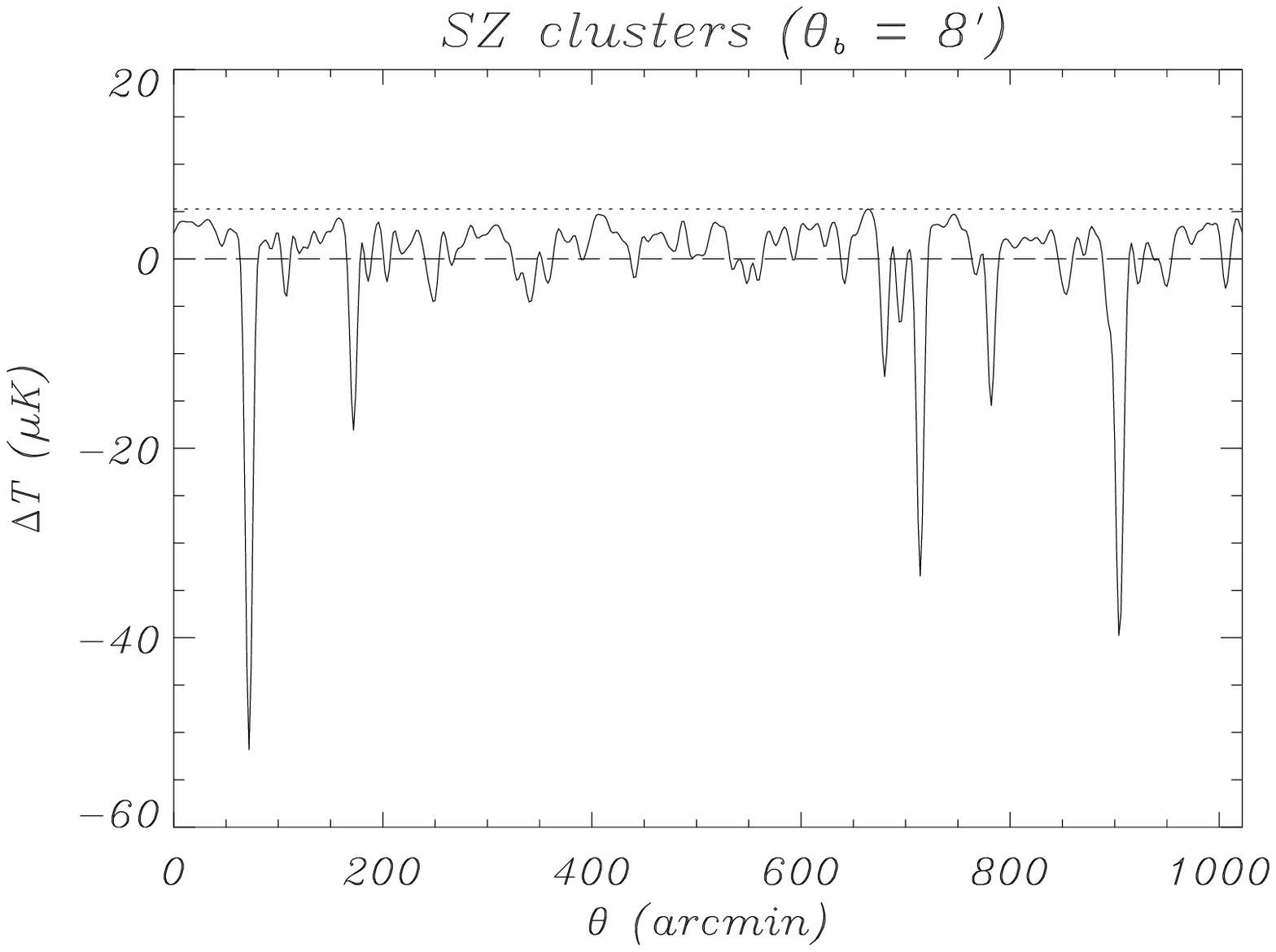}
\includegraphics[width=\columnwidth]{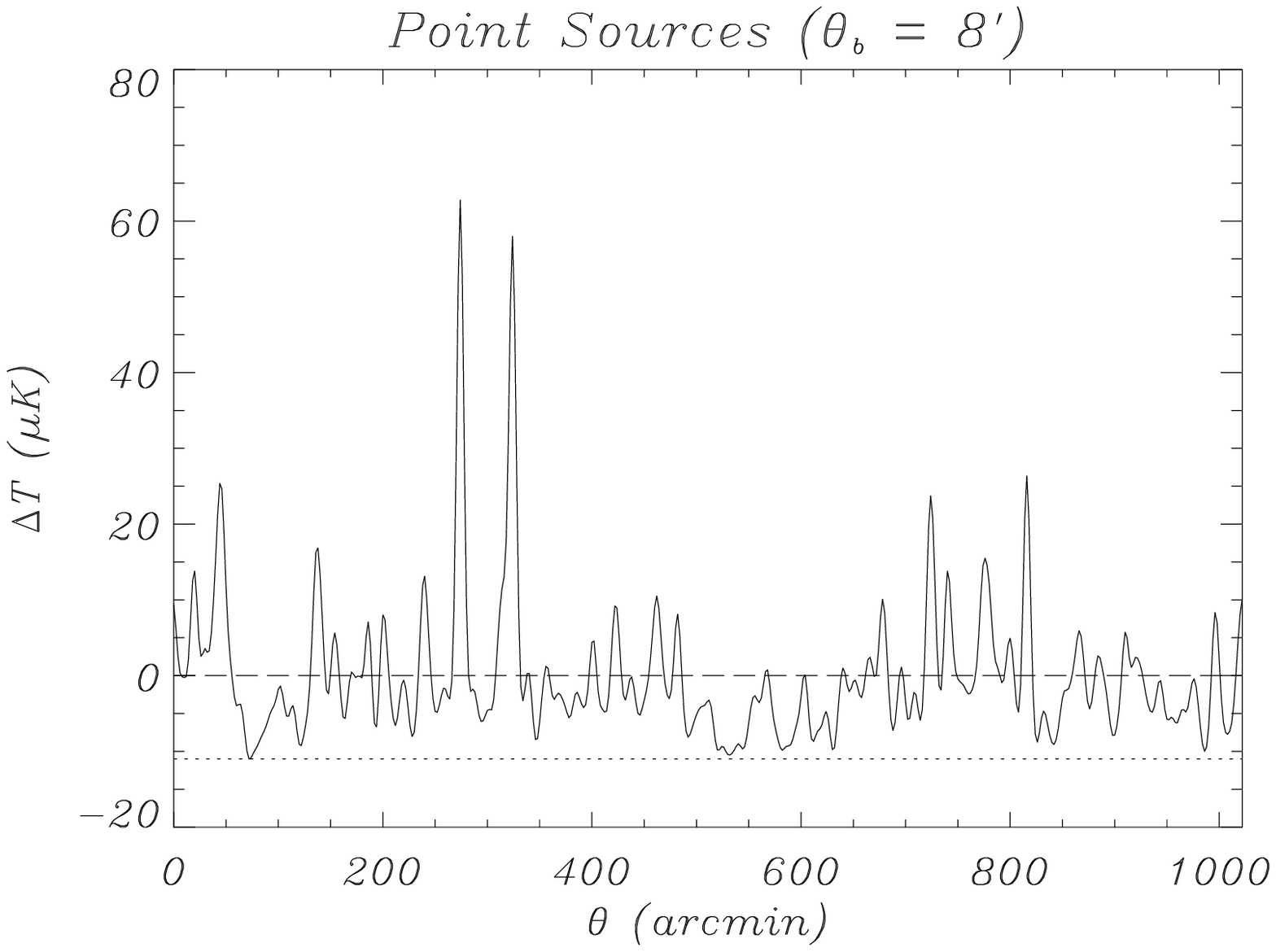}
\caption{One dimensional map of a single realisation of SZ clusters
(upper panel) and point sources (lower panel), observed with a
gaussian beam of $8'$ full-width half-maximum, and no noise. 
Dashed lines show the average (zero)
level of fluctuations, while dotted lines show the original zero level
before subtracting the mean to the map. 
With the same level of fluctuations, the 
$\cl$ analysis does not permit to distinguish between these
two cases. Therefore, we need to use the skewness, or to proceed with
an analysis of the asymmetry of the $P(D)$ curve in order to
separate these two cases.}
\label{typical}
\end{figure}

Here, we propose a general model-independent method to determine 
if the measured power excess in a {\sl single-frequency} map 
is (mainly) due to 
point sources or SZ clusters. To this end, we use the fact
that for frequencies below $217$ GHz ($\lambda > 1.25$ mm), 
the thermal SZ effect produces
negative features in the maps, while the point sources 
produce positive peaks. We illustrate this fact with 
figure \ref{typical}, where we show two simulated one-dimensional maps, one of 
SZ clusters observed at $\nu = 30$ GHz, and the other one of point
sources. Dotted lines show the original (without sources of any kind) zero level
of fluctuations, while dashed lines show the observed (average) zero 
level once the mean of the map has been subtracted. 
With the same level of fluctuations (same rms at the 
observed scale), a power spectrum analysis is not able to
distinguish these two cases, so we need to use an 
statistic carrying information about the sign of the 
subjacent signal (e.g. the skewness) to suggest  
the nature of the objects 
producing this excess of power. 
The existence of negative skewness at $\lambda > 1.25$ mm,
while positive skewness at $\lambda < 1.25$ mm, is a clear
prediction for SZ clusters.

\begin{figure}
\includegraphics[width=\columnwidth]{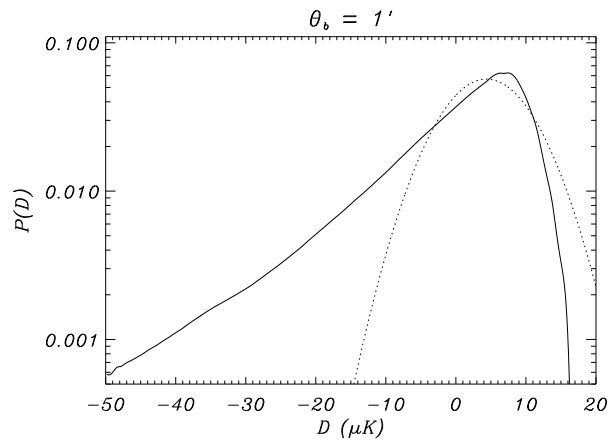}
\caption{Example of the strong non-gaussianity of the P(D) function
for SZ clusters. We present the P(D) function for a SZ map
in the Rayleigh-Jeans region of the spectrum, where clusters
are ``negative'' sources. For comparison, it is also shown the
best gaussian fit to this P(D) curve ($\sigma = 6.1~\mu K$). This
curve will be explained in detail in section \ref{sec:general}.}
\label{figura_bonita}
\end{figure}

\begin{figure}
\includegraphics[width=\columnwidth]{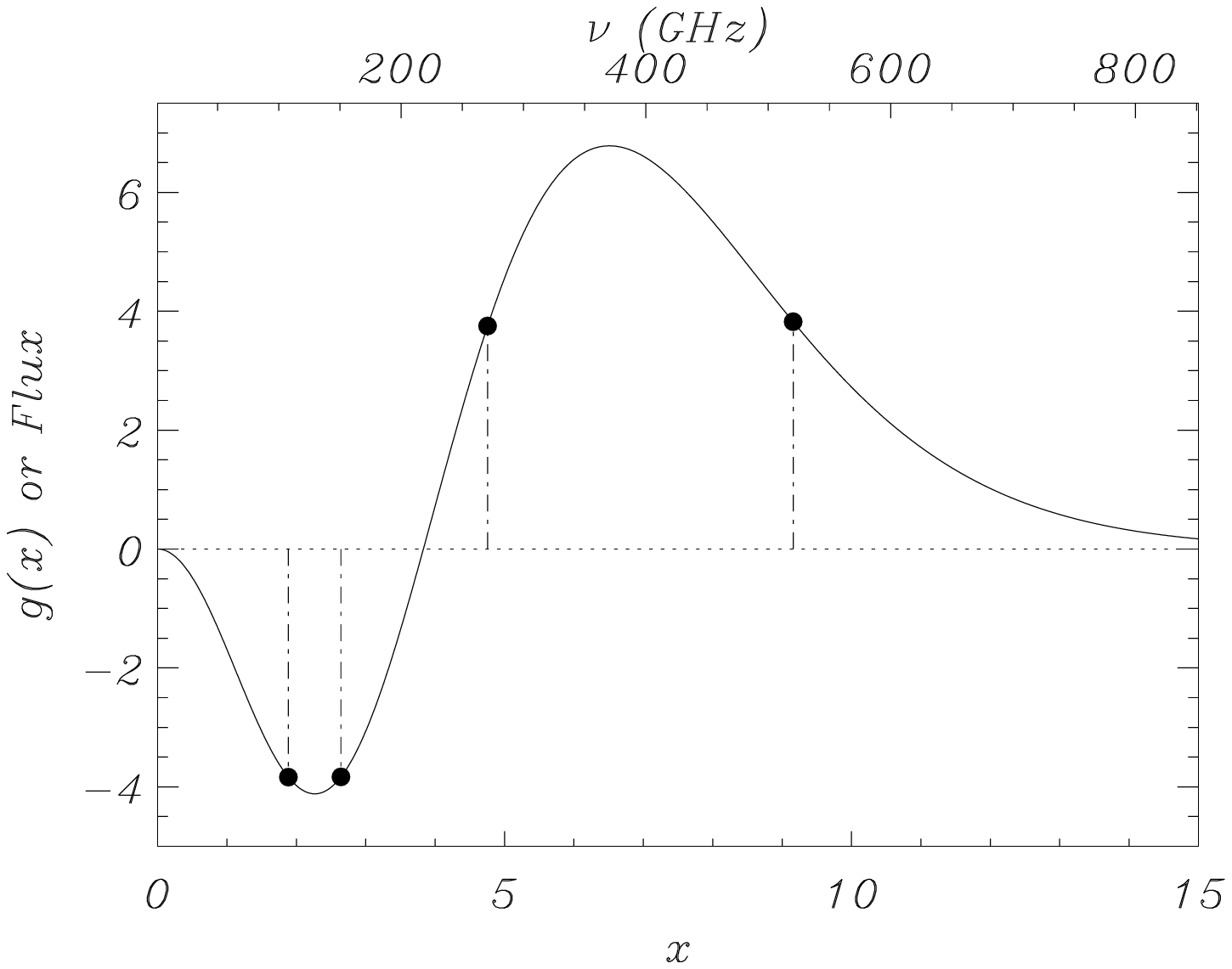}
\includegraphics[width=\columnwidth]{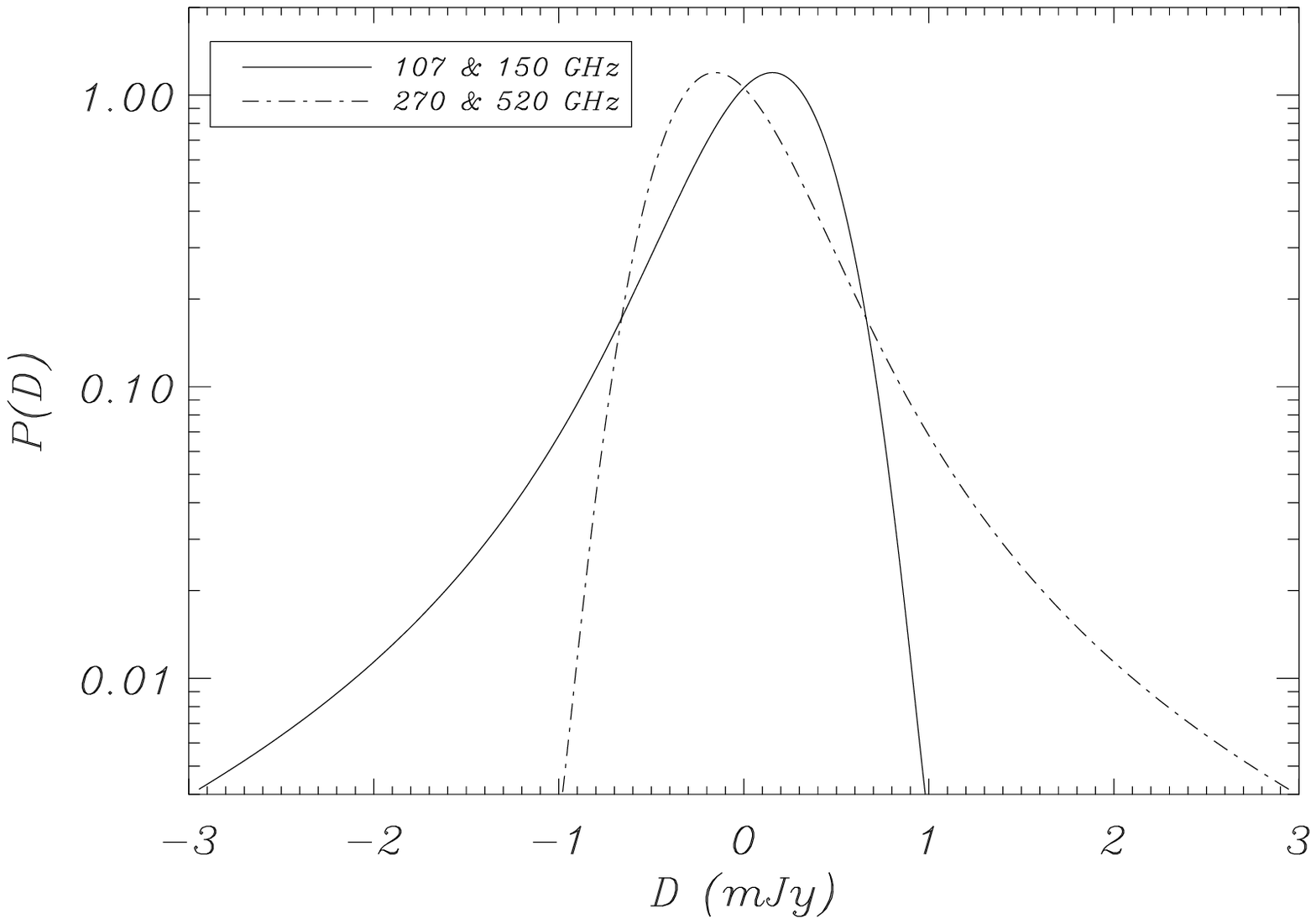}
\caption{P(D) function for SZ clusters at different frequencies.
First panel shows the $g(x)$ function (see equation (\ref{s2:g(x)})), and 
four frequencies ($\nu = $ 107, 150, 270 and 520 GHz) 
where this function (and so the
flux density) takes 
the same absolute value. Second panel shows the P(D) function
for these four cases, using a simple truncated power-law to model the 
cluster source counts (see section 4), 
with values $n(S)$ = 28 (S/1Jy)$^{-2.5}$ sr$^{-1}$
Jy$^{-1}$ at 150 GHz, truncating at $S_0 = 0.1$ mJy, 
and with angular resolution $\theta_b = 1'$. The P(D) function
is presented relative to its average value, so the distribution 
for the cases $\nu=107$ GHz and 150 GHz is symmetric around zero respect
to the other two cases.}
\label{compara_pd_sz}
\end{figure}

We investigate here the discrimination between positive and 
negative sources using the probability distribution function (PDF) 
for the observed flux. From a given map, {\it the PDF function can 
be obtained easily as an histogram of the (normalised) number of
pixels within a given flux interval}. 
This tool has 
been widely used in radio astronomy when studying 
the statistical properties of a background 
of point sources \citep{scheuer57,cavaliere73,condon74}, because in 
that case the shape of this function is strongly related with the statistical
properties of the sources (i.e. theirs spatial
distribution).  
In this context, this function is known as the 'deflection 
probability distribution', or the $P(D)$ curve. 
This '$P(D)$ formalism' has been 
successfully applied to study the diffuse X-ray background 
\citep{scheuer74,fabian75,cavaliere76,condon78}, 
as well as to determine the contribution of discrete point sources
to CMB maps \citep{franceschini89,toffolatti98}. 
For the CMB, if we assume the standard inflationary
scenario, then the primordial fluctuations are
gaussian, so the $P(D)$ itself is a gaussian, as well as for the 
standard instrumental noise. 
However, the main characteristic of this $P(D)$ curve for point 
sources (\citealt{franceschini89}) or for SZ clusters (\citealt{cole88}) 
is its non-gaussianity. 
Typical curves for a $P(D)$ distribution of point 
sources or SZ clusters will exhibit long tails 
(see Figure \ref{figura_bonita}).
The point is that at $\lambda > 1.25$ mm, 
sources will produce a positive tail, while
SZ clusters will give a negative one. It is important to mention
that at $\lambda < 1.25$ mm, both AGNs and SZ-clusters will
produce positive tails. Then it is necessary to use other 
characteristics of both populations (frequency spectra, etc) to
distinguish them. As an illustration, 
Figure \ref{compara_pd_sz} demonstrates P(D) for 
SZ sources at four frequencies, $\nu$ = 107 and 
150 GHz (where clusters are giving negative 
signal) and 270 and 520 GHz (where the signal 
from clusters is positive, and exactly opposite 
in sign to the previous cases).

\section{SZ clusters as negative sources}
\label{sec:2}

It is possible to consider clusters
of galaxies as ``extended sources'' with a peculiar 
spectrum given by \citep{rashid80}
\begin{equation}
F_{\nu_z} (\theta) = B_{\nu_z}(T_r) 
\frac{\delta I_{\nu} (\theta)}{I_{\nu}} = 
\frac{2 (k T_{r})^3}{(hc)^2} \frac{k T_e}{m_e c^2} g(x)
\tau_T(\theta)
\label{dI_sz}
\end{equation}
where $\tau_T$ is the optical depth for Thompson scattering, 
$x=h\nu/k T_{cmb}$ is the dimensionless frequency, with 
$\nu_z = \nu (1+z)$ and $T_{r} = T_{cmb} (1+z)$, so that 
$x$ does not depend on redshift $z$, and
\begin{equation}
g(x) = \frac{x^3}{e^x-1} f(x) = 
\frac{x^4 e^x}{(e^x-1)^2} [x coth(x/2)-4]
\label{s2:g(x)}
\end{equation} 
is the spectral shape factor. Note that this shape factor includes
the term from $B_\nu$, the Planck function, and $\delta I_\nu/I_\nu$ 
and $f(x)$ are 
the formulae for the CMB spectrum distortions due to Comptonization 
from \cite{zeldovich69}.
From here, two different but equivalent approaches can be used to estimate
the spectral luminosity of the cluster.
We can obtain the spectral luminosity by just
integrating the change of the CMB intensity due to scattering by
individual electrons of temperature $T_e(r)$ over the cluster volume 
\citep{rashid80}
\begin{equation}
L_{\nu_z} (x,z) = 8 \pi \frac{(k T_{cmb})^3}{(h c)^2} \sigma_T g(x) (1+z)^3 
4\pi \int \frac{k T_e(r)}{m_e c^2} n_e(r) r^2 dr
\label{L_sz}
\end{equation}
This expression, for the case of 
isothermal intergalactic gas, is proportional to the total amount of 
electrons in the cluster,  because in that case 
$\frac{4\pi}{3} \int n_e(r) r^2 dr = M_{IGG} / (\mu _e m_p)$, 
where $M_{IGG}$ is the mass of the
hot intergalactic gas, and $\mu _e$ is the mean molecular weight 
per electron. 
The important point here is that clusters increase 
rapidly their spectral luminosities with 
redshift ($L_{\nu_z} \propto (1+z)^3$).
Taking into account the luminosity distance to the source 
$d_L(z)$, we can obtain the spectral flux as $S_\nu =
\frac{L_\nu(z)}{4\pi d_L^2(z)}$. 

On the other hand, according to \cite{korolev86}, we can use
the central value of the Comptonization parameter for the cluster 
\begin{equation}
y_C = \int \frac{k T_e(l)}{m_e c^2} \sigma_T n_e(l) dl
\end{equation}
For a given $y_C$, the surface brightness of the cluster does not
depend on redshift. Then, the flux from the cluster is equal to
\begin{equation}
S_\nu (x) = 2 \pi \theta_0^2 \frac{(k T_{r})^3}{(h c)^2} g(x) y_C 
Y(\theta / \theta_0)
\end{equation}
where $\theta_0$ corresponds to the angular dimension of the cluster
core radius, and the $Y(\theta / \theta_0)$ function takes
into account the angular dependence of $y$ over the cluster image.
One important conclusion from this point of view of the 
problem is that clusters with given physical parameters 
should have a minimum flux at 
some redshift, due to the well-known redshift dependence of the
angular dimension of the cluster with a given core radius. 
In the Universe with $\Omega = 1$ we
have minimum angular dimension 
at $z=1.25$ \footnote{For a cosmological model with $\Omega_m=0.3$ and
$\Omega_\Lambda=0.7$, this happens at $z=1.60$.}, and at higher $z$, 
both angular distance and the flux will increase: clusters 
with given physical parameters have minimum flux where its
angular dimension is minimum. 
It is important to note that
\begin{enumerate}
\item for an experiment with an angular
resolution $\theta_b$ larger than the 
core radius of clusters ($\theta_b \gg \theta_0$), they will be
unresolved objects, and therefore they will appear as 
point sources for us;
\item according to the dependence of $g(x)$ from x, these
point sources will have ``positive'' flux at $\lambda < 1.25$ mm, 
and ``negative'' flux at $\lambda > 1.25$ mm.
\item for given physical parameters, clusters show a minimum flux at
that redshift $z$ where the angular dimension is minimal.
On the other hand, the observed source counts depend on the luminosity
function. For instance, 
if we use the  Press-Schechter 
formalism (\citealt{press74}, hereafter PS), then we have a
divergence of the (comoving) number density of objects at 
low masses (i.e. low fluxes). 
Nevertheless, we know that cooling and feedback play an
important role in the SZ predictions, so we do not expect to
find bright SZ clusters with masses below few times $10^{13} M_\odot$. Then, 
it is necessary to introduce a low mass cutoff in PS formalism in
order to derive realistic SZ predictions, as it has been done by several 
authors (e.g. \citealt{deluca95,komatsu99,molnar00}).
Thus, we expect to observe a minimum flux and a minimum angular 
dimension for SZ clusters.
\end{enumerate}

Several authors have studied the contribution
of the SZ effect to the power spectrum of CMB fluctuations 
at small scales, both theoretically
(e.g. \citealt{cooray01}) or using simulations
(e.g. SWH, \citealt{zhang02}). 
In any case, the main observational emphasis is
put on the power spectrum ($\cl$), because it is easier to measure
than, for example, the bispectrum. When determining if the
excess of power at low scales detected by CBI and BIMA is due
to SZ clusters, the comparison has been done in terms of the 
power spectrum \citep{dawson02,bond02,komatsu02}.
However, when we are working with the power spectrum, we are losing
information about the sign of the fluctuations, 
so we need to find the skewness, or 
to proceed with an analysis of the asymmetry of the $P(D)$ function.

\section{Statistical description of randomly distributed 
positive/negative sources}
\label{sec_pd}

The formalism relating the (differential) source counts 
and the PDF (or $P(D)$ function) of the observed deflection 
$D$ at a given point, due to a population of poissonian-distributed 
unresolved sources, was first discussed by
\cite{scheuer57}, and extended by \cite{condon74}. 
This analytical approach will permit us to understand much more deeply
the observed shapes of the $P(D)$ curves, and to relate them with 
the underlying source counts.

Using the standard notation from radio astronomy, let $n(S)$ be the
differential counts per solid angle, at a given frequency $\nu$, and
let $b(\theta, \phi)$ the response of a radio telescope to a
point source (normalised to 1 at the peak). Let $s=S b(\theta, \phi)$
be the response of the instrument to a source of flux density $S$
located at a given distance $(\theta, \phi)$ of the beam centre.
Then, the mean number of source responses of 
flux between $s$ and $s+ds$ in the beam, $R(s)$, is  given by
\begin{equation}
R(s) = \int  n \Bigg[ \frac{s}{b(\theta, \phi)} \Bigg] 
\frac{d\Omega}{b(\theta, \phi)}
\end{equation}
The relationship between $n(S)$ (or $R(s)$) and the
$P(D)$ function, for
the case of a pencil beam antenna, is given in terms of the 
characteristic functions of $R(s)$ and $P(D)$, which
can be written as $r(w)$ and $p(w)$, respectively. 
The equations are
\begin{equation}
r(w) = \int_{-\infty}^{+\infty} R(s) \exp (2\pi i w s) ds
\end{equation}
\begin{equation}
p(w) = \exp [ r(w) - r(0) ]
\end{equation}
and the $P(D)$ function can be obtained as the inverse Fourier transform
\begin{eqnarray}
\nonumber
P(D) = \int_{-\infty}^{+\infty} p(w) \exp (-2\pi i w D) dw = \\
\int_{-\infty}^{+\infty} \exp ( r(w) - r(0) - 2\pi i w D) dw
\label{ec:pd}
\end{eqnarray}
This relation can be also employed in the case of tracking
interferometers (\citealt{fomalont88,fomalont93}), 
using the CLEANed map.
For the case of a phase switch interferometer, the above relations
still holds, but replacing Fourier transforms by Bessel transforms 
\citep{scheuer57}.

It should be noted that these expressions are general, and therefore
valid for the case of negative sources. Indeed, from the previous 
equations it follows that \emph{if a population of positive sources 
$n(S)$, with $S>0$, is described by the function $P(D)$, then
the population of sources $n(|S|)$, with $S<0$, is 
described by $P(-D)$}. Thus, in this section, we will restrict
ourselves to study distributions of positive sources ($n(S)=0$ for
$S<0$), given that we can obtain the corresponding distribution
for negative sources with the transformation $P(D) \rightarrow P(-D)$.

All the equations described through this section are implicitly assuming 
non-resolved objects. \cite{rowan74} have
studied the modifications introduced by extended sources, showing
that the condition for unresolved objects applies until 
$\theta_b \approx \theta_s$, where $\theta_s$ is the typical size
of the source, and $\theta_b$ the beam size. 

\subsection{Analytical cases}
Let us consider here several particularly simple but useful cases for the $n(S)$
function, which can be treated analytically. 
These cases will be used later. For a more general case of source
counts, we can use a Monte Carlo method to work out the P(D)
distribution
(e.g. \cite{hewish61}). 
First of all, we will consider a power-law shape, 
$n(S) = K S^{-\beta}$, with $S>0$. 
As a second case, we will also consider a
truncated power-law at a certain flux density $S_0$, i.e.
$n(S) = 0$ for $S<S_0$, and $n(S) = K S^{-\beta}$, for $S>S_0$. 
Hereafter, we will assume that $S$ is given in $Jy$, so 
we are implicitly writing $n(S) = K (S/1Jy)^{-\beta}$, and 
the units for $K$ are Jy$^{-1}$ sr$^{-1}$.
We will also assume a gaussian antenna pattern, described as
$b(\theta) = \exp (-\frac{1}{2} (\theta / \sigma_b)^2)$, where 
$\sigma_b$ is the width of the beam, and $\theta_b=\sqrt{8\log 2}
\sigma_b$ its full-width half maximum (FWHM).

In these cases, the $R(s)$-function can be analytically obtained. Following 
\cite{condon74}, we define the effective solid angle ($\Omega_e$) as 
\begin{equation}
\Omega_e = \int [b(\theta,\phi)]^{\beta-1} d\Omega = 
\frac{\pi \theta_b^2}{4 (\beta-1) \log 2}
\end{equation}
We then obtain $R(s) = K \Omega_e s^{-\beta}$ for a pure
power-law, and  
\begin{equation}
R(s) =  \left \{
\begin{array}{ll}
K \Omega_e s^{-\beta}, \qquad & s>S_0 \\
K \Omega_e S_0^{-\beta} \frac{S_0}{s}, \qquad & s<S_0 \\
\end{array}
\right.
\label{r_s_truncated}
\end{equation}
for a power-law truncated below $S_0$ and the gaussian beam. 
From these equations, it is 
straight-forward to obtain numerically the $P(D)$ function. 
These problems have been studied by \cite{scheuer57} and 
\cite{condon74} for the power law case\footnote{For the 
power law case, Condon (1974) gives an analytic expression
for $p(w)$, which is valid for $2<\beta<3$, so in these cases
the numerical calculation is even easier.}, and
by \cite{scheuer74} (analytically) and 
\cite{hewish61} (numerically) for the truncated case. 

Finally, it is also interesting to consider the case of a power-law 
source counts with an upper cut-off in flux, $S_c$. 
This is the expression to use in order
to  compute the $P(D)$ function from a map where the brightest sources
have been subtracted down to the flux $S_c$. 
Therefore, if $n(S) = K S^{-\beta}$ for $S_0 < S < S_c$, 
and $0$ elsewhere, then
\begin{equation}
R(s) =  \left \{
\begin{array}{ll}
K \Omega_e s^{-1} (S_0^{1-\beta} - S_c^{1-\beta}), \qquad & s < S_0 \\
K \Omega_e s^{-\beta} (1-(\frac{s}{S_c})^{\beta-1}) , \qquad & S_0 < s < S_c\\
0, \qquad & s > S_c
\end{array}
\right.
\label{r_s_truncated_twice}
\end{equation}

\subsection{Confusion noise}
\label{sec_pd_confusion}
Apart from the $P(D)$ function, it is also interesting to 
characterise the properties of the source population by the
moments of the $R(s)$ distribution.
For a pure power-law expression for the number-flux-density relation, 
it is clear that the $n$th-moment of the R-distribution
is
\begin{equation}
<s^n> = \int_{0}^{s_c} s^n R(s) ds = 
\frac{K \Omega_e s_c^{n+1-\beta}}{n+1-\beta}, \qquad n=1,2,3,...
\label{s^n_powerlaw}
\end{equation}
where $s_c$ represents the cutoff value for 
point-source subtraction. 
For the truncated power-law case, the $n$th-moment 
of this distribution is 
\begin{equation}
<s^n> = 
\frac{K \Omega_e s_c^{n+1-\beta}}{n+1-\beta} - 
\frac{\beta-1}{n(n+1-\beta)} K \Omega_e S_0^{n+1-\beta}
\label{s^n_powerlaw_truncated}
\end{equation}
The second moment of the $R(s)$ function ($\sigma_c^2 = <s^2>$)
has been extensively used to characterise the 
'confusion noise', i.e. the noise due to the presence
of faint unresolved sources inside the beam \citep{scheuer57}. 
Normally, the adopted criterion for the detection of sources is such
as the intensity $q$ times the sigma of the map, so 
the minimum subtraction threshold can be written as $s_c = q \sigma_c$, 
being $q=3-5$ the usual values. Thus, inserting 
this condition in equation (\ref{s^n_powerlaw}), we obtain for 
the power-law case:
\begin{equation}
\sigma_c(q) \equiv  \Bigg[ \frac{q^{3-\beta}}{3-\beta} K \Omega_e
\Bigg]^{1/(\beta -1)}
\label{ec:sigmacq}
\end{equation}
where we explicitly write that $\sigma_c$ depends on $q$ for this
choice of the subtraction threshold. Note that this threshold can be
decreased if we use measurements at higher angular resolution. We will 
discuss this in section \ref{sec:roleofradiosources}.

For the power-law source counts, it is easy to show that the
flux at which we have one source every $X$ beam areas, $S_X$, can
be written as
\begin{equation}
S_X ({\it Jy}) = \Bigg( K \Omega_e X \Bigg)^{1/(\beta -1)}
\label{ec:sigmax}
\end{equation}
From these last two equations, we immediately see that 
for the flux level $s_c = q \sigma_c(q)$, we have one
source every $q^{2}/(3-\beta)$ beam areas. For the particular 
case of $\beta = 2.1$ and $q=5$, this expression takes
the value $\sim 28$, so we recover the well-known result that 
if we have more than one source of a given flux every
30--40 beam areas (it depends on $\beta$), 
we will be limited by confusion noise.
We should remind here that the $P(D)$ function provides information
(roughly) down to the flux $S_1$ at which we have one source
per beam area \citep{scheuer74}.

\subsection{Scaling of P(D) with frequency}
To conclude this section, we will derive the scaling of $P(D)$ with
frequency for SZ clusters. This scaling can be derived from the
one for the source counts, which is given by
\begin{equation}
n(S;\nu) = \Bigg| \frac{g(x_0)}{g(x)} \Bigg| \;  n \Bigg( S \frac{g(x_0)}{g(x)}; \nu_0 \Bigg)
\end{equation}
where $x_0 = h \nu_0 / k T_{cmb}$. 
This equation is similar to the one obtained by \cite{condon84} for 
the scaling of the differential source counts of point sources with 
power-law spectra ($S \propto \nu^{-\alpha}$)
\footnote{Indeed, this relation holds for all cases 
where the frequency dependence of the observed flux of 
the object can be factorised, i.e. $S_\nu = g(\nu)
\Phi$, where $\Phi$ does not depends on $\nu$.}.
From here, we obtain 
\begin{equation}
P(D;\nu) = \Bigg| \frac{g(x_0)}{g(x)} \Bigg| \;  
P \Bigg( D \frac{g(x_0)}{g(x)}; \nu_0 \Bigg)
\label{P_D_scaling}
\end{equation}
We see that clusters of galaxies should be described by a single
$P(D)$ function, which is the same at all frequencies (but rescaled)
if $\theta_b$ is the same, 
and which is equivalent to the PDF for the $y$ parameter. However,  
our description permits the use of the main characteristic of the
effect, the existence of negative sources. Therefore, 
if we compare data from two frequencies, one 
above $\nu = 217$ GHz and the other one below, then in the first case
the $P(D)$ for clusters will exhibit a positive tail while in 
the second case, a negative tail, 
being in both cases described by the same (rescaled) $P(D)$
function. 
If we now use this expression to derive the moments of the $P(D)$ 
function, we obtain 
\begin{equation}
< D^n ; \nu > = \Bigg[ \frac{g(x)}{g(x_0)} \Bigg]^n <D^n ; \nu_0>, 
\qquad n = 1,2,3,...
\end{equation}
Therefore, we explictily see that the normalised moments of a 
map of thermal SZ clusters (with no noise), $<D^n>/\sigma^n$,  
are exactly the same in magnitude for all frequencies, 
but we have a change of
sign in all the (normalised) odd-moments when we cross 
the frequency $\nu = 217$ GHz.

\section{Source counts for radio sources and clusters}

As we have seen in the last section, the shape of the $P(D)$ function
for (positive/negative) sources provides a means of determining 
the underlying source counts. Therefore, we will discuss here 
which are the typical source counts both for radio sources and SZ
clusters. 

\subsection{Radio point sources}

\begin{figure}
\includegraphics[width=\columnwidth]{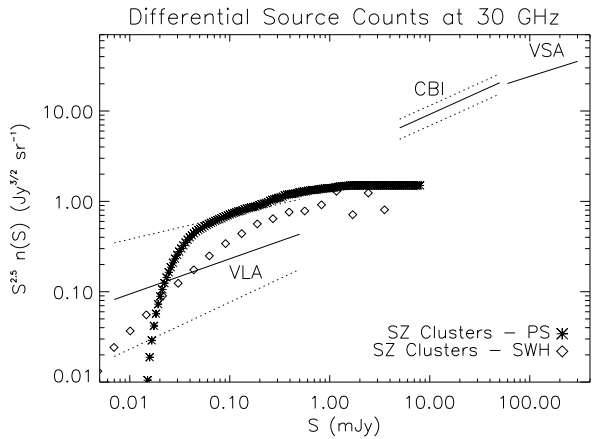}
\caption{Differential source counts for SZ clusters 
at 30~GHz, normalised to the Euclidean slope. 
We show the results from a Press-Schechter prescription with
$M_{min} = 5 \times 10^{13} h^{-1} M_\odot$ and $\sigma_8=0.9$ 
(see section \ref{sec_scounts} for details), as well as the source counts
from the paper of SWH (squares) using hydrodynamic simulations.  
Both source counts for clusters compare well in shape around 1 mJy, 
although the hydrodynamic simulations show a $\sim$20\% less objects
at these fluxes, and do not show an strong cutoff at low fluxes. 
For comparison, we also present here the 
differential source counts for radio sources 
at 30 GHz from several experiment:
VSA (Taylor et al. 2002), at 34 GHz; CBI (Mason et al. 2002) at 31
GHz; and VLA (Fomalont et al. 2002) at 8.4 GHz. The source counts
for this last experiment have been extrapolated up to 30 GHz 
using their mean spectral index $\alpha=0.75$. 
From the simple observation of these source
counts, we expect that experiments with
high angular resolution are going
to be dominated by radio sources, if they do not consider a
source subtraction strategy. }
\label{review_scounts}
\end{figure}

Several authors (e.g. \citealt{fischer93}) have studied the contribution to
the confusion limits from different populations of extragalactic 
sources. In the context of CMB measurements at frequencies 
close to $\nu \sim 30$ GHz, radio sources are 
known to produce the main contribution
to the confusion noise. Because of this reason, recent 
experiments have used a source subtraction technique, and
thus have produced source counts 
for these radio sources. 
These curves are well fitted by power-laws 
in the flux density region around a few mJy.
Typical values for the $K$ and $\beta$ parameters, at
frequencies around 30 GHz, are 
$n(S) \approx 54~(S_{34~GHz}/1Jy)^{-2.15}$ sr$^{-1}$ Jy$^{-1}$, for
$S_{34~GHz} > 60$ mJy (from the VSA experiment, \citealt{taylor02}), 
and 
$n(S) \approx (92 \pm 23)~(S_{31~GHz}/1Jy)^{-2.0}$ sr$^{-1}$ Jy$^{-1}$, for
$S_{31~GHz} > 5$ mJy (from the CBI group, \citealt{mason02}). 
Apart from these source counts, we can also extrapolate
the $\mu$Jy source counts at 8.4 GHz from
VLA observations \citep{fomalont02} up to 30 GHz. Using
the spectral index $\alpha = 0.5$ ($S \sim \nu^{-\alpha}$), we
obtain $n(S) \approx (8.4 \pm 0.8)~(S_{30~GHz}/1Jy)^{-2.11\pm0.13}$ 
sr$^{-1}$ Jy$^{-1}$. 
All these source counts are summarised in Figure \ref{review_scounts}.

We should mention that the contribution of
radio sources to the power (and thus to the skewness) 
is decreasing with frequency in 
power $\nu^{\alpha+2}$, where $\alpha$ is the spectral index of radio
sources. Therefore, the contribution of radio sources to the 
observed map becomes less important at higher frequencies 
($\nu \gg 30~GHz$), while the contribution of clusters does not
depend on frequency in the Raileigh-Jeans (RJ) region of the spectrum. 
It is important to remind that other populations of
sources (e.g. thermal dust emission from galaxies) contribute to the 
source counts at higher frequencies. However, we expect that
future experiments like ALMA will show us those populations very 
precisely. 

Given these source counts, the shapes  
of the corresponding distribution functions are
characterised by long positive tails, as we have seen in the
previous section. Inclusion of the clustering 
effects of sources \citep{barcons92} broadens the shape of the
$P(D)$, but the important point here is that
the long positive tail is still maintained.

\subsection{SZ clusters}

\cite{korolev86} discussed count curves
for thermal SZ clusters, and showed that they differ strongly from the
case of radio sources. 
\cite{deluca95} have derived the source counts
for the thermal SZ effect using the Press-Schechter 
mass function and assuming unresolved single-type clusters. 
For the scaling of the temperature with the mass of
the cluster, they use $T_e \sim M^{2/3} (1+z)$, so $S_\nu \propto
M^{5/3}$. These numbers are in agreement with those obtained 
from recent X-ray observations \citep{mohr99,ettori02}, and have been 
shown to fit simultaneously optical and X-ray cluster data \citep{diego01}. 
\cite{deluca95} show that typical curves can be well fitted 
by Euclidean power-laws $n(S) = K  |S|^{-2.5}$ down to a few $mJy$, 
with typical values of $K \sim 0.44$ sr$^{-1}$ Jy$^{-1}$ 
(extrapolated down to 30 GHz $\equiv$ 1 cm), and introducing a 
low-flux cutoff of $S_0 \approx 0.1$ mJy. 
SWH, using hydrodynamical simulations, came 
to similar conclusions, although their results do not show
an exact power-law behaviour, and the low-flux cutoff is one
order of magnitude smaller. 
For illustration, we 
show in Figure \ref{review_scounts} these source counts, and 
our derived source counts for
PS clusters, together with the radio source counts described in the last
subsection. The qualitative behaviour is the same pointed above.
These curves will be described in detail in Section \ref{sec:general}.
Here, we will point out two general aspects of any modelling of SZ clusters.

First, we should stress that the value for $S_0$ in each model
(semi-analytic or numeric) depends on
the chosen mass cutoff, $M_{min}$, i.e. the minimum mass of an object
contributing to the SZ effect (see discussion at the end of
section \ref{sec:2}). 
This minimum mass ($M_{min}$) is related to the minimum flux ($S_{min}$) 
observed in a given cosmology. 
Using equation (\ref{dI_sz}), and assuming that the gas in the cluster 
is isothermal, we obtain the expression for the total SZ decrement $S_{tot}$
for a galaxy cluster, as a function of its mass 
\begin{equation}
S_{tot} = \frac{2(k T_{cmb})^3}{(h c)^2} 
\frac{g(x) \sigma_T}{d_A^2(z)} 
\frac{k T_e}{m_e c^2} \frac{M f_g}{\mu _e m_p}
\end{equation}
where $d_A(z)$ stands for the angular diameter distance, and $f_g$ is the
gas mass fraction. Using the scaling $T_e = T_{e0} (M/M_0)^{2/3}
(1+z)$ pointed above, we can derive for the Rayleigh-Jeans region of 
the spectrum that
\[
S_{tot}^{RJ} = -1.9 \times 10^{5} h^{-1} \frac{f_g}{\mu _e} 
\Bigg( \frac{\nu}{30~GHz} \Bigg)^2 
\Bigg( \frac{T_{e0}}{9 \times 10^7~K} \Bigg) \times \]
\begin{equation}
\times \Bigg( \frac{M_0}{10^{15} h^{-1} M_\odot} \Bigg)
\Bigg( \frac{M}{M_0} \Bigg)^{5/3} (1+z) 
\Bigg( \frac{1~Mpc}{d_A(z)} \Bigg)^2 
\qquad Jy
\end{equation}
From here, it is straight-forward to infer the minimum flux for a
given $M_{min}$. As an example, if we use the standard 
values $h=0.7$, $\Omega_m=0.3$, $\Omega_\Lambda=0.7$, and we assume 
a constant value $f_g = 0.1$, then we find that for a given mass
we have the minimum flux at $z \approx 0.98$, and its value is
\[
|S_{min}^{RJ}| \approx 20 \Bigg( \frac{M_{min}}{M_0} \Bigg)^{5/3} \qquad mJy
\]
at $30~GHz$. Typical values for $S_{min}$ can be seen in
Table \ref{table0}. \cite{korolev86} were considering $T_e$
independent on redshift, so $S_{min}$ in their case was reached
simultaneously with the minimum of angular diameter.  
Here we use a dependence of $T_e$ with redshift ($\propto (1+z)$), 
and hence the minimum is reached when the function $(1+z)/d_A^2(z)$ 
takes its minimum.

\begin{table}
\caption{Dependence of the observed flux cutoff for clusters with
the minimum mass of a cluster, for $\nu = $ 30 GHz (see details in text).}
\label{table0}
\begin{tabular}{@{}cc}
\hline
$M_{min}$ ($M_\odot$)& $|S_{min}|$ (mJy)\\
\hline
	$10^{14}$  & 0.238  \\
	$5\times 10^{13}$  & 0.075 \\
	$10^{13}$  & 0.005  \\
	$3\times 10^{12}$  & 0.0007 \\
\hline
\end{tabular}
\end{table}

Finally, we will mention that 
the inclusion of large scale clustering in the simulations has been 
studied by several authors (e.g. \citealt{cole88,zhang02}). 
As we pointed in the last subsection, clustering broadens the 
shape of the $P(D)$ function respect
to the poissonian case. In any case, these 
studies confirmed that 
the PDF for y parameter is always characterised by a long positive tail,
which corresponds to a negative tail  
when $\lambda > 1.25$ mm, independently of whether 
we include clustering or not.

\section{Contribution of Thermal SZ sources to the bispectrum of CMB 
angular fluctuations}

A detailed study of the non-gaussian aspects of 
the thermal SZ effect can be found in \cite{cooray01}. 
Here we will show that the result about the change in sign
from the previous section holds for the
bispectrum, or indeed any odd-moment of the distribution.

We first decompose the temperature anisotropy in the spherical harmonics
basis, so we have 
$\delta T/T_{cmb} (\hat{n})= \sum_{\ell m} a_{\ell m} Y_{\ell m}(\hat{n})$.
From here, the bispectrum is usually defined as 
$B_3(\ell_1 m_1, \ell_2 m_2, \ell_3 m_3) = < a_{\ell_1 m_1} a_{\ell_2
m_2} a_{\ell_3 m_3}>$ (e.g \citealt{luo94}). 
For the case of a thermal SZ sky, the temperature anisotropy will be
given by $\delta T/T_{cmb} = \hat{f}(x) y$, where 
$\hat{f} = x coth (x/2) - 4$. In this equation, all the frequency
dependence is factorised in the $\hat{f}(x)$ function. Given that the
decomposition in the $Y_{\ell m}$-basis is unique, we can conclude
that the $a_{\ell m}^{SZ}$-coefficients will satisfy the
relation $a_{\ell m}^{SZ} = \hat{f}(x) y_{\ell m}$, where the $y_{\ell m}$
quantities correspond to the coefficients of 
the decomposition of the $y$ function. In this way, we can
write
\begin{eqnarray}
\nonumber
B_3^{SZ}(\ell_1 m_1, \ell_2 m_2, \ell_3 m_3 ; \nu ) = \\
\Bigg[ \frac{ \hat{f}(x) }{ \hat{f}(x_0) } \Bigg]^3 
B_3^{SZ}(\ell_1 m_1, \ell_2 m_2, \ell_3 m_3 ; \nu_0 )
\end{eqnarray}
where we explicitly see the change in sign when we pass
through $\nu = 217$ GHz. A similar relation also holds for 
all the higher odd-moments of the $a_{\ell m}^{SZ}$ quantities.

Therefore, we expect an overall change of the sign of the
contribution of SZ clusters to the bispectrum 
when comparing two maps, one observed at 
$\lambda < 1.25$ mm and the other one at $\lambda > 1.25$ mm.
However, for the case of SZ clusters, we would expect a larger
value of the non-gaussian features in real space. The reason 
is that clusters are localised objects in real space, but
when averaging modes in Fourier space, the resulting non-gaussianity
is diluted. As an illustration of this fact, 
\cite{zhang02} show that the kurtosis is (roughly) twice larger 
in real than in Fourier space when comparing angular
scales around $\ell \sim 1000$, while for $\ell \ga 6000$,
kurtosis in Fourier space goes rapidly to zero. 
Hence, it is better for the detection of negative skewness to
work directly with real-space statistics,  
with the advantage that they are easier to infer
from data. This approach has been used by several authors, and
in particular \cite{cooray01} gives the 
relationship between the bispectrum and the
skewness of the map smoothed on some scale with a given
window function. The important point here is that this skewness, 
filtered at some scale, will exhibit the same sign-dependence 
with frequency.

Summarising these sections, the main characteristics of SZ clusters
are the long negative tails of their brightness distributions at frequencies
below 217 GHz, the absence of sources in the vicinity of $\lambda =
1.25$ mm, and the existence of a positive contribution at 
$\lambda < 1.25$ mm.
In addition, skewness (or any odd-moment of the map) will
retain the net sign of the effect. 

\section{Estimators which discriminate the sign of the sources}
\label{sec:estimators}

In this section, we will be interested in the problem of determining
if an excess of power in a map is due to positive of negative source.
Therefore, we will not be interested in identifying individual features in the
maps, but in an average contribution. As has been pointed out, we will
use here the P(D) function. This function, for a given map, can be 
estimated by selecting a reasonable flux interval $\Delta D$, and 
computing an {\it histogram} (number of pixels with a flux
between $D-\Delta D/2$ and $D+\Delta D/2$).

Let $P_{s}(D)$, $P_{SZ}(D)$, $P_{cmb}(D)$, and $P_{n}(D)$ be
the distribution functions for the point sources, the SZ clusters, the
CMB and the instrumental (plus atmospheric) noise, respectively. 
The observed $P(D)$ function will
then be given by their convolution, 
\begin{equation}
P(D) = P_{s}(D) \star P_{SZ}(D) \star P_{cmb}(D) \star P_{n}(D)
\label{convol}
\end{equation}
The (primordial) CMB distribution function is assumed to be a Gaussian, 
although for the considered angular scales 
($\ell \ga 2000$), it is expected to 
produce a negligible contribution compared with the SZ or with the
noise. We will discuss this point in detail in section \ref{sec:addcmb}.
The noise is also assumed to be gaussian distributed. This is
a reasonable assumption for single dish radio-telescopes, and 
drift-scan interferometers, but also can be used for CLEANed images
of tracking interferometers.
Thus, the expected non-gaussianity in the $P(D)$ is introduced 
by sources (positive or negative), whose distributions are
characterised by skewed shapes. 
Therefore, if we want to
detect this asymmetry, and in particular, its sign, one
could use one of the following estimators:
\begin{itemize}
\item {\bf Asymmetry} ($A$) of the observed $P(D)$ distribution. 
This quantity can be estimated directly as the difference in area
between the positive and negative regions:
\begin{equation}
A \equiv \int_{D_p}^{+\infty} P(D) dD \; - \; \int_{-\infty}^{D_p} P(D) dD  
\label{asim}
\end{equation}
where $D_p$ stands for the value at which the $P(D)$ function peaks. 
Previous equation assumes that $P(D)$ is normalised to
unit area, i.e. $\int P(D) dD = 1$, so $A$ 
directly gives the fractional difference in area.
It should be noted that $D$ is usually quoted with respect to the 
deflection around the mean level ($\overline{D}$), so once we have
the $P(D)$ function, we make $D \rightarrow D - \overline{D}$, 
with $\overline{D} = \int D P(D) dD$. 
\item {\bf Non-gaussianity of the wings}. 
If we obtain the $P(D)$ function, we could 
test the presence of point sources/SZ clusters even in the
case when they produce a mutually cancelling asymmetry.
This can be done by comparing the positive/negative tail
of the distribution with the one expected from gaussian 
noise. This excess could be quantified as: 
\begin{equation}
\Delta_{+} \equiv  \int_{D_p}^{+\infty} [ P(D) - G(D)] dD 
\label{a_wings}
\end{equation}
for the positive tail, where G(D) is the expected 
distribution if we only have noise (normally assumed to be 
gaussian), and a similar equation for the negative one.
\item {\bf Skewness} of 
the observed map. This cumulant has
information about the overall sign of the features producing
the deviation from gaussianity. This quantity can be estimated
using the third centred moment of the data:
\begin{equation}
E[M_3] = \frac{1}{N_{pix}} \sum_{i=1}^{N_{pix}} (x_i - E[\overline{x}])^3
\label{skew}
\end{equation}
where $E[...]$ means that this is an estimator of the quantity inside 
brackets, $N_{pix}$ is the number of pixels of the map, $x_i$ is the
measured flux density at pixel $i$, and
$E[\overline{x}]=\frac{1}{N_{pix}} \sum _i x_i$ is the standard estimator 
for the mean of the distribution. From here, the skewness
is obtained as $Skew \equiv M_3 / \sigma^{3}$, where $\sigma$ is 
the rms of the data.
Equation (\ref{skew}) is a biased estimator of the third moment 
of the population, but for large $N_{pix}$ it converges to 
the true $M_3$ value. Assuming an underlying gaussian PDF, the variance
of this estimator (to lowest order in $1/N_{pix}$) is 
$Var(E[M_3]) = \frac{6}{N_{pix}} Var(x)^{3}$ (see \cite{kesden02}).

If the $P(D)$ is known, any moment of the distribution can be derived
from it, and in particular, the skewness can be written as
\begin{equation}
Skew = \frac{\int D^3 P(D) dD}{ \Bigg( \int D^2 P(D) dD \Bigg)^{3/2} }
\end{equation}
\item {\bf Bispectrum} of the observed map. We will concentrate here on
the quantities $B_\ell \equiv B_{\ell \ell \ell}$, which in the
case of statistical isotropy are related to the $B_3$ function defined
above as 
\begin{equation}
B_3(\ell_1 m_1, \ell_2 m_2, \ell_3 m_3) = \Bigg(
\begin{array}{ccc} \ell_1 & \ell_2 & \ell_3 \\ m_1 & m_2 & m_3
\end{array} \Bigg)
B_{\ell_1 \ell_2 \ell_3}
\end{equation}
where the $(...)$ is the Wigner $3-j$ symbol. In particular, we will
be interested in the dimensionless bispectrum, 
$I_\ell \equiv B_\ell / \cl^{3/2}$ (e.g. \cite{ferreira98}). 
The absolute value of this
quantity will be the same for any frequency, while it will 
change its sign when we are observing above or below $\lambda = 1.25$ mm.
A fast and efficient method to compute the angular bispectrum up to
$\ell \sim 100$ for 
maps on the sphere is described in \cite{komatsu02b}, and
applied to COBE data. However, for the case of small patches of 
sky (as is the case for CBI or BIMA), we can use 
the flat-sky approximation, and the estimator described 
in \cite{santos02}. 
\end{itemize}

Any of the above estimators is able to detect an excess of 
positive unresolved sources over negative ones (or viceversa).
However, the study of the $P(D)$ function is preferable to the
computation of skewness, given that it contains much more
information. Unfortunately, obtaining the $P(D)$ function for
sources/clusters from noisy data requires more integration time that
just the detection of skewness, as we will see in section \ref{sec:noise}.

\section{Source counts, P(D), skewness and bispectrum of SZ clusters}
\label{sec:general}

P(D) analysis gives much more information 
than the power spectrum about the 
sources of CMB fluctuations, and 
even than skewness or the bispectrum. 
Therefore, we will describe below the expected P(D)
functions, skewness and bispectrum 
for simulated maps of SZ clusters. It is
clear that observers should make a lot of effort to 
get all the information about P(D),
but the results which we will describe here 
will make it easier to understand
our predictions in the following subsections.

\begin{figure}
\includegraphics[width=\columnwidth]{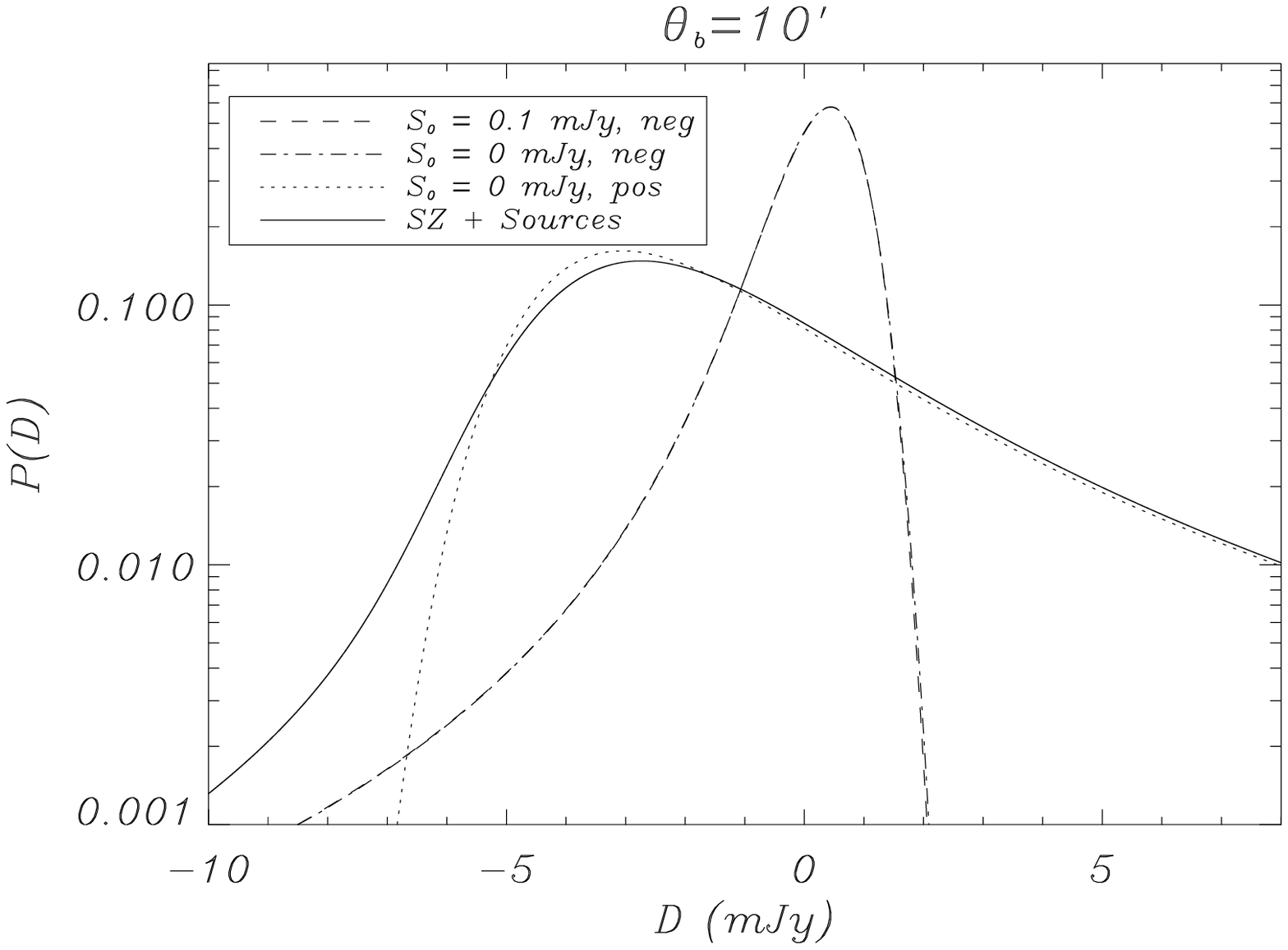}
\includegraphics[width=\columnwidth]{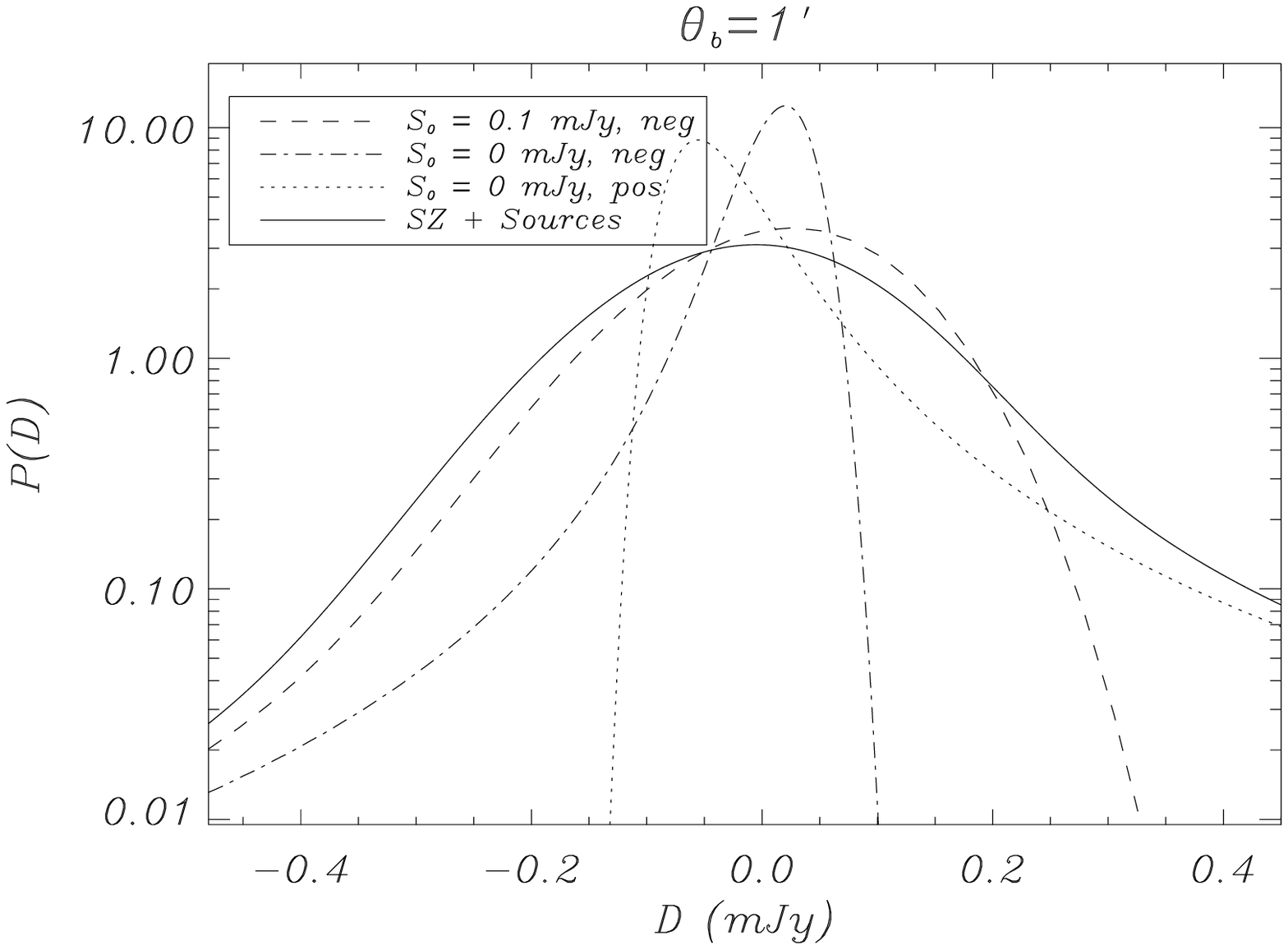}
\caption{$P(D)$ functions derived from analytical modelling of
source counts, for: 
(a) ``negative'' sources described with a power-law with parameters 
$K=1$ Jy$^{-1}$ sr$^{-1}$, $\beta=2.5$, and a low-flux cut-off
of $S_0 = 0.1$ mJy (dashed line); 
(b) same as before, but without considering the cut-off ($S_0 = 0$, dot-dashed line); 
(c) ``positive'' sources following a power law with $K = 54$ 
Jy$^{-1}$ sr$^{-1}$ and $\beta=2.15$ (dotted lines);
(d) sum of the maps of clusters (a) and sources (c), so the P(D) is the 
convolution of those two cases. 
The beam is assumed to be gaussian, with a $\theta_b$ of $10\arcmin$ 
in the upper panel, and $1\arcmin$ in the lower one. 
The deflection $D$ is refered to $\overline{D}$ in all cases, 
so all these curves give $\overline{D}=0$. 
Comparing cases (a) and (b), we can see the effect of 
a flux density cut-off on the shape of the $P(D)$ function. A cut-off
becomes of importance when we observe a low angular resolution. 
For these two values of $\theta_b$, the asymmetry introduced
by radio sources dominates, because for these values of the
spectral indices they are more numerous at a given flux, although for
the low value of $\theta_b$ (lower panel), clusters increase their relative 
contribution. 
However, figure shows that studying the $P(D)$ function provides 
information of both positive and negative contributions.}
\label{truncated}
\end{figure}

In Figure \ref{truncated} we present the results
of simple modelling of SZ thermal effect following the 
model where ``negative''
clusters are assumed to have truncated power law source 
counts and ``positive" radio sources have power law source 
counts up to very low fluxes.
In both cases it is possible to compute P(D) numerically from
equations in Section \ref{sec_pd}. 
For definiteness, we use here $S_0 = 0.1$ mJy, which roughly 
corresponds to $M_{min} = 5 \times 10^{13} M_{\odot}$. 
We are implicitly 
assuming here that no source subtraction has been carried
out on the map, so the width of the P(D) functions
is directly related to the source confusion in each case.

We consider here two cases: $\theta_b=10'$, which is roughly the angular
resolution of current experiments (like CBI), and
$\theta_b=1'$, which corresponds to the angular resolution 
of future experiments dedicated to measure clusters (ACT, AMI, AMIBA,
APEX and 8m-South Pole Telescope will have angular resolutions
of 1.7\arcmin, 1.5\arcmin, 2\arcmin, 0.8\arcmin and 1.3\arcmin, respectively).

For the beam width $\theta_b$ = 10\arcmin, the negative sources with
truncated source counts produce practically no difference with the
P(D) for the ``negative'' source counts with the same slope
extended to zero flux. This is because for this case, the 
typical flux cutoff is well below the flux which gives the
maximum contribution to the P(D). 
We see immediately that the ``positive" radio sources 
alone have completely different P(D) distribution 
than "negative" SZ-clusters.  For
the beam width $\theta_b$ = $10\arcmin$ contribution 
of positive sources in the total 
power is larger than the contribution of ``negative'' SZ clusters, 
therefore the positive wing of P(D)
is similar to the P(D) for sources only.  However, the negative 
wings are very different.

For the beam width $\theta_b$ = 1\arcmin, the negative tail of the 
P(D) function of clusters is becoming more important than in the last
case. 
The reason for this dependence is the following: 
once we have specified the 
shape of the source counts, and we define the experiment (i.e. we
specify $\theta_b$), then the P(D) function is completely defined, 
and it has the dominant contribution coming from the flux range between  
the flux at which we have one source per beam area, 
and the flux at which we expect one source every 30--40 beams. 
Thus, for smaller beam areas we are sampling lower fluxes where, 
according to figure \ref{review_scounts}, clusters become more
numerous compared with radio sources (at a given flux). 

We would also mention that if we go deep enough in $\theta_b$, we would
reach the low flux cut-off for clusters, 
and positive sources will dominate again.
The existence of a low flux cut-off 
is clearly seen in the figure for $\theta_b=1'$, 
where the shape of the $P(D)$ function 
is sensitive to the cutoff flux, and thus is completely
different from the shape that we would expect if we had no truncation
flux. Therefore, we can see that $P(D)$ at these angular resolutions 
will also give us information about the truncation in flux for 
SZ clusters, and hence, about the low mass cutoff.

In the following subsections, we will explain in more detail 
these general aspects by using simulations of SZ clusters, and
we will derive the $P(D)$ function from these simulated maps. 
Through this subsections, we will illustrate the
$P(D)$ curves for the angular resolution of $\theta_b$ = 1\arcmin.

\subsection{Modelling clusters using Press-Schechter prescription}
\label{subsec:ps}

Statistical properties of the population 
of SZ clusters has been extensively studied in the literature (see
SWH, fig.4 in that paper,  
for a recent review). For the case of the power spectrum, the 
published estimates show differences of an order of magnitude,
although these differences can be understood due to the different
scaling relations and mass ranges considered in each case. 
Therefore, we will be interested here 
in determining the qualitative behaviour
of these new quantities ($P(D)$, skewness and bispectrum). 
For this purpose, we will use a simple modelling
of clusters, based on a Press-Schechter prescription. 
If someone is interested in predictions in agreement with
hydrodynamic simulations, then it is possible to use 
refined versions of the PS formalism, as that described in
\cite{sheth99}, or in \cite{jenkins01}. 
In PS, the comoving number density of bound objects
of total mass $M$ at redshift $z$, is given by 
\begin{equation}
\frac{d n(M,z)}{dM} = -\sqrt{ \frac{2}{\pi}} \frac{\overline{\rho}}{M}
\frac{d \sigma(M,z)}{dM} \frac{\delta_c}{\sigma^2(M,z)} 
\exp \Bigg[ \frac{-\delta_c^2}{2\sigma^2(M,z)}\Bigg]
\label{ec:ps}
\end{equation}
where $\overline{\rho}$ is the mean comoving background density, 
$\sigma(M,z)$ is the variance of the linear density fluctuation field 
filtered on some mass M, 
$\overline{\rho}$ is the mean comoving background density, and
$\delta_c$ is the linear density contrast of a perturbation
that has virialized. 

In our modelling, we will assume the value of $\delta_c=1.686$ 
(see e.g. \cite{molnar00}) and the scaling of $\sigma(M,z)$ with
mass from \cite{viana99}. 
As it was discussed in section 2, it is neccesary to introduce
a mass cutoff because $dn/dM$ diverges at low masses. 
Following \cite{komatsu99}, we will use
$M_{min} = 5 \times 10^{13} h^{-1} M_\odot$, and 
$M_{max} = 5 \times 10^{15} h^{-1} M_\odot$. Changing the upper limit has
little effect on the predictions because the PS function falls
exponentially. The effect of changing the lower
limit is discussed below. 

For illustration, we have adopted 
here the 'concordance' model of \cite{ostriker95},
which has $\Omega_{tot}=1$, with $\Omega_m=0.3$ and 
$\Omega_\Lambda = 0.7$, $n=1$, $h=0.67$ (where 
$H_0 = 100 h$ km s$^{-1}$ Mpc$^{-1}$), and the normalisation 
$\sigma_8 = 0.9$. For the scaling relation of the gas
temperature with the mass and the redshift of the cluster, we
use the scaling pointed out in section 2, with
\begin{equation}
k T_{gas} = 7.75 
\Bigg( \frac{M}{10^{15} h^{-1} M_\odot} \Bigg)^{2/3} (1+z) \; keV
\end{equation}
and we will adopt a $\beta$-model for the intra-cluster
gas with $\beta=2/3$. The relationship between the virial radius ($r_v$) 
and the core radius ($r_c$), and their scalings with mass 
and redshift $z$ are those obtained assuming spherical 
collapse (see e.g.\citealt{atrio99,molnar00}), and an entropy-driven
model with $\epsilon=0$ for the core evolution (e.g. \citealt{bower97}).
The parameters of our cluster model are $r_v(z=0) = 1.3 h^{-1}$ Mpc, 
$r_c(z=0) = 0.13 h^{-1}$ Mpc, and $n_c(r=0,z=0) = 2 \times 10^{-3}$
cm$^{-3}$. In Appendix \ref{app1} we study the dependence of
our results on the scaling assumptions, as well as 
the dependence on the normalisation $\sigma_8$.

We then generate 15 
realisations of a $1^\circ$-side map of SZ clusters 
using a Press-Schechter law, and a single class of clusters.
We have chosen these values to
allow direct comparison with the results of 
of the hydrodynamic simulations of the thermal SZ effect
described in SWH\footnote{These simulations are quoted in 
that paper as 134A, and have been corrected for the h factor.},  
which correspond to 15 maps of the same previous size,  
of the Comptonization parameter $y$ due
to structure in the same $\Lambda$CDM model 
between $z=0$ and $z=19$. 
SWH computed the angular power-spectrum 
of the SZ effect from these maps, 
as well as the source counts of thermal SZ sources.
The mean Comptonization parameter in our 15 Press-Schechter realisations
is $<y> = 2.1 \times 10^{-6}$.

\subsection{Source counts for clusters following a Press-Schechter prescription}
\label{sec_scounts}
\begin{figure}
\includegraphics[width=\columnwidth]{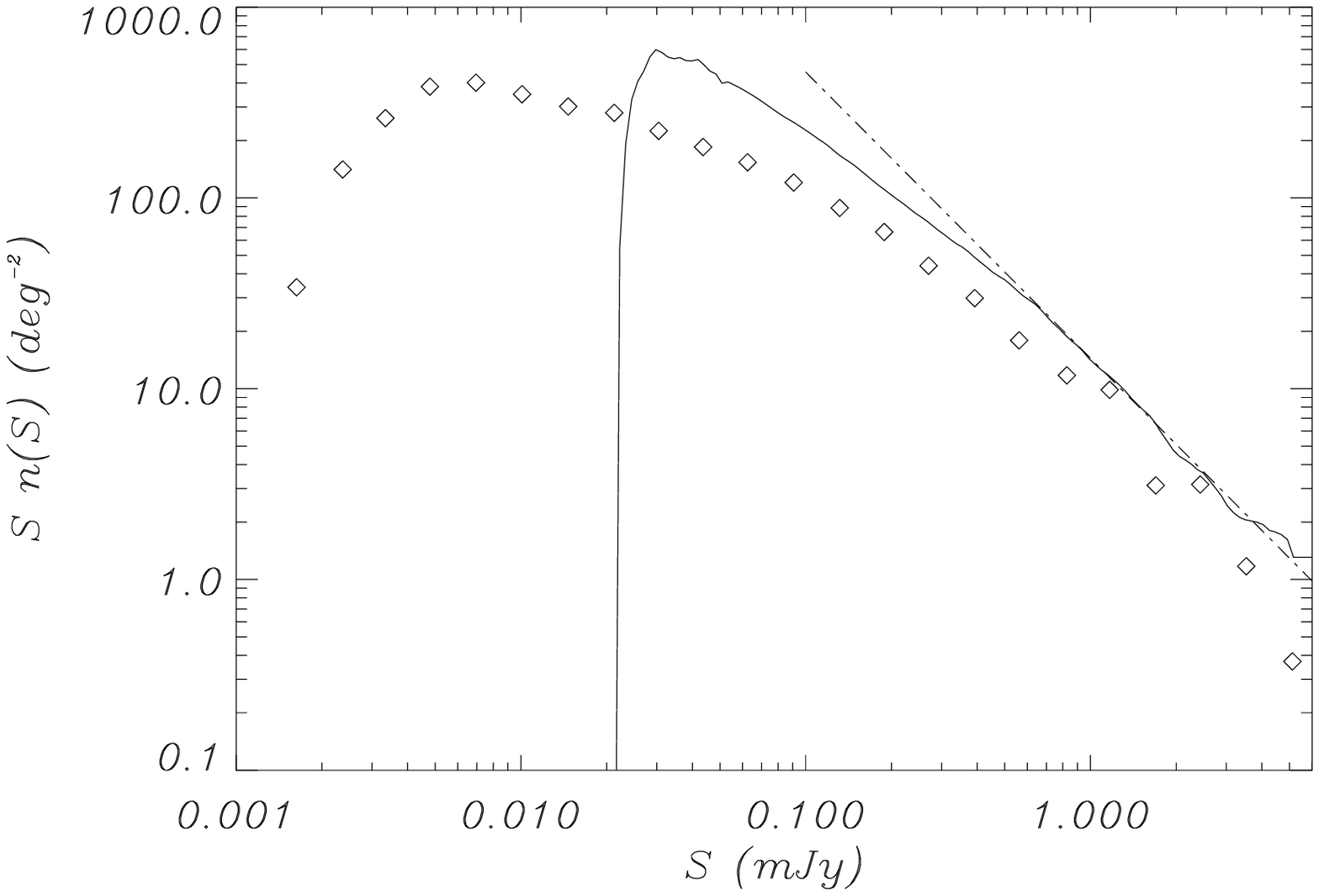}
\includegraphics[width=\columnwidth]{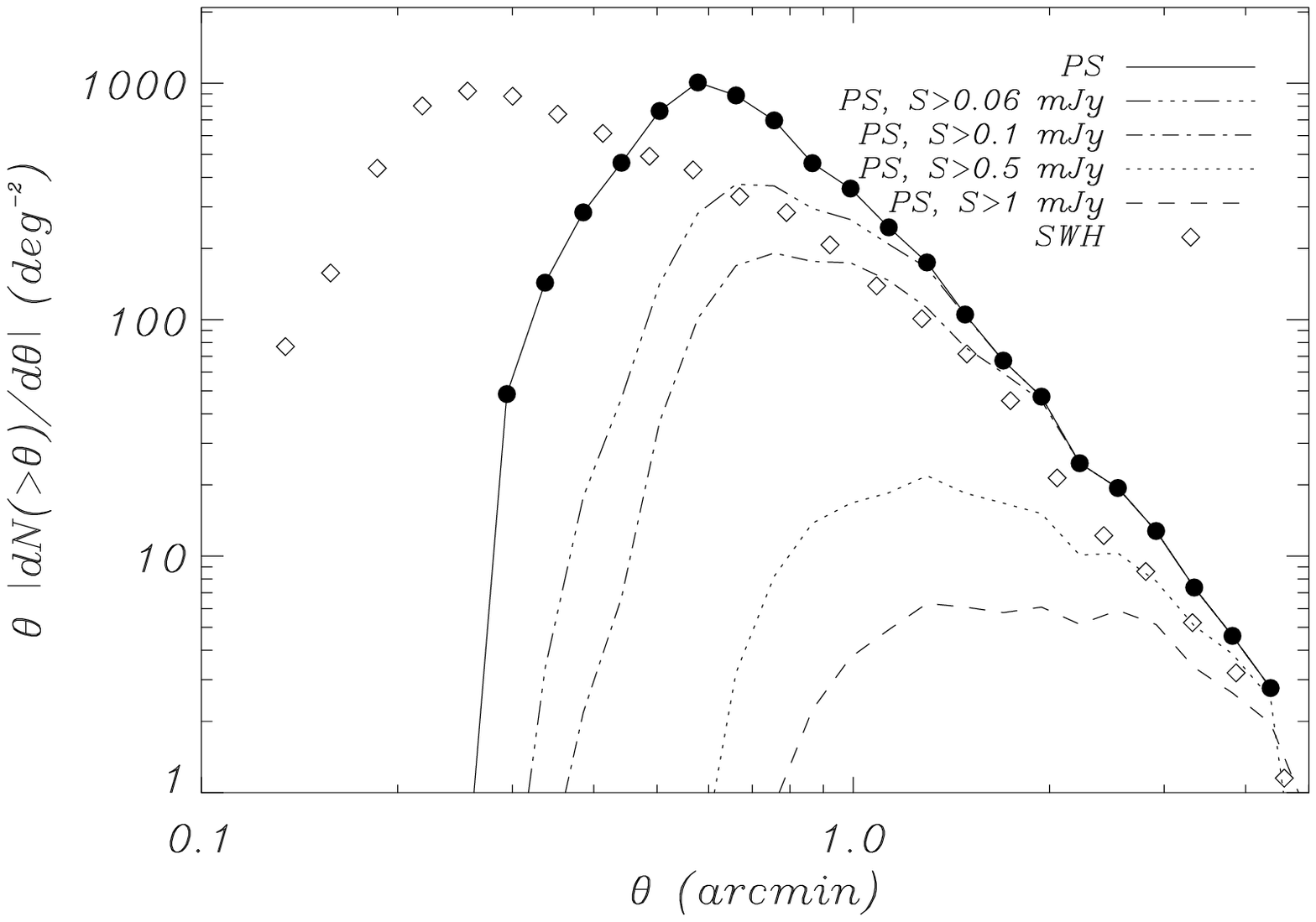}
\caption{Top: Differential source counts at 30 GHz from the 15 simulated maps
following the PS prescription (solid line). As expected, the behaviour
at fluxes $S \ga 1$ mJy is well described by a power-law with
$\beta=2.5$, and we have a strong cutoff at low fluxes, corresponding
to the mass cutoff $M_{min} = 5 \times 10^{13} h^{-1} M_\odot$.   
In this case, this power law is described by 
$n(S) \approx 1 (S/1Jy)^{-2.5}$ sr$^{-1}$ Jy$^{-1}$. For 
comparison, we also show (open squares) the results from SWH. 
These curves are also presented in Figure 4, in the context of 
the source counts for radio sources. 
Bottom: Differential source counts from the 15 maps 
following a PS prescription, 
as a function of the source angular radius. We again use the 
coordinates $|\theta dN/d\theta|$, where we explicitly see the
typical size of the most common sources. 
We present the source counts for the whole dataset (solid circles 
with solid line), and considering only sources with a flux at $30~GHz$ 
greater than a certain threshold. We can see that the most intense 
sources are also the largest ones. 
We also include the results from SWH (open squares) showing that
the hydrodynamical simulations have a larger amount 
of small sources, and a wider curve. }
\label{sourcecounts_press}
\end{figure}

When generating the previous 15 realisations, we keep 
the total flux of all the simulated clusters, as well as 
the core and virial radius, so 
we are able to find out the source counts for our
maps. These source counts are presented in 
figure \ref{sourcecounts_press}. In the top panel,
we show these counts as a function of the flux, 
for the frequency of 30 GHz. 
As expected from the result of \cite{deluca95}, the slope 
of the source counts at fluxes greater than $\sim 1$ mJy 
corresponds to an Euclidean power-law ($\beta=2.5$).
However, the amplitude of our curve is 
different from theirs. Our source counts are
well fitted by $n(S) \approx 1 (S/1Jy)^{-2.5}$ sr$^{-1}$ Jy$^{-1}$
at 30 GHz, and by 
$n(S) \approx 28 (S/1Jy)^{-2.5}$ sr$^{-1}$ Jy$^{-1}$ at 150 GHz, while
their source counts at 150 GHz is  
$n(S) \approx 8.6 (S/1Jy)^{-2.5}$ sr$^{-1}$ Jy$^{-1}$.
This difference is due to the fact that they introduce an extra 
normalisation factor in the PS mass function to fit 
the observed mass function in X-rays, and that changes the total
amplitude. 

Our prediction for the source counts compares
well with that of SWH (open squares), although it is clear
that the hydrodynamical simulations do not show an strong
cut-off in flux, but they have
much more 'small' (low mass) objects. This excess of small objects respect
to the PS result will be responsible for a larger power at small
angular scales, as it has been discussed in SWH. 
We can see better this point in the bottom panel of figure 
\ref{sourcecounts_press}, where we show the source counts as
a function of the angular radius of the cluster. 
We define here the angular radius as the radius which 
contains half of the total flux of the cluster. 
It is interesting to see that 
if we only consider clusters with a flux density greater
than a given value, then the most brightest clusters
in the PS maps are the largest ones in size. 
From this figure, we immediately see that for the angular
resolutions of the upcoming SZ experiments ($\theta_b \approx
1\arcmin$), more than 90\% of the clusters will be unresolved
objects (or even more, if we look at the SWH source counts).  
\begin{figure}
\includegraphics[width=\columnwidth]{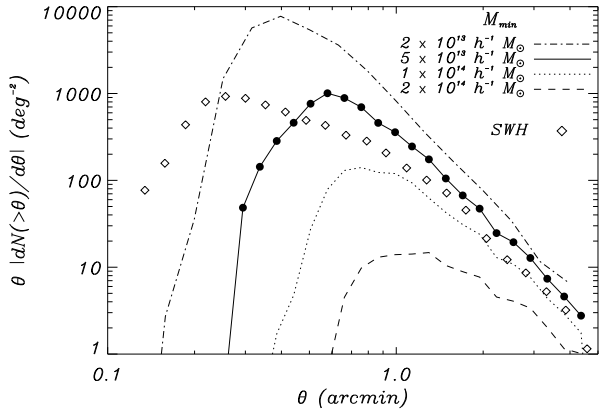}
\caption{Dependence of the differential source counts (as a function
of the source angular radius) with the selected minimum mass 
for the Press-Schechter mass function. 
Curves for $M_{min} \ge 5\times10^{13} h^{-1} M_\odot$ 
have been obtained from the 15 PS realisations used 
in figure \ref{sourcecounts_press}. The curve for 
$M_{min} = 2\times10^{13} h^{-1} M_\odot$ has been derived, just for 
illustration purposes, from an small realisation 
of 3 square degrees using this new low mass cutoff. We also show 
the source counts from SWH. }
\label{mmin_press}
\end{figure}
We would like to stress that the coordinates that we are 
using to plot the source counts are specially suitable, because
they directly give us the total amount of sources at a 
given angular size/flux interval. In these coordinates, we clearly
see that our PS modelling with $M_{min} = 5 \times 10^{13} h^{-1} M_\odot$ 
produces a peak of objects with fluxes around 0.04 mJy
and sizes around $0.6'$, and below these quantities we have
an strong cutoff (both in size and flux), as expected 
from the discussion on section 2. 
It is important to note that for the angular size of $\theta=0.6'$ (close to
the resolution of APEX), we have around $\sim 1000$ sources 
per square degree, so given that 
the number of beam areas inside a square degree is $\sim 10^4$ for
$\theta_b = 0.6\arcmin$, we have that at this angular 
scale we will obtain 1 source every
10 beams. These values are inside the confusion
(e.g. \cite{scheuer57}), so the $P(D)$ function is a suitable tool to 
study them. For larger values of $\theta_b$ (e.g. $1.3\arcmin$ for the
South Pole Telescope, or $\sim 10\arcmin$ for present day experiments), 
we will be even more inside confusion due to 
these small clusters.  

The natural question which follows now is 
to explore if this conclusion 
holds for different values of $M_{min}$.
For this purpose, we should study the dependence of the source 
counts as a function of the angular size using different 
values for $M_{min}$ in the simulations. This can be
done using our simulations by selecting clusters with
masses above the new threshold. 
This is done in figure \ref{mmin_press}. As
we would expect, the $\theta$ value at which we have the
maximum amount of sources ($\theta_{peak}$) roughly scales as $r_v$. 
Fitting our data, we have:
\[
\theta_{peak} \approx 0.6\arcmin 
\Bigg( \frac{M_{min}}{5 \times 10^{13} h^{-1} M_\odot}\Bigg)^{1/3}
\]
We can also infer from here the total amount of sources (per square
degree) that we have at that angular scale ($N_{peak}$), obtaining 
\[
N_{peak} \approx 1000 
\Bigg( \frac{M_{min}}{5 \times 10^{13} h^{-1} M_\odot}\Bigg)^{-2.8}
\qquad deg^{-2}
\]
Using these two last equations, we conclude that for the angular resolution 
$\theta_{peak}$, we expect 1 source every $10 \times (M_{min}/5\times10^{13}
h^{-1} M_\odot)^{2.13}$, which is smaller than 30 only for
$M_{min} < 8.4 \times 10^{13} h^{-1} M_\odot$.  

Finally, we can see from figure \ref{mmin_press} that all
future experiments (with $\theta_b \approx 1\arcmin$) will
be confusion limited at this angular resolution 
if $M_{min} \la 1.1 \times 10^{14} h^{-1} M_\odot$. 
For all of them, a $P(D)$ analysis
will be suitable, and in addition, this analysis will provide information 
about the low mass cutoff.

\subsection{Predictions of the Press-Schechter approximation for P(D)
function, skewness and bispectrum} 
\label{sec:7.3}

We will be now interested in obtaining the qualitative behaviour of
the $P(D)$ function, the skewness 
and the bispectrum for SZ clusters following PS mass function.
Therefore, and through this section and the next one, we will
restrict ourselves to the computed simulations, with 
$M_{min} = 5 \times 10^{13} h^{-1} M_\odot$.

\begin{figure}
\includegraphics[width=\columnwidth]{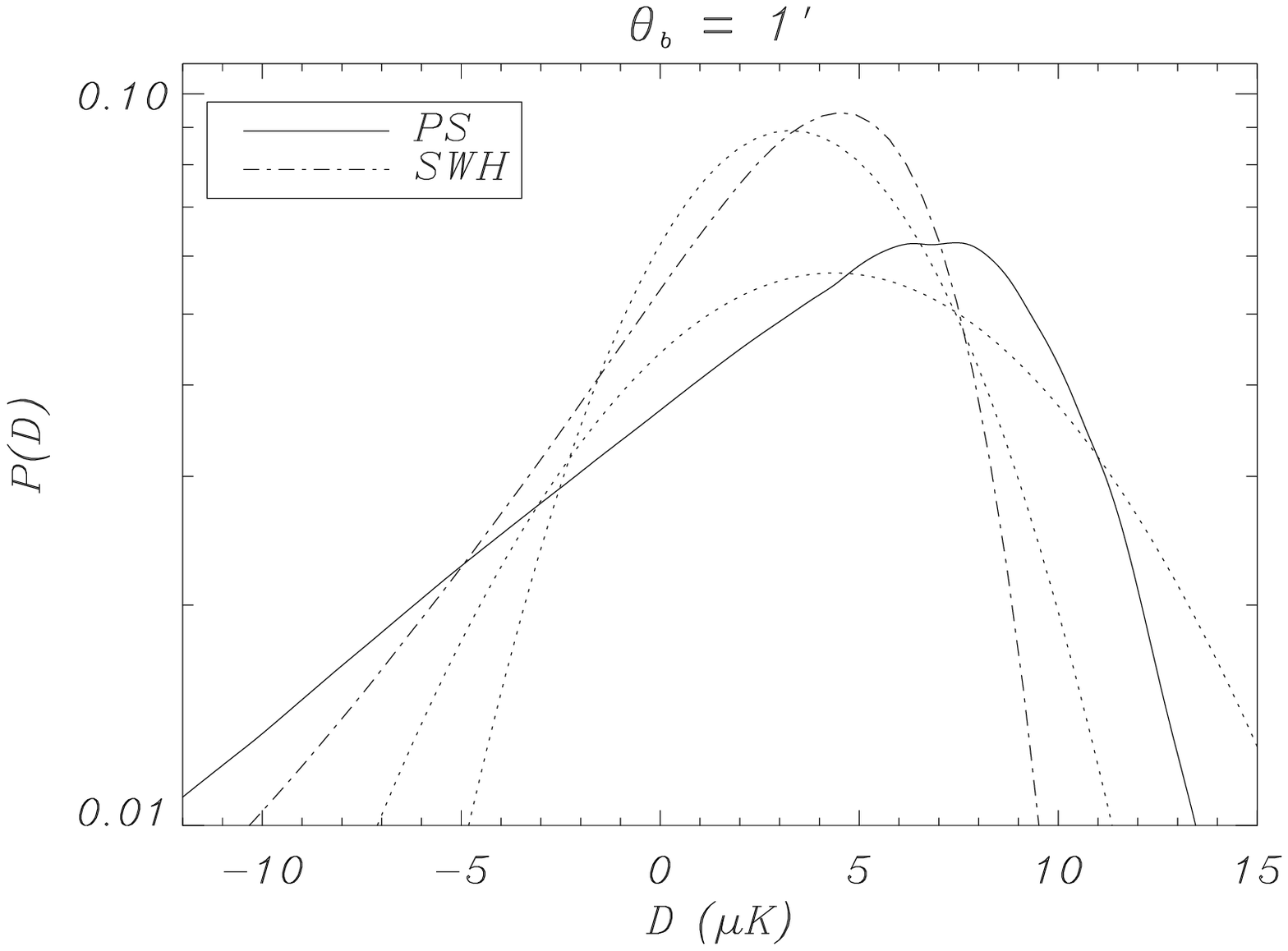}
\includegraphics[width=\columnwidth]{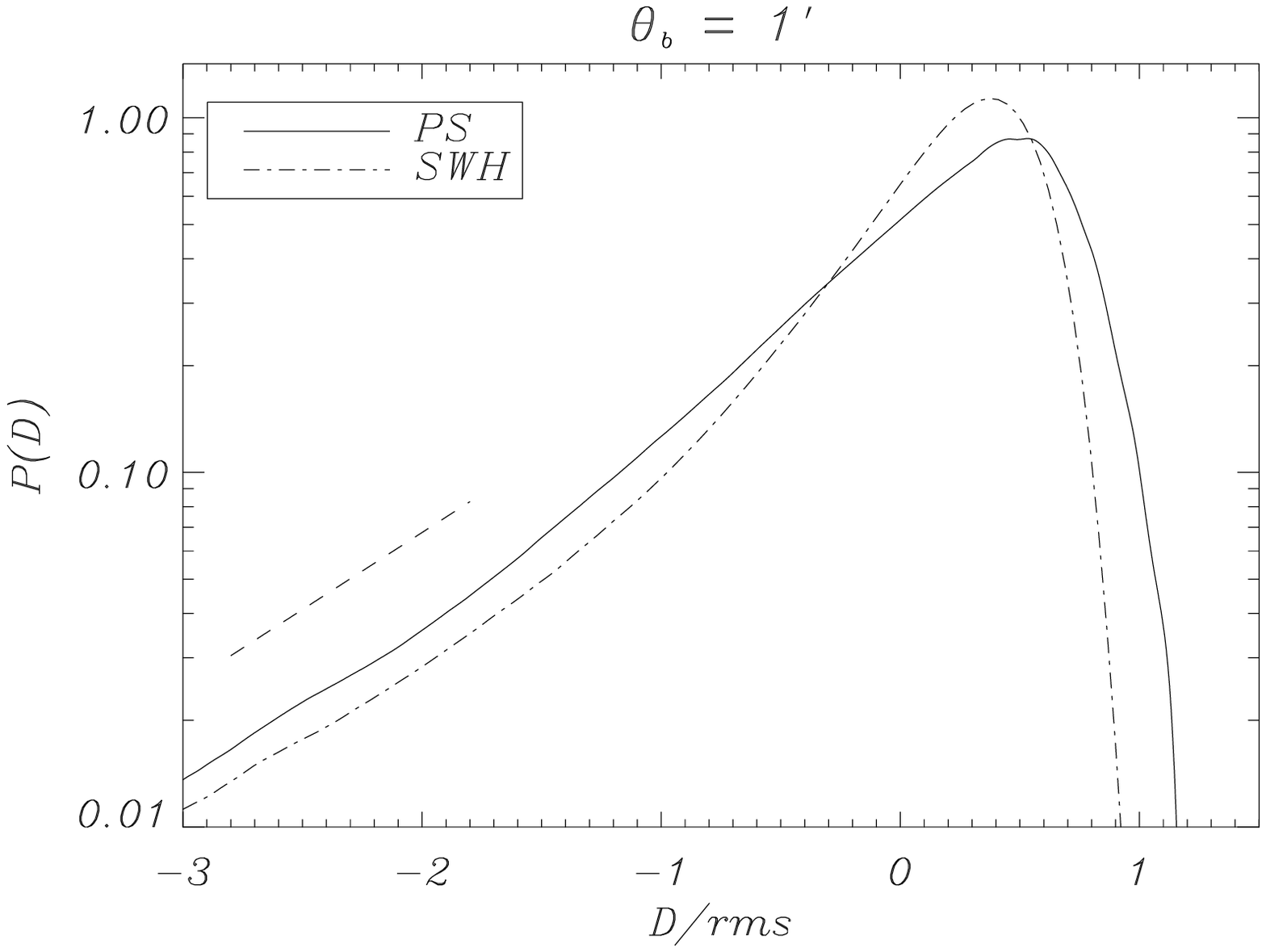}
\caption{First panel: $P(D)$ function for a single map of 
SZ clusters following a Press-Schechter prescription (solid 
line), and for an hydrodynamic simulation from 
Springel et al. 2001 (dot-dashed line). 
Both maps are $1^\circ$ on a side, 
with a pixel size of $0.12'$; they have been smoothed with 
a gaussian beam of $\theta_b = 1'$, and 
are given for the Rayleigh-Jeans region of the spectrum. 
These curves can be transformed into PDF curves for y parameter
by the transformation $D=-2y T_{cmb}$, so the $P(y)$ function
will exhibit a positive tail. The best gaussian
fit for both curves is shown with dotted lines, being 
their widths $\sigma = 6.1 \mu K$ 
and $\sigma = 3.9 \mu K$, respectively. 
Second panel: same functions but computed dividing the map
by the rms prior to the $P(D)$ computation. 
When rescaling by the rms, both curves show the same 
qualitative behaviour, with a nearly linear tail 
in these coordinates (log P(D) vs D). Long dashed line
shows a straight line in these coordinates for a slope
of unity. }
\label{pd_examples}
\end{figure}

\begin{itemize}
\item {\bf P(D) function}. In figure \ref{pd_examples} we
present the P(D) analysis of the maps generated 
in both PS and hydrodynamical simulations using a beam size of $\theta_b = 1'$. 
Usually, the maps of SZ thermal effect are presented as maps
of y-parameter. However, observers are seeing the brightness distribution 
on the sky.  Therefore, we present deviations on the map for the 
Rayleigh-Jeans temperature.  The temperature in the Rayleigh-Jeans region
of the spectrum is connected 
with the y-parameter by the simple relation  
$\Delta D = -2 y T_{cmb}$ \citep{zeldovich69}, 
but the P(D) graphs are valid for any frequency: we 
only need to recompute D and P(D) according to 
formula (\ref{P_D_scaling}), but using $\hat{f}$ instead of $g$. 
Therefore, our plots for $P(D)$ are easy to convert into 
PDF for $y$ parameter. This opens the 
possibility for comparing our results for the PS approximation with those of
other authors (e.g. \cite{seljak01}).

The P(D) curves for both PS and hydrodynamic simulations
show very broad non gaussian negative wings. To demonstrate
this, we are presenting on figure \ref{pd_examples}a 
the best fit gaussian curves for both P(D) functions.  
It should be noted that the width of the best fit gaussian 
($\sigma = 6.1 \mu$K and $\sigma = 3.9 \mu$K, respectively)
is smaller than the rms of the maps
($rms(PS) = 13.9 \mu$K and $rms(SWH) = 12.0 \mu$K), 
because the fitting reproduces
the central part of the curve, while the rms is larger because
of the negative tail. If these curves were gaussian, both the 
$rms$ and the $\sigma$ value would be the same. 
The two P(D) curves have sufficient differences, specially
in the slope of the negative tail. 
Both of them fall more rapidly than the gaussian fit, and
the PS prescription produces an slightly broader curve (because
of the slightly larger rms).
We will not discuss these differences 
in detail because it is well known from
previous studies that these two methods are giving significantly
different results. However, 
it's impressive that if we include the 'normalisation' $D/rms$,
dividing the maps by their rms before the $P(D)$ computation,  
the resulting $P(D)$ curves are similar in shape, showing a 
practically identical slope in the intermediate asymptotic 
region (see lower panel on fig. \ref{pd_examples}).

\item {\bf Intermediate asymptotic for the P(D) function}.
In a broad region, the normalised P(D) 
curve could be described by a simple analytical
formula: the powerful left wing 
of the P(D) distribution is close to the straight line 
in the coordinates we used in
our figure ($\log (P(D))$ versus $D$). 
Therefore, we see that $P(D) \sim exp (aD)$  
in a sufficiently broad region of $-3 < D/rms < 0$. 
This is an impressive feature of P(D) which 
might help to identify SZ clusters 
from the noise of the observed maps, given that this behaviour
is completely different from a gaussian. 
We should stress that both the Press-Schechter
approximation and the hydrodynamic simulations produce practically the same  
"{\sl intermediate asymptotic}". 
In addition, this 'a' factor has an strong dependence with the
normalisation $\sigma_8$, as we show in the Appendix \ref{app1}, where we find
that $a \sim \sigma_8^{4.5} \theta_b^{-0.2}$. 
The numerical coefficients for this expression 
are also shown in that appendix.

It's well known that the $P(D)$ distribution for 
sources with power law source 
counts $n(S)=KS^{-\beta}$ gives a simple power law asymptotic 
$P(D) \sim D^{-\beta}$ for high values of the deflection, 
because it that case 
the distribution becomes dominated by strong, 
well-resolved sources. 
To look for this asymptotic, we presented the graph of the derivative 
$d \log (P(D))/ d \log(-D)$ versus $D/rms$ in Figure \ref{dlogP_dlogD}.  
In these coordinates the asymptotic 
$P(D) \sim D^{-\beta}$ will be the horizontal line with 
d log (P(D))/ d log(-D) = -2.5.
Figure \ref{dlogP_dlogD} shows that  
$d \log (P(D))/ d \log(-D)  \approx aD + B$ in the range 
$-2.5 < D/rms < 0$, which corresponds
exactly to the intermediate asymptotic 
described above P(D)$\sim \exp(aD)$. The limiting value of $-\beta =
-2.5$ is reached only for very large deviations.  

\begin{figure}
\includegraphics[width=\columnwidth]{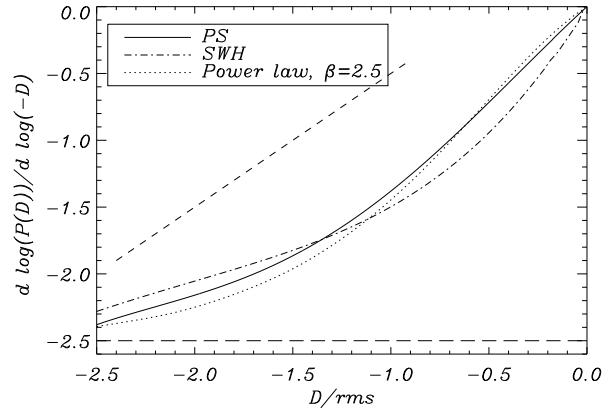}
\caption{Slope of the negative tail of the 
$P(D)$ function for SZ clusters following Press-Schechter prescription.
It is shown the values for $d \log P(D) / d \log(-D)$, for the
average of the 15 realisations of SZ clusters, in the Rayleigh-Jeans region
of the spectrum, and using a gaussian beam with $\theta_b = 1'$ (solid
line). The dot-dashed line shows the same 
calculation but for the SWH simulations, while the dotted-line 
shows the analytic calculation for a power-law source counts 
with $n(S) = (S/1Jy)^{-2.5}$ sr$^{-1}$ Jy$^{-1}$. 
The horizontal axis is plotted in
terms of the rms of the map. The slope of this tail tends 
to the value of -2.5 (horizontal long-dashed line), 
as expected for a power-law differential
source counts with $\beta=2.5$. However, in this 'intermediate' region,
the $P(D)$ curve follows a nearly exponential law ($P(D) \propto 
\exp(aD/rms)$), with $a \approx 1.0$. This behaviour is 
completely different from that of a gaussian function. }
\label{dlogP_dlogD}
\end{figure}

Finally, it is important to notice that this 'intermediate
asymptotic' behaviour can be also obtained semi-analytically, by  
using a power-law source counts with $\beta=2.5$ inside 
equation (\ref{ec:pd}), as it is
also shown in figure \ref{dlogP_dlogD}. Therefore, 
this intermediate asymptotic is closely 
related with the PS prescription, which is the 
ultimate responsible
of the Euclidean type power-law behaviour 
of the source counts for clusters ($N \sim S^{-3/2}$). 

\item {\bf Skewness}. 
\begin{figure}
\includegraphics[width=\columnwidth]{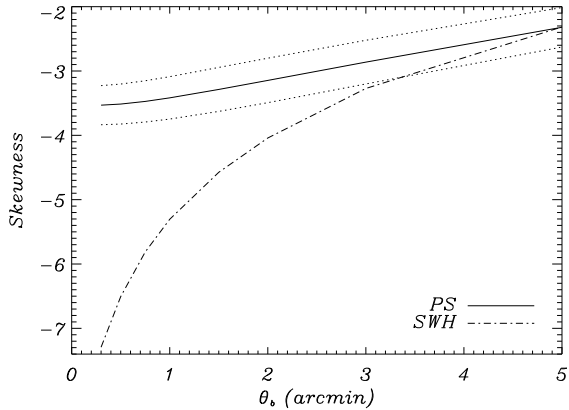}
\caption{Negative skewness due to SZ clusters as a function of the 
angular scale. These values have been obtained from 
15 (1\degr -side) simulated maps of the thermal 
SZ effect using a Press-Schechter
prescription (see details in text).  
For each angular scale, the skewness is obtained
after convolving the map with a gaussian beam of FWHM equal to 
$\theta_b$. The solid line 
shows the average value of the skewness over these 15 realisations,
while dotted lines show the 1-sigma error from this
ensemble of maps. The dot-dashed line shows the same result, 
but obtained from the 15 maps of SWH. The hydrodynamic simulations
show an excess of skewness with respect to the PS modelling at
small angular scales. }
\label{skew_vs_theta}
\end{figure}
 We investigate the dependence of the skewness introduced by
SZ clusters as a function of angular scale, both in real and
Fourier spaces. In the real space, to obtain the
skewness at a certain scale, we smooth
the SZ map using a gaussian filter of FWHM $\theta_b$, and measure
the skewness on the smoothed map. The results for modelling 
using Press-Schechter approximation are shown in
Figure \ref{skew_vs_theta}. Error bars correspond to 
the field-to-field variance from this ensemble of maps. 
In the same figure, a dot-dashed line shows the
same calculation, but for the hydrodynamic simulations of
SWH. It can be seen that the shape of the curve 
and the values from these two panels are similar for beam sizes
larger than $\theta_b \sim 2'$, although for very small angular
scales, the hydrodynamical simulations produce a larger
skewness, suggesting that it 
may be even easier to detect. This discrepancy at low angular
resolution between PS and hydrodynamical simulations
has been already discussed by many authors when predicting
the power spectrum, as 
we discuss in the following item. 
\item {\bf Bispectrum}. 
We are presenting the 'skewness' in Fourier space by computing the
dimensionless bispectrum ($I_\ell$), using the 
estimator described in \cite{santos02}.
For completeness, we also
derive the power spectrum ($\cl$), although this computation 
for the case of a Press-Schechter prescription has
been done by several authors (e.g. \citealt{atrio99,molnar00}).
Here, we explicitly shown that SZ clusters provide a negative
contribution to the bispectrum. 
Similar curves to the ones presented here for 
the skewness and bispectrum can be found in \cite{cooray00}.
The first panel on Figure \ref{bispectrum_press} shows the power
spectrum results for the PS case, in the Rayleigh-Jeans region
of the spectrum. To allow comparison, it is also included an 
standard $\Lambda$CDM power spectrum from primordial fluctuations, and the 
measurements from CBI (\cite{mason02}) and BIMA (\cite{dawson02}). 
The dot-dashed line shows the result from SWH, which has 
a similar amplitude to our PS modelling, although it peaks
at larger $\ell$ with a wider shape. This qualitative behaviour is
common to all the hydrodynamical simulations (see SWH for a review of
the recent predictions), so they have more power at larger angular
scales.

\begin{figure}
\includegraphics[width=\columnwidth]{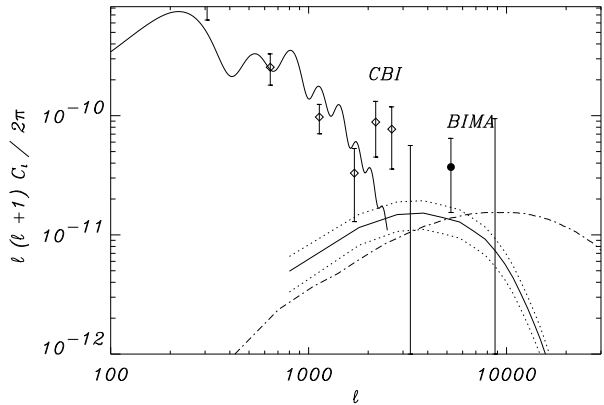}
\includegraphics[width=\columnwidth]{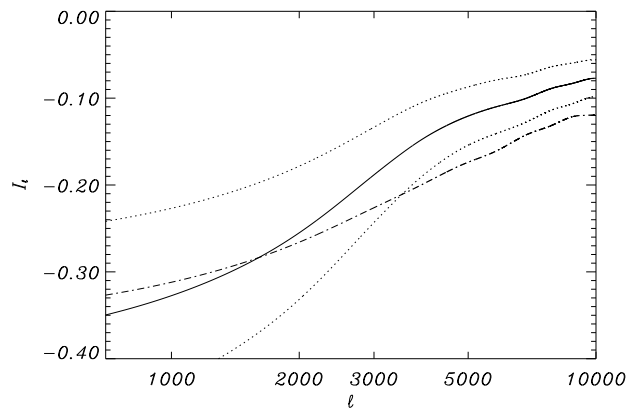}
\caption{Power spectrum and bispectrum for the thermal SZ effect
in the Rayleigh-Jeans region of the spectrum, where the
SZ clusters have the same spectrum as the primordial 
fluctuations (but are negative). First panel: Angular power spectrum (in
$(\Delta T/T)^2$ units) for the
SZ effect averaged over 15 simulations using the 
Press-Schechter prescription. Dotted lines shows the 1-sigma
field-to-field dispersion. It is also shown a standard $\Lambda$CDM
model, and the reported measurements of CBI (Mason et al. 2002, open
squares), and BIMA (Dawson et al. 2002, filled circles). 
The dot-dashed line shows the result
using the 15 hydrodynamic simulations of SWH.
Second panel: Angular bispectrum ($I_\ell = B_\ell / \cl^{3/2}$) for 
the same 15 simulations using PS prescription. Again, the
dotted lines shows the 1-sigma
field-to-field dispersion, and the dot-dashed line the result
using the 15 maps of SWH.
As expected, 'skewness' in Fourier space is also negative in this
frequency range, but smaller than in real space. 
Both in the power and in the skewness, the hydrodynamic simulations 
show an excess of signal (in absolute value) respect to the PS modelling at 
large $\ell$. }
\label{bispectrum_press}
\end{figure}

The second panel in Figure \ref{bispectrum_press} shows the 
angular bispectrum for our PS modelling. As expected, 
'skewness' both in real and Fourier space is negative, but 
the non-gaussianity in Fourier space is smaller than in real 
space. When comparing our results with those using the SWH
simulations, we find again that at small angular scales (high
$\ell$) the hydrodynamical simulations produce a larger
skewness. 
\end{itemize}

\section{Role of Radio Sources}
\label{sec:roleofradiosources}

We discuss now the influence of radio sources on the 
observed P(D), skewness and bispectrum. 
The formalism to describe the confusion noise
introduced in a map due to radio sources is widely known, 
and has been already presented in section \ref{sec_pd_confusion}. 
This formalism is easy to extend to any moment of the 
observed map, in particular the skewness.
Hence, if the population of sources at our observing 
frequency is described by the differential
source counts $n(S) = K (S/1Jy)^{-\beta}$ sr$^{-1}$ Jy$^{-1}$, then
the confusion 'skewness' on our
map is given by 
\begin{equation}
Skew_c = \frac{<s^3>}{<s^2>^{3/2}} = \frac{(3-\beta)^{3/2}}{4-\beta}
(K \Omega_e)^{-1/2} s_c^{\frac{\beta-1}{2}}
\label{skewc}
\end{equation}
Note that this equation, with a minus sign, 
is also valid for negative clusters following a power-law source counts. 
We see that the contribution of radio sources 
to the skewness will depend on our source subtraction threshold,
$s_c$, i.e. on the most brightest sources remaining in the map. 
In order to distinguish the signal coming from SZ clusters,
it is necessary to decrease these quantities below the level of
the SZ signal. 

It is obvious that 
when adding two maps (sources and SZ clusters in this case), both the 
power ($<D^2>$) and the third-moment ($<D^3>$) of the resulting 
distribution are the sum of the individual quantities, {\sl assuming 
uncorrelated maps}. Let $\sigma_1$ and $\sigma_2$ be the 
rms (power) of each of these two families of sources, and
let $Skew_1$ and $Skew_2$ be the (dimensionless) skewness for each one 
of them. The rms of the combined map is $\sqrt{\sigma_1^2 + \sigma_2 ^2}$,
and the skewness is
\begin{equation}
Skew \equiv  \frac{ (Skew_1) ~ \sigma_1^3 + 
(Skew_2) ~ \sigma_2 ^3}{ (\sigma_1^2 + 
\sigma_2 ^2)^{3/2}}
\label{sum_skew}
\end{equation}
Combining these two last equations, we can infer the required 
flux threshold for the source subtraction in order to have
an overall negative skewness in a map containing sources and 
clusters. However, it
is also well-known that using a single map we are unable to
subtract sources down to an arbitrary flux level, because
of the intrinsic confusion noise introduced by sources.
As we have seen in section \ref{sec_pd}, the minimum subtraction 
threshold is usually taken to be $s_c = q \sigma_c$, with $q=3-5$. 
Inserting this condition in equation (\ref{skewc}),
we have 
\begin{equation}
Skew = q (3-\beta)/(4-\beta)
\end{equation}
Using $q=5$ in this equation, and in the equation 
for the confusion noise (\ref{ec:sigmacq}), we obtain 
the minimum contribution of radio-sources
to any experiment which does not consider any source
subtraction strategy. 
In the particular case of $\beta \sim 2$, which is 
the case for the observed radio sources at 30 GHz,
we have $Skew = q/2$, so the minimum skewness
due to sources that we expect without source
subtraction is $\sim 1.5-2.5$ for $q=3-5$, and the minimum
confusion noise
\[
\sigma_c = 2.86 (q/3) (K/90 Jy^{-1} sr^{-1} ) (\theta_b/10')^2 mJy
\]
If we now convert this into temperature using the
Rayleigh-Jeans approximation, we 
obtain 
\[
\sigma_c = 10.8 (q/3) (K_{30GHz}/90 Jy^{-1} sr^{-1} ) (\nu/30 GHz)^{-2}\mu K
\]
which is independent of the beam size because $\beta=2$. 
Observing at a single frequency and with a single instrument,
allows us to go down to q=3 at the most. If we want to go 
deeper, we need to use information from an instrument 
with a better angular resolution to decrease 
the effective value of $q$, as we see from 
equation (\ref{ec:sigmacq}).

\begin{figure}
\includegraphics[width=\columnwidth]{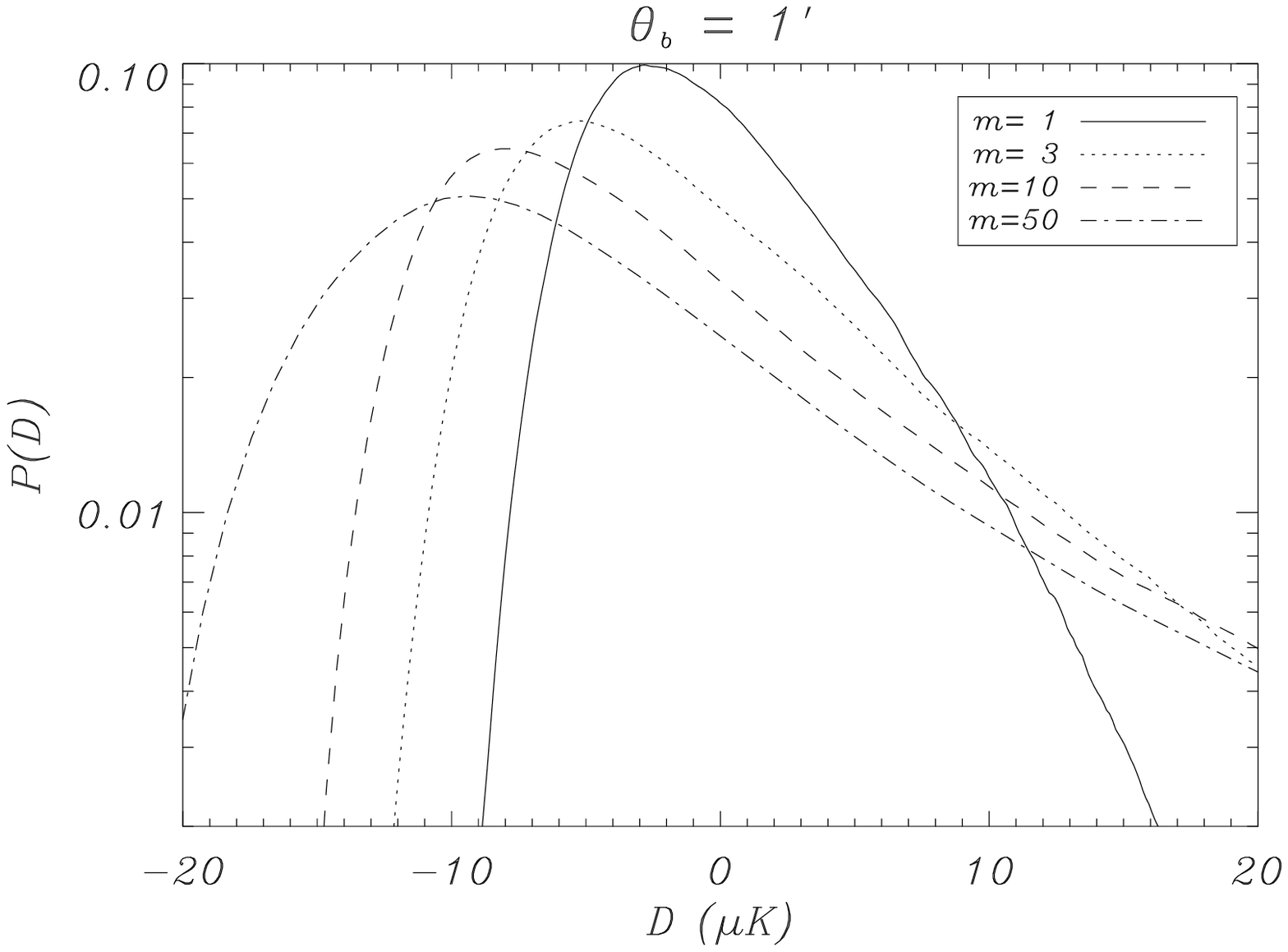}
\includegraphics[width=\columnwidth]{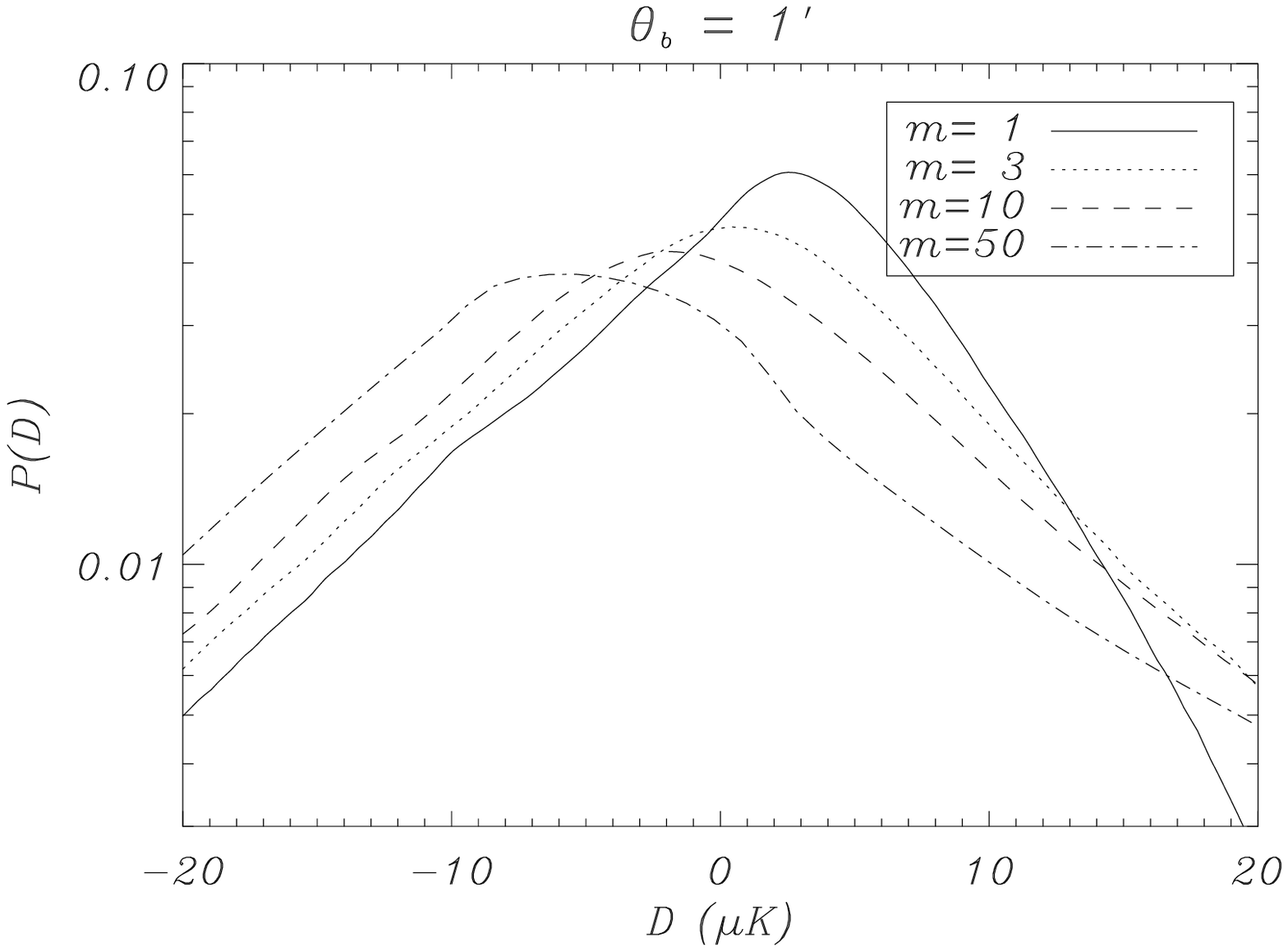}
\caption{Top: Effect of the source subtraction on the P(D) function
of sources. We parameterise the source subtraction limit as  
$s_c = m ~ \sigma_c(q=5)$, where $\sigma_c(q=5)= 16.6 \mu K$ in
this particular case. Here, 
we consider the following values for $m$: 1, 3, 10 and 50. 
All these calculations correspond to radio sources
described by parameters $K=92$ sr$^{-1}$ Jy$^{-1}$ and $\beta=2$, and
using a gaussian beam of $\theta_b=1'$. 
Bottom: Same as in the top panel, but adding a realisation
of SZ clusters following a PS prescription. In all cases the negative
tail is visible, but the asymmetry and the skewness are positive for
$m$ values greater than 3, so the detection of the 
SZ component will require a $P(D)$ analysis 
if we do not consider a source subtraction strategy. Simply
measuring the skewness is not enough to see the negative contribution.}
\label{pd_sources}
\end{figure}

We illustrate in figure \ref{pd_sources} 
how a source subtraction technique affects the observed $P(D)$
function of a map containing sources and SZ clusters. 
Radio sources are modelled here with a power-law source counts
with parameters $K=92$ Jy$^{-1}$ sr$^{-1}$ and $\beta=2$.  
We have considered different source subtraction thresholds, parameterised in
terms of $m \times \sigma_c(q=5)$. Without source subtraction at all,
we can not decrease $m$ below 5. 
The main result is that, as we would expect, 
for small values of $m$ we delete the strong
positive deviations, so $P(D)$ becomes more narrow and permits
to look for a negative tail connected with SZ clusters. 
In particular, we see that the negative tail is always 
visible, but the asymmetry and the skewness are positive 
for $m>3$, so a simple analysis
of the skewness will not be able to detect a negative
contribution if we do not subtract radio sources from the map.

\begin{figure}
\includegraphics[width=\columnwidth]{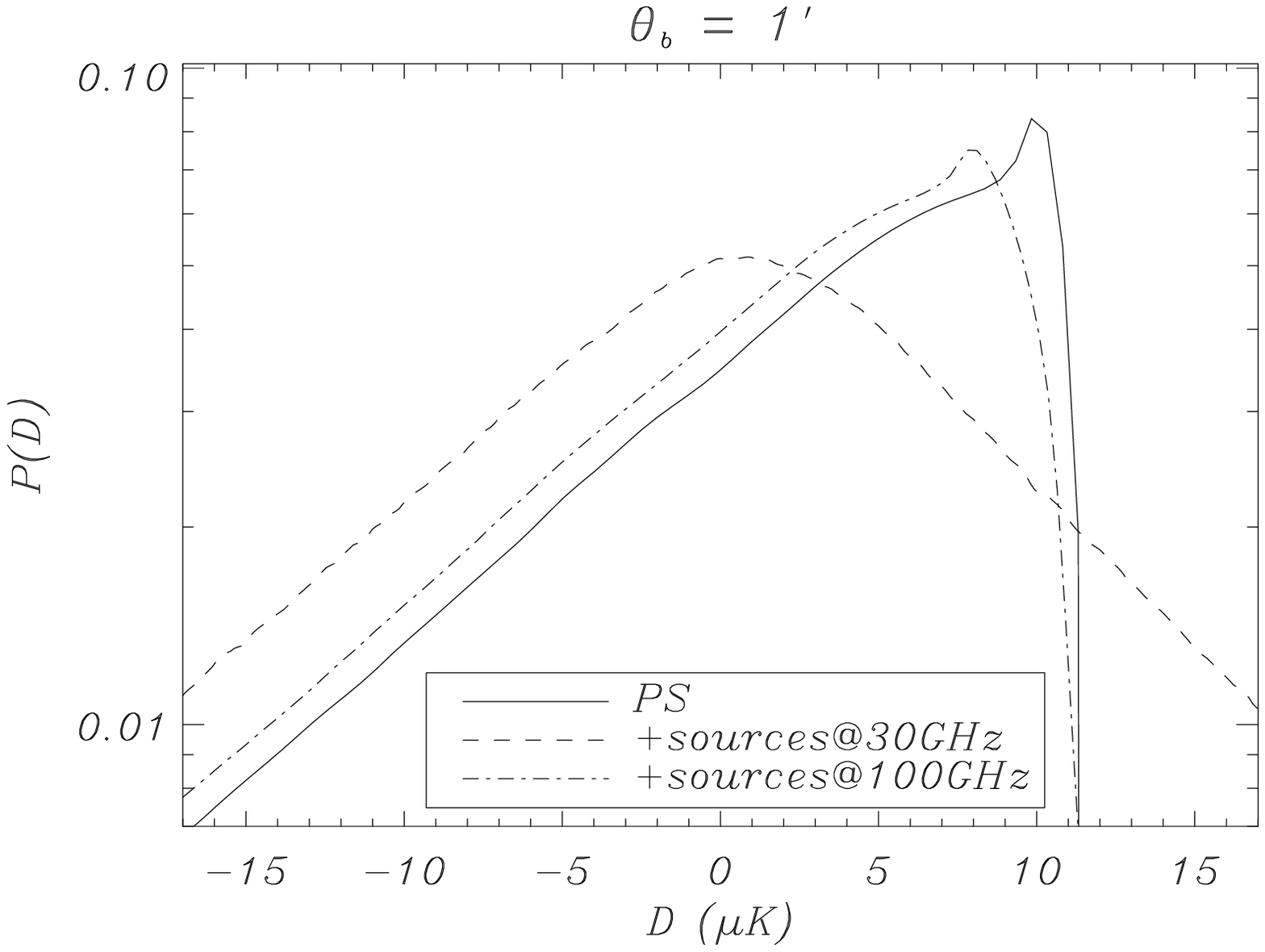}
\includegraphics[width=\columnwidth]{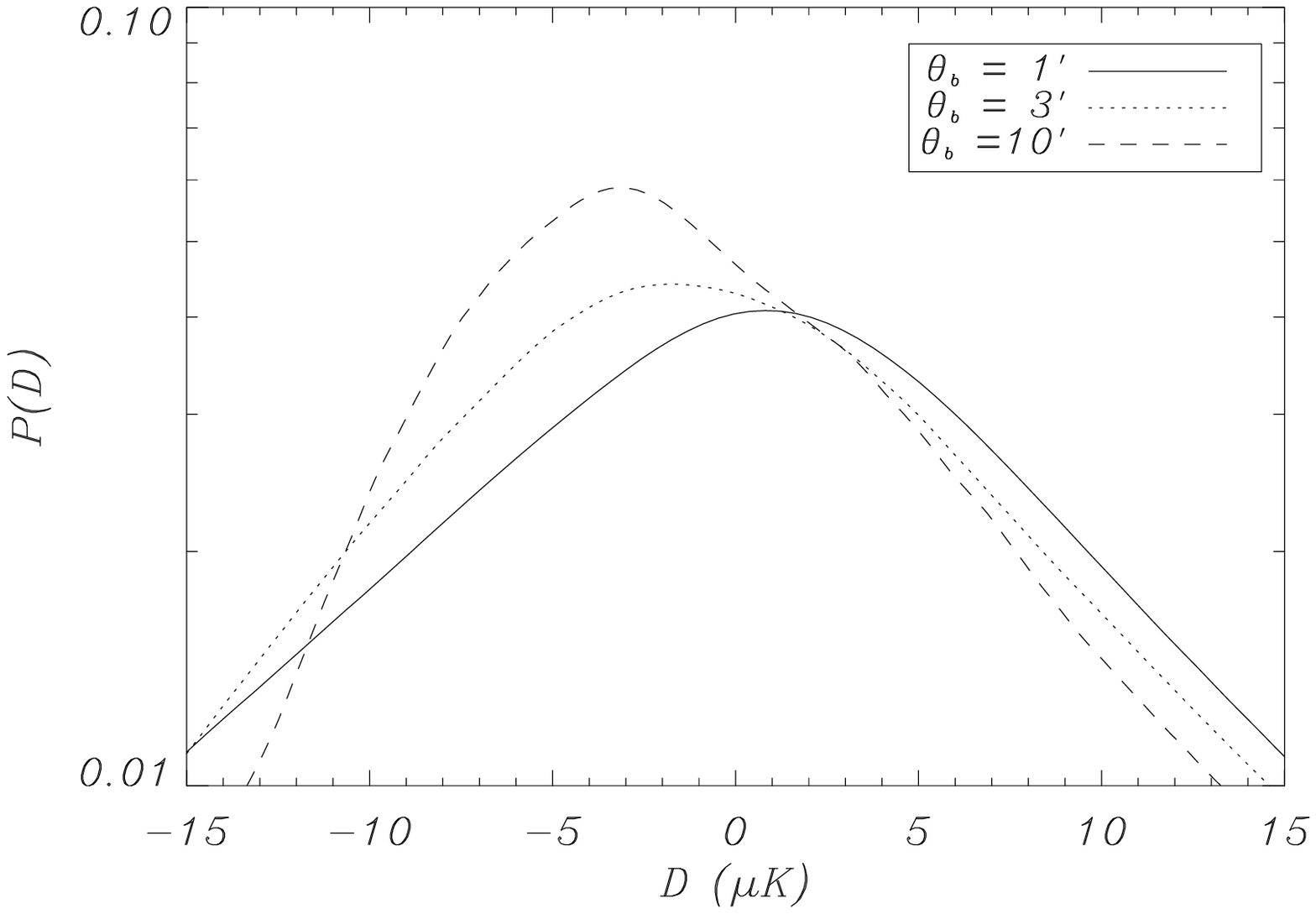}
\caption{Top: P(D) function for the 15 (1\degr -side) PS 
realisations plus sources,
assuming that we are able to subtract radio sources down
to $s_c = 5 \sigma_c(q=5)$. We calculate sources using
the parameters $K=92$ sr$^{-1}$ Jy$^{-1}$ and $\beta=2$, and we
present the results both at 30 GHz and at 100 GHz. 
The extrapolation of radio sources up to 100 GHz 
has been done using an spectral index $\alpha = 0.5$.  
We can see that, without a source subtraction strategy,
radio sources become of the same importance as, or even much more
important than clusters at 
this angular scale for $\nu=30~GHz$, while they practically disappear
at 100 GHz. 
Bottom: P(D) function for clusters and sources as
a function of the beam size. We use the 15 PS realisations, plus
sources as in the top panel, and we compute the P(D) function
at 30 GHz. As expected from the simple inspection of figure
\ref{review_scounts}, at larger beam sizes sources are more
numerous, so their tail dominates the asymmetry. 
At smaller beam sizes, clusters become of importance. }
\label{pd_sources2}
\end{figure}

We will now discuss how this picture changes with the 
observing frequency and the beam size, assuming that 
we do not subtract sources at all, so their contribution 
will be at least $q=5$ (i.e. $s_c = 5 \sigma_c(q=5)$).
We present in figure \ref{pd_sources2} 
the P(D) function for the 15 PS 
realisations, with and without adding to the maps
a simulation of sources with $K=92$ Jy$^{-1}$ sr$^{-1}$ and
$\beta=2$. 
Just for illustration, it is also computed the P(D) function extrapolating
the radio sources up to 100 GHz using $\alpha=0.5$. As it has been
pointed out above, at higher frequencies we would expect 
other populations of sources, so the real $P(D)$ function might show a
bigger positive wing. 
We can see that at these angular scales ($\theta_b=1'$), and without a source
subtraction strategy, clusters can be (at the most) of the same
importance of radio sources at 30 GHz. 
This fact is well-known, and 
several planned experiments that will operate 
close to these frequencies (30~GHz) also have 
designed a source subtraction strategy to eliminate the radio source
confusion (e.g. AMI, \citep{kneissl01}). 

In the bottom panel of figure \ref{pd_sources2} we illustrate the
dependence of the shape of the $P(D)$ as a function
of the beam size ($\theta_b$). 
As we would expected from the simple inspection of figure
\ref{review_scounts}, and as we pointed out at the beginning of
section \ref{sec:general} (see figure \ref{truncated}), 
at larger beam sizes radio sources are more
numerous than clusters, so their tail dominates the asymmetry. 
At smaller beam sizes, clusters become of importance. 

It is important to note here that 
all these calculations have been done assuming uncorrelated sources,
but in principle one could have two types of correlations:
spatial correlations between sources (SZ clusters) themselves, and 
correlations between the positions of the SZ clusters and the radio
sources. 
In the first case, spatial correlations of sources/clusters can be
modelled to first order \citep{barcons92} as an extra convolution with  
another gaussian. Thus, the above tools can still be applied, although
it will required a much more detailed study. 
On the other hand, it is well-known that clusters of galaxies 
may contain radio point sources (e.g. \citealt{birkinshaw99,cooray98}), so
the signal from SZ clusters could be diluted
\citep{holder02b,lin02}. This point can be easily checked by 
introducing spatial correlations between clusters and radio sources.
In the most unfavourable case of having a radio source inside
each cluster, we obtain from the previous simulations 
that skewness is underestimated by about $\sim 20$ \%.
This number is in agreement with the result 
obtained by \cite{holder02b}, who showed  
that the rms fluctuations of the thermal SZ effect could be 
underestimated due to correlations between clusters and radio 
sources by as much as 30\% at the observing frequency
of 30 GHz and $\ell \ga 1000$.

\section{Detection of a negative contribution of clusters}
\label{sec:noise}

The performance of the estimators described on section \ref{sec:estimators}
for detecting a negative contribution of clusters 
depends on the particular 
shape of the source counts for both radio sources and SZ clusters. 
In this subsection, we will concentrate on showing that 
the skewness may be used to distinguish the nature
of the subjacent fluctuations, using for that purpose a toy-model
for sources and clusters. 
Once we are able to detect skewness
in the map, the $P(D)$ function will show an asymmetry.
However, the determination of the $P(D)$ function of the underlying sources 
will require additional observational 
effort than simply measuring the skewness. 
Thus, the skewness will give us the
minimum signal-to-noise level required to distinguish between
clusters and radio sources.

\begin{table}
\caption{Detectability of the skewness when adding white gaussian noise
to SZ maps.}
\label{table1}
\begin{tabular}{@{}cccccc}
\hline
$\sigma_{noise}/\sigma_{SZ}$ & rms ($\mu K$) & $E[\sigma_{SZ}]$ ($\mu K$) & $E[Skew]$ & $\sigma_{Skew}$\\
\hline
        0.3   &   15.10 & 14.46  &   -3.320  &  0.003 \\
        1.0   &   20.43 & 14.44  &   -1.340  &  0.003 \\
        2.0   &   32.32 & 14.43  &   -0.340  &  0.003 \\
        5.0   &   73.67 & 14.15  &   -0.030  &  0.003 \\
       10.0   &   145.35& 14.84  &   -0.001  &  0.003 \\
       20.0   &   289.78& 18.67  &    0.0004 &  0.003 \\
\hline
\end{tabular}
\medskip

We consider an SZ simulated map following 
PS prescription, covering 3 square degrees with a pixel
size of $0.12'$ and a beam size of $\theta_b=1'$, 
with $\sigma_{SZ} = 14.46 \mu K$, and we add different noise levels. 
The obtained P(D) are shown in Figure \ref{pd_plus_noise}, while we
present here the observed power ($rms$) in the map, the
infered signal due to SZ (estimated as $E[\sigma_{SZ}] = \sqrt{rms^2 -
\sigma_{noise}^2}$ ), and the measured skewness in the map (without
applying any filtering technique), with its 
variance ($\sigma_{Skew}$).
\end{table}

\begin{figure}
\includegraphics[width=\columnwidth]{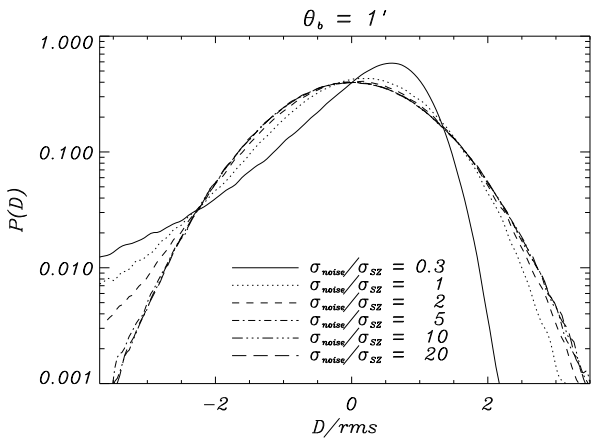}
\caption{Shape of the P(D) function for clusters when
adding instrumental gaussian white noise. 
A single map of 3 square degrees of SZ clusters 
following PS prescription has been used, with $\sigma_{SZ} = 14.46~
\mu K$. We add to this map several realizations of white gaussian noise,
with the amplitude quoted in the figure, and then we 
compute the P(D) function. The measured
excess of power and skewness in each particular case are quoted 
in Table \ref{table1}. }
\label{pd_plus_noise}
\end{figure}

We illustrate these facts by computing the $P(D)$ functions and
the skewness for the complete set of 15 SZ maps based on PS 
modelling, observed with a gaussian beam of $\theta_b=1'$,
and adding different noise levels per pixel ($\equiv 
\sigma_{noise}$),
quoted in terms of the power due to clusters ($\sigma_{SZ}$). When
then consider the following cases for $\sigma_{noise}/\sigma_{SZ}$: 0.3, 
1, 2, 5, 10 and 20. We will also assume that the noise is
gaussian and uncorrelated, as a first approximation. 
The resulting P(D) curves are shown in figure \ref{pd_plus_noise}, 
while the measured rms in the map, the recovered excess of
power, and the observed skewness, are quoted in
Table \ref{table1}. 
We can see that the tail of the P(D) function has practically 
disappeared at $\sigma_{noise}/\sigma_{SZ} = 5$, although is still
possible to measure skewness in the map at a high significance.
When we go to $\sigma_{noise}/\sigma_{SZ} = 10 $, 
the skewness is no longer visible, 
although it is possible to detect an excess of power.
At $\sigma_{noise}/\sigma_{SZ} = 20$, anything can be detected.

If we now include sources in our maps, the determination
of how deep we need to integrate in order
to get information about the skewness can be done in the
following simple way. 
We define here the signal-to-noise ratio per pixel, $S/N$, as 
the quotient $\sigma_{signal}/\sigma_{noise}$. 
We quote this quantity because it is easy to infer from the map.
In addition, it is also straight-forward to convert these
values into integration time $t$ for a given experiment, 
because we usually have $\sigma_{noise} \propto \sqrt{t}$.
In the previous expression, $\sigma_{signal}$ corresponds to
the power introduced by clusters and sources all together.
Then, if we want a $q$-sigma detection of skewness, it
is straight-forward to derive from equation (\ref{sum_skew}) that
we need a signal-to-noise ratio better than
\begin{equation}
\Bigg( \frac{S}{N} \Bigg) \ge 
\Bigg[ \Bigg( \frac{Skew^2 N_{pix}}{6q^2} \Bigg)^{1/3} -1 \Bigg]^{-1/2}
\end{equation}
where $Skew$ is the combined skewness of sources and clusters (\ref{sum_skew}), 
and $N_{pix}$ is the number of pixels in the map, i.e. 
$N_{pix} = 4\pi f_{sky}/ \Omega _p$, with
$f_{sky}$ the fraction of sky covered, and $\Omega _p$ the pixel 
solid angle.
For the typical values of skewness found in 
our simulations, 
the required signal-to-noise  
ratio for the detection of skewness due to both point 
sources and SZ clusters is around 25-40\% times larger than 
the required ratio for the detection of an excess of power.
In other words, we need (the well-known result of) 1.5-2 times more integration 
time to detect skewness than to detect an excess of power. 

We will mention here that these values have 
been obtained directly from the maps, without applying any filtering.
These numbers can be improved if we apply to our maps an appropriate
filter to enhance the contribution of sources over the noise level,
previously to the skewness or P(D) computation. We
will discuss this issue in the next section, where we also consider
the effect of primary anisotropies on the maps.

\section{Including primary anisotropies}
\label{sec:addcmb}

The computations in the last two sections were done assuming that the
primordial CMB gives a negligible contribution compared with that of the SZ
signal and the noise. However, this is only true at arcminute scales,
where the primordial anisotropy is damped to very small amplitudes,
well-below the level of the expected SZ anisotropy (see 
the top panel of figure \ref{bispectrum_press}).
Therefore, only those experiments which are observing 
at the high multipole region of the angular power spectrum are 
going to directly observe the cluster contribution. 
This is the 
case of interferometers with long baselines (e.g. AMIBA, see
\citealt{zhang02}, or BIMA \citep{dawson02}, which has
a primary beam of $6.6\arcmin$), or experiments with small fields of view. 
As it is well-known, if we are observing small patches on sky, we are
sensible to multipoles greater than $\ell_{min} \approx 2\pi/
\theta_{map}$, being $\theta_{map}$ the size of our map (see
e.g. \cite{hobson96} for a detailed study). Therefore, for small
fields of view we are implicitly filtering out the lowest multipoles,
and we can directly apply the formalism described in the last section. 

However, if our experiment is sensitive to low multipoles ($\ell \la
2000$), then it is necessary to add  a new step in our pipeline, which 
consists in filtering these low multipoles, if we want
to remove the contribution of primordial anisotropies. 
This will be the case for experiments which will cover thousands of
square degrees in sky (e.g. 8m South Pole telescope), or
experiments covering the whole sky (e.g PLANCK satellite). 
This step is common to all techniques which are aimed to extract
sources/clusters from a map using a single frequency observation. 
Many filters can be found in literature, going from matched filters
(e.g. \cite{tegmark96,tegmark98}), pseudo-filters (e.g. \cite{sanz01})
or wavelets (e.g. \cite{cayon00}). 
Any of these methods can be used to pre-process the maps, reducing the
relative contribution of primary anisotropies, and also that of
the noise. The important point is
that under linear filters, gaussian data will still be gaussian, and
a distribution of sources still will show an skewed shape.

\subsection{Effect of primary anisotropies on the observed P(D)}
For illustrating this issue, 
we have performed a new simulation of SZ clusters, covering a
larger area ($2048 \times 2048$ pixels of 0.5\arcmin side each, so the map
covers $\sim 291$ square degrees), using the same parameters 
as for the previous simulations, but with $M_{min}=10^{14} M_\odot$. 
The reason of considering such a
large area is because future experiments will cover hundreds (or
thousands) of square degrees on sky. 
We add to this map a CMB realisation 
following the power-spectra 
plotted in the top panel of figure \ref{bispectrum_press}. 
The rms of the simulated SZ map 
and the CMB realisation are $5.7 \mu K$ and $112.3 \mu K$,
respectively. 
Finally, we also add white gaussian noise with an 
amplitude $10 \mu K$ (per pixel of 0.5\arcmin), and we smooth 
the resulting map with a gaussian beam of $\theta_b = 1\arcmin$.
Using this map, we have examined two different filters: 
\begin{itemize}
\item a Hanning (or high-pass) filter, which removes the 
contribution of all multipoles below a certain value $\ell_{0}$, but
linearly increasing from 0 to 1 in the range $[\ell_{0}-\Delta \ell, 
\ell_{0}+\Delta \ell]$ to avoid the ringing associated to a sharp
cut. We have used here $\ell_{0} = 2000$, and $\Delta \ell = 50$. 
\item a matched filter optimised to detect point-source like objects 
\citep{tegmark96}, so $W_\ell \propto (B_\ell
C_\ell^{TOT})^{-1}$. Here, $W_\ell$ and $B_\ell$ are the 
coefficients in a Legendre polynomial expansion of the filter and 
of the beam, respectively, and $C_\ell^{TOT}$ is the sum of
the power spectrum of the components to be removed (noise and CMB in
this case). 
\end{itemize}
\begin{figure}
\includegraphics[width=\columnwidth]{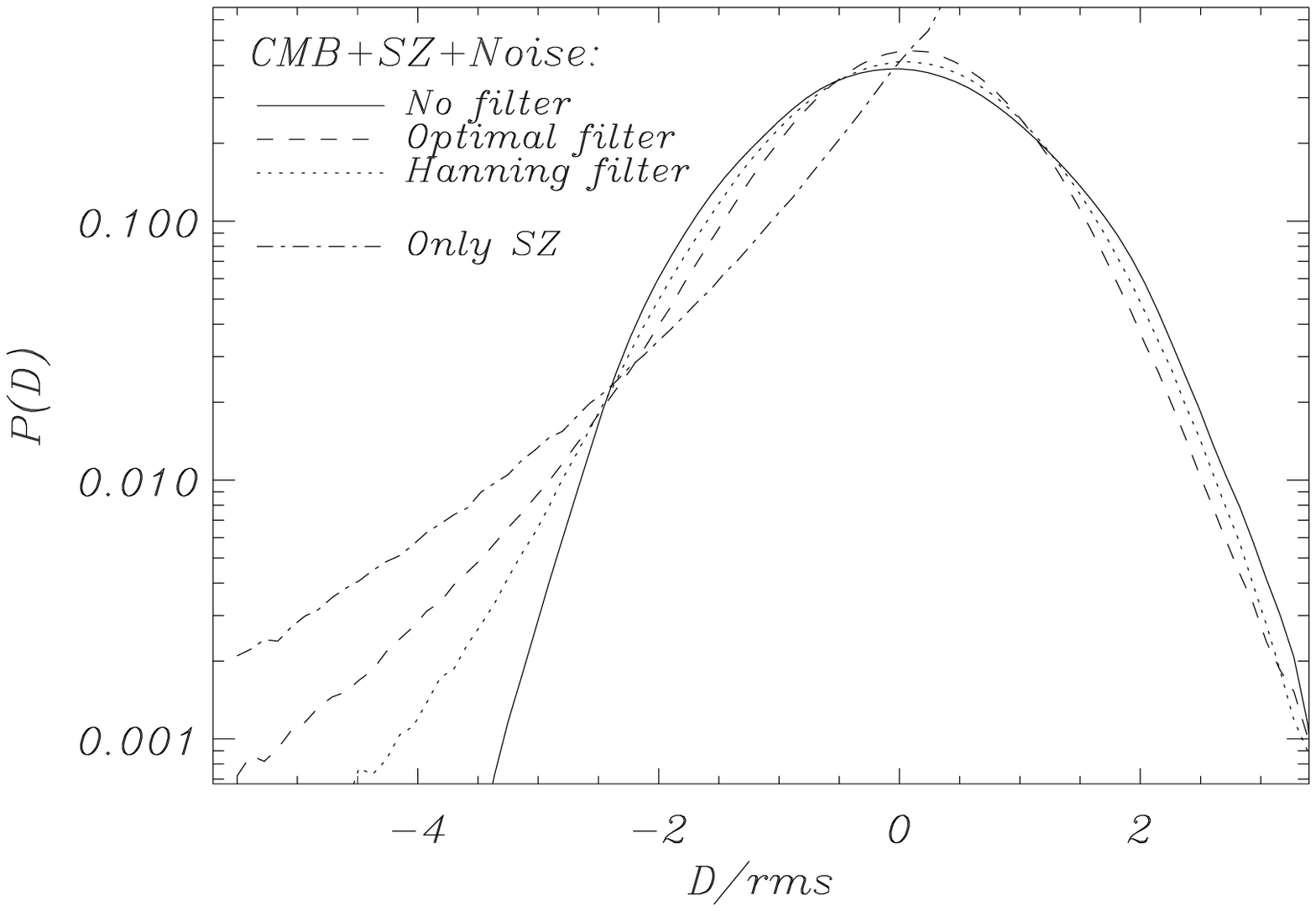}
\includegraphics[width=\columnwidth]{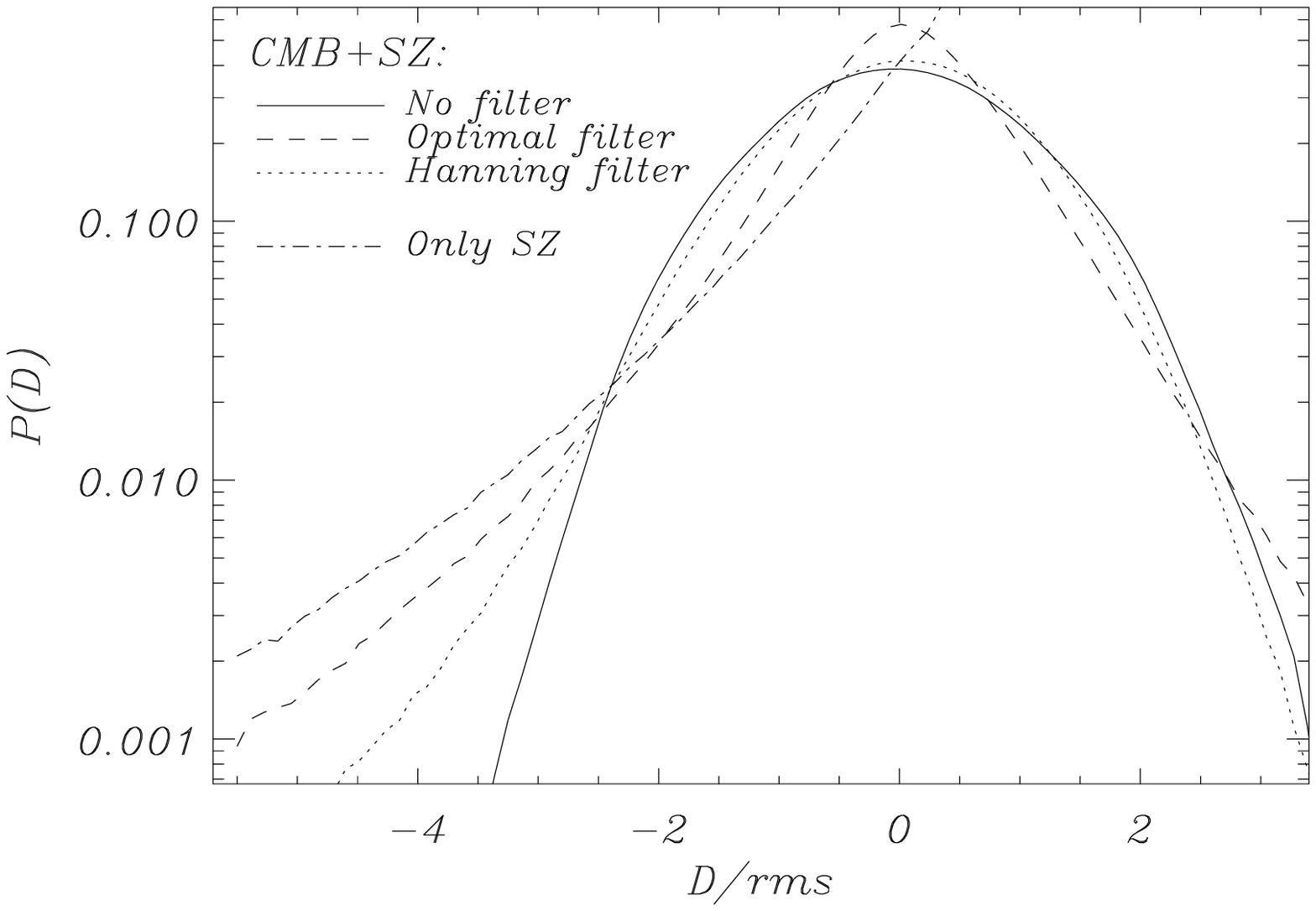}
\caption{Top: P(D) function from a Press-Schechter SZ 
realisation of a 291 square degrees patch of the sky, observed
with a gaussian beam of $\theta_b = 1\arcmin$, for four cases:
(a) adding CMB primary anisotropies, and white gaussian noise of 10 $\mu K$ per
pixel (solid line);
(b) filtering the previous map with the optimal matched-filter
described in Haehnelt \& Tegmark (1996) (dashed line);
(c) filtering the map with a Hanning (high-pass) filter (dotted line);  
(d) considering the SZ map alone (dot-dashed line).
Bottom: Same as the previous panel, but without including noise. 
In order to use the same scale for plotting all curves, we have 
divided each map for its $rms$ prior to the P(D) computation. 
In this case, $rms(SZ) = 4.9\mu K$, $rms(SZ+CMB) = 112.2\mu K$
and $rms(SZ+CMB+Noise) = 112.3\mu K$. }
\label{filters}
\end{figure}
In Figure \ref{filters} we show the predicted $P(D)$ function from our
simulation when adding CMB primary anisotropies, with and without
noise, and the predicted $P(D)$ after filtering 
the maps with the previous filters. 
For comparison,
it is also shown the $P(D)$ function for the SZ map alone. 
In order to use the same scale for plotting all the curves, 
we have divided each map by its rms prior to the $P(D)$ computation. 
We can see that a simple high-pass filter is able to remove the main 
contribution from the primordial CMB, 
so the $P(D)$ function becomes dominated by the
contribution of SZ clusters, and the negative tail is visible. 
However, an small residual signal of the CMB fluctuations
still persists, even in the case where we do not consider noise.
Nevertheless, this residual can be modelled as an extra gaussian
noise: a $P(D)$ analysis does not distinguish two signals if 
both of them are gaussian. 
When using the optimal filter, the $P(D)$ function shows its tail more 
clearly. Therefore, when pre-processing our maps with a filter, we are
able to reduce the problem to the case discussed in the previous
section, where we only have SZ signal, sources and noise, and
hence the negative tail is showing us the presence of clusters. 
For completeness, we report the number of pixels which are 
above 3 sigmas in these maps, to illustrate the significance of the 
``detection'' of the tail. For the case which includes noise, these
numbers are 7500, 21503, 40475 and 76345 for the non-filtered, 
the Hanning filtered, the optimal filtered and the pure SZ maps, 
respectively. These numbers have a sampling error of 106 pixels.
For a gaussian distribution, we should expect to have 11324 pixels 
(0.27\% for $3\sigma$) out of $2048^2$ in total.

\begin{figure}
\includegraphics[width=\columnwidth]{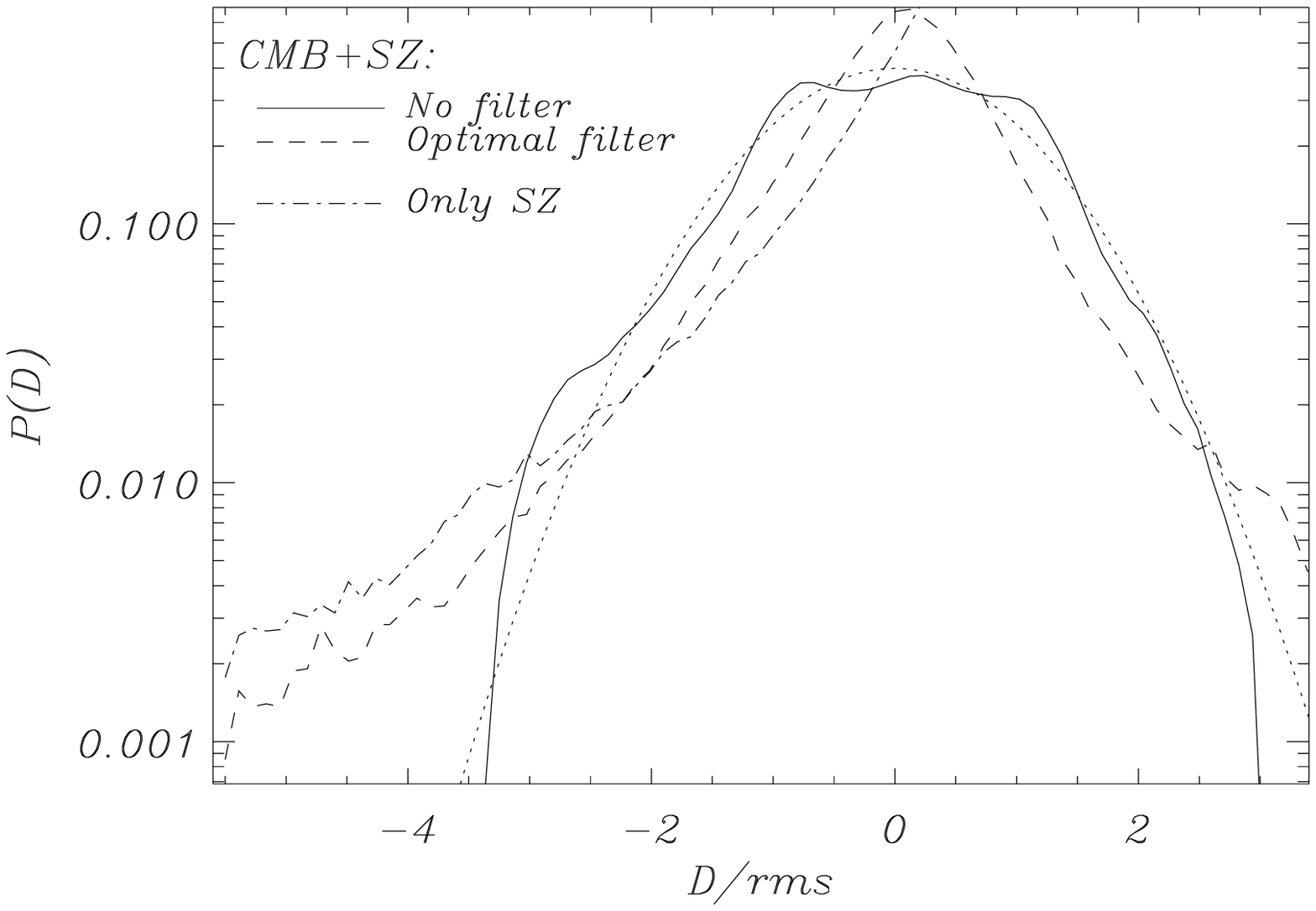}
\includegraphics[width=\columnwidth]{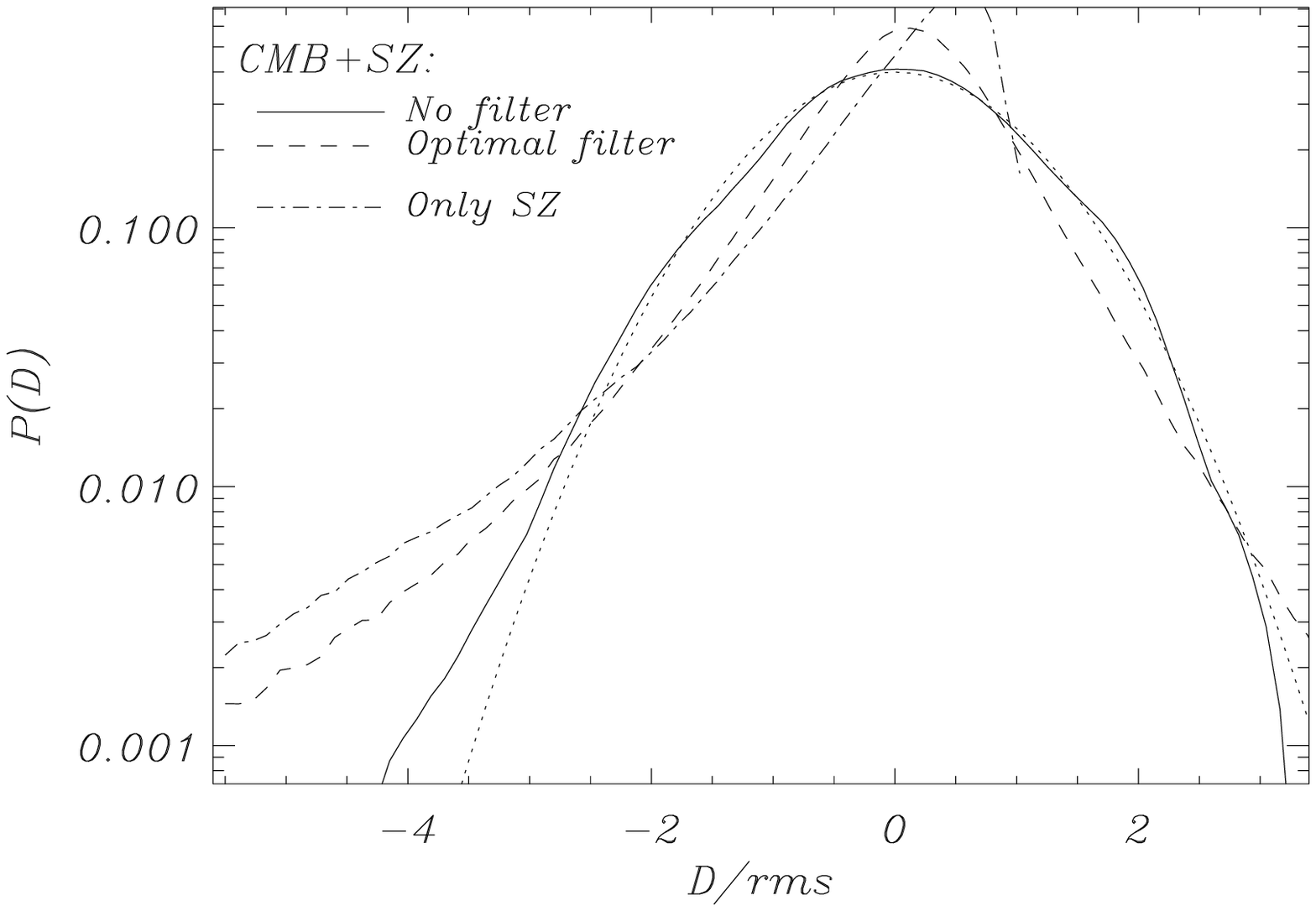}
\caption{Effect of the CMB sample variance in the observed P(D)
function. 
Top: P(D) function from a single Press-Schechter SZ map of
1\degr-side, observed with a gaussian beam of $\theta_b = 1\arcmin$,
for the cases (a), (b) and (d) discussed in figure \ref{filters}. 
We can see that the observed P(D) function is distorted due to the 
sample variance contribution of the primordial CMB anisotropies. 
The dotted line corresponds to a gaussian with a width equal to 
the observed rms in the map, i.e. the P(D) function that we would
observe if all the signals in the map were gaussian and we had
no sample variance. 
Bottom: same as previous panel, but averaging 15 realisations.
In this case, the sample variance is reduced and a negative
tail is marginally detected even without filtering. 
In this case, $rms(SZ) = 12.3\mu K$ and $rms(SZ+CMB) = 54.5\mu K$ 
in the top panel and $rms(SZ) = 13.9\mu K$ and $rms(SZ+CMB) = 74.2\mu
K$ in the bottom one.}
\label{filters2}
\end{figure}
It is also interesting to demonstrate what we would expect if a given
survey is covering an small area on sky, but large enough to have an
important contribution from primary anisotropies.  This would be the
case of an hypothetic experiment covering 1\degr-side patch  on sky,
so we have used here the simulations described  in section
\ref{subsec:ps}.  In this case, we have the additional problem of the
sample variance: the observed P(D) function will show deviations from
the expected gaussian curve.  Therefore, filtering in this case is
absolutely necessary.  We illustrate this point with figure
\ref{filters2}, where we show the P(D) function from a single
1\degr-side map, containing SZ signal, CMB and no noise, and also the
same plot for the average of 15 maps of the same type.   We can see
that, although the P(D) function for the map without filtering is
marginally showing a negative tail when we average 15 maps,  the
sample variance is completely deleting this tail when we consider a
single map.  This point can also be illustrated if we use those 15
$1\degr$-side simulations to compute the uncertainty introduced by the
sample variance in the number of pixels above the 3$\sigma$
threshold. This number was found to be 0.5\%, 
whilst the expected number of pixels
for a gaussian distribution above 3$\sigma$ is 0.27\%.  However, even
in this case the filtering is able to remove the main contribution of
the primary anisotropies, and hence the negative tail is seen.

\subsection{Effect of primary anisotropies on the observed skewness/bispectrum}
\begin{figure}
\includegraphics[width=\columnwidth]{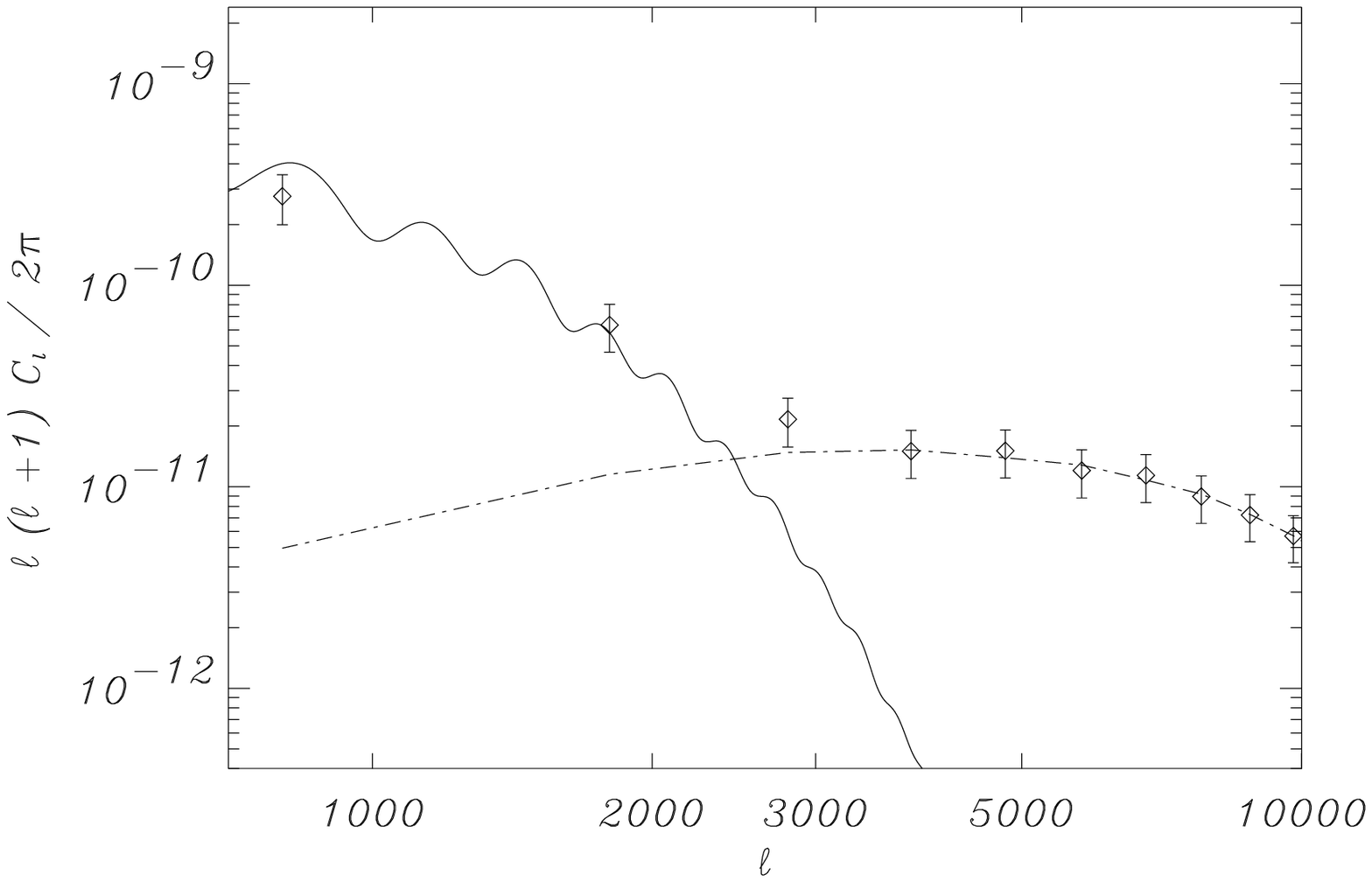}
\includegraphics[width=\columnwidth]{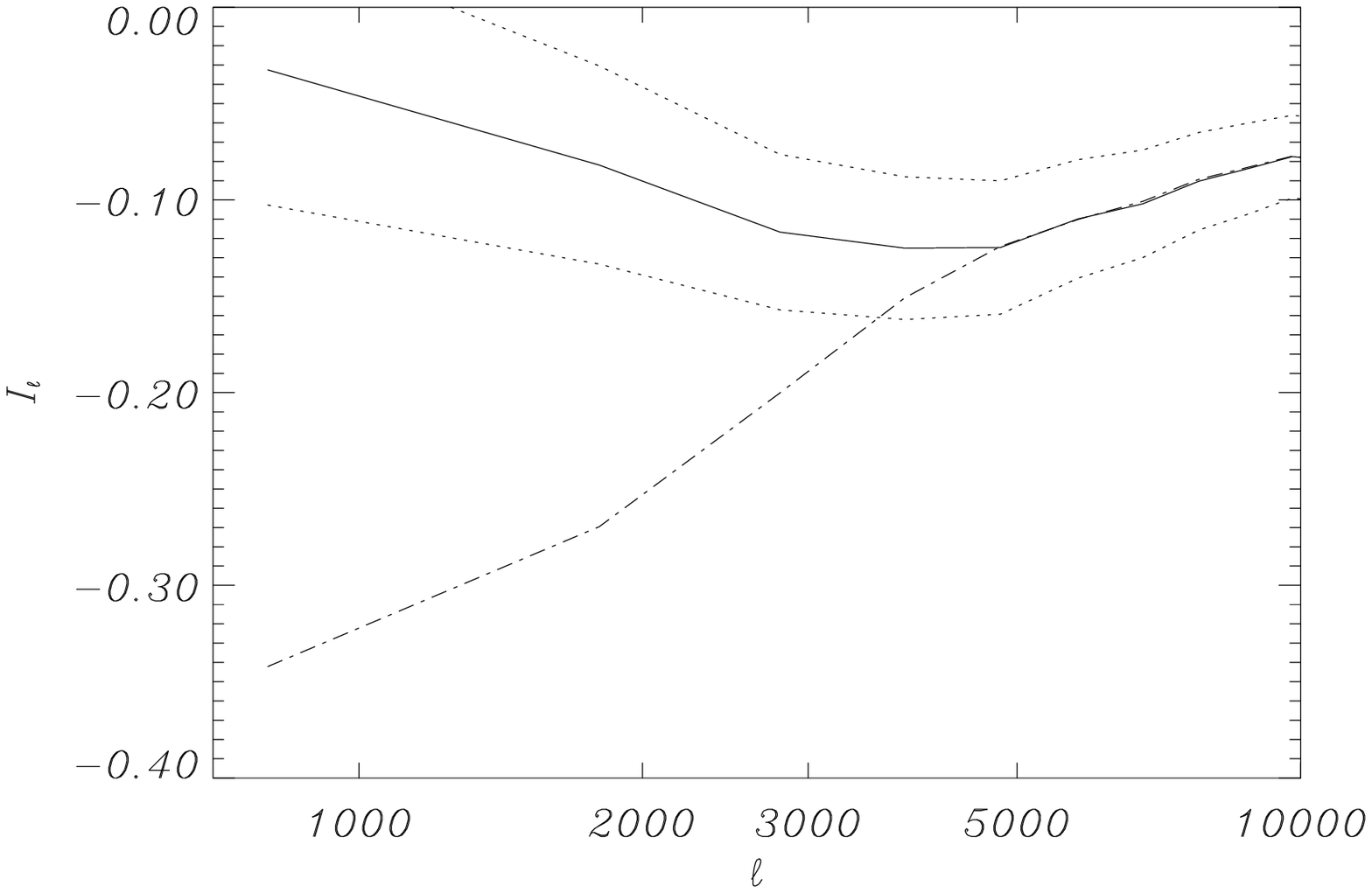}
\caption{Effect of primordial CMB anisotropies on the
observed power spectrum/bispectrum. We use the same 15 realisations
as in figure \ref{bispectrum_press}, but adding a CMB realisations to 
each one of them, following the plotted $\Lambda$CDM model. Here we 
do not apply any filtering to the maps. 
Top: The recovered power spectrum from the 15 realisations
(diamonds). Error bars correspond to the field-to-field variance. 
At low $\ell$-values, the power spectrum traces that of the 
CMB, while in the high $\ell$ region, the spectrum traces that
of the SZ maps (dot-dashed line).
Bottom: Bispectrum for the same 15 realisations (solid line). Dotted
lines show the 1-sigma field-to-field variance. We see that at high
$\ell$, the bispectrum follows that of the SZ signal (dot-dashed
line), but a low-$\ell$ (large scales) its goes to zero because the
power becomes dominated by that of the primordial CMB, which has
zero skewness. }
\label{bisp_sz_cmb}
\end{figure}
\begin{figure}
\includegraphics[width=\columnwidth]{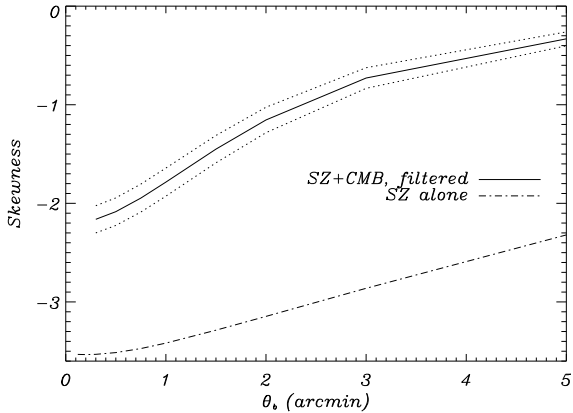}
\caption{Negative skewness for SZ clusters in the presence of primary CMB
anisotropies, as a function of the angular scale. We present the
skewness from the same 15 PS maps as in figure \ref{skew_vs_theta}, 
with (solid) and without (dot-dashed) adding 
primordial CMB anisotropies. When the
CMB is added, the maps are filtered, prior to the skewness
computation, using a Hanning filter with 
$\ell_0 = 2000$, as described in the text. Although the skewness in
real map is diluted due to the residual power from the CMB in the
filtered map, it is still detectable. }
\label{skew_sz_cmb}
\end{figure}

We conclude this section showing the expected skewness and bispectrum
from our 15 PS simulations taking into account a CMB component in the
maps. It is important to notice here that for the bispectrum
is not necessary to use any filter, because it directly gives
the 'skewness' at each angular scale. These results
for the power-spectrum and the bispectrum are shown 
in figure \ref{bisp_sz_cmb}. From here, we can conclude that if the
relative contribution of SZ clusters to the observed bispectrum is
larger than the one from radio sources, then we expect to directly
see their negative signal in the bispectrum without pre-processing the maps. 

To obtain the skewness in real-space, we proceed as in section
\ref{sec:7.3}, but pre-processing the maps with any of the
filters described in the last subsection. The obtained results
from averaging over the same 15 PS maps are shown in figure \ref{skew_sz_cmb}.
The main conclusion is that the residual CMB power in the
filtered map is diluting the SZ skewness, although it is still
clearly seen. This is exactly what we would expect, using equation 
(\ref{sum_skew}). From simple inspection of the top panel of
figure \ref{bisp_sz_cmb}, we expect to have 
a residual CMB power after filtering with a Hanning filter
of $\ell_0 = 2000$ which is similar, or slightly smaller than the SZ
power. Therefore, we would expect the skewness to be
reduced (roughly) by a factor $2^{3/2} \approx 2.8$, as we observed
from the simulations.

\section{Conclusions}

In this paper we have discussed five statements:

\begin{itemize}
\item The contribution of SZ clusters to the map noise at $\lambda > 1.25$ mm
does not depend on the wavelength, and has a strong and peculiar
non-gaussian negative tail in the $P(D)$ function.
\item This contribution has characteristic negative skewness (or
bispectrum) at $\lambda > 1.25$ mm. This fact 
can be used by current single-frequency experiments, such as CBI or
BIMA, or by future experiments, such as ACT, AMI, AMIBA, APEX, or the 8-m 
South Pole telescope,  
to distinguish if the detected excess of power at 
small angular scales is due to SZ clusters. In addition, 
the detection of skewness only requires a factor 1.5 or 2 more
integration time than the detection of an excess of power. Once the 
skewness is detected, the $P(D)$ function starts to show an
asymmetry. 
\item Any multi-frequency experiment would have
noise due to clusters of galaxies with $P(D)$ at $\lambda < 1.25$ mm
equal to $P(-D)$ at $\lambda > 1.25$ mm.
\item Skewness and bispectrum will have different signs in these two 
spectral regions.
\item When dealing with real maps where primordial CMB fluctuations
are present, it is necessary to use filters to remove the contribution
of large angular scales. Only in that case we can 
detect the presence of clusters/sources in the $P(D)$ function. 
\end{itemize}

\section*{Acknowledgments} 

We would like to thank Volker Springel for permission to use the 
maps obtained in \cite{springel01}. 
We also thank S.D.M. White, 
C. Hern\'andez-Monteagudo, A. Banday, M. Bartelmann, S. Zaroubi 
and P. Carreira for useful comments. 
JARM acknowledges support from the CMBNet European Network. 
We would like to thank also the anonymous referee for his 
suggestion of including a discussion about the P(D) function
in the presence of primary CMB anisotropies.

\newpage

\appendix

\section{Dependence of the P(D) function on the scaling assumptions
and the normalisation $\sigma_8$}
\label{app1}

The scaling relation for the gas temperature 
used in this work was obtained from fittings
to X-ray observations. However, we could also have used that
one derived from the spherical collapse model, which takes
the form \citep{eke96}:
\begin{equation}
k T_{gas} = \frac{7.75}{\beta} 
\Bigg( \frac{M}{10^{15} h^{-1} M_\odot} \Bigg)^{2/3} (1+z)
\Bigg( \frac{\Omega_0}{\Omega(z)} \frac{\Delta_c}{178} \Bigg)^{1/3} \; keV
\end{equation}
where $\Delta_c$ is the ratio of the mean density inside the virial
radius to the critical density, and $\beta$ is the ratio 
of the specific galaxy kinetic energy to specific gas
thermal energy. This scaling has been used by other authors (e.g. 
\citealt{komatsu99,molnar00}). We will consider this scaling here to
recompute the simulations, and we will show that the qualitative 
behaviour of the P(D) function and the skewness remains unchanged. 
We will adopt here $\beta=1$, the same parameters describing
the cluster ($r_{v0} = 1.3 h^{-1}$ Mpc, 
$r_{c0} = 0.13 h^{-1}$ Mpc, and $n_{c0} = 2 \times 10^{-3}$
cm$^{-3}$), and the same cosmological model as in the main text 
($\Omega_m=0.3$, $\Omega_\Lambda = 0.7$, $h=0.67$, and $\sigma_8 = 0.9$).
We integrate the PS mass function in the same range as in the main
text. 

Concerning the core evolution, in this work 
we have used an entropy-driven model
with $\epsilon = 0$, which is described as:
\begin{equation}
r_c(M,z) = r_{c0} \Bigg( \frac{M}{10^{15} h^{-1} M_\odot}
\Bigg)^{-1/6} 
\Bigg( \frac{\Omega_0}{\Omega(z)} \frac{\Delta_c}{178} \Bigg)^{-1/12}
(1+z)^{(3\epsilon-1)/4}
\end{equation}
but without using the $\Delta_c$ factor. Therefore, we now repeat the
computation taking it into account. 
A detailed study of how the 
power spectrum changes when assuming different core 
evolution models (a self-similar collapse, or an
entropy-driven model with different values for the $\epsilon$
parameter), can be found in \cite{komatsu99}.

\begin{figure}
\includegraphics[width=\columnwidth]{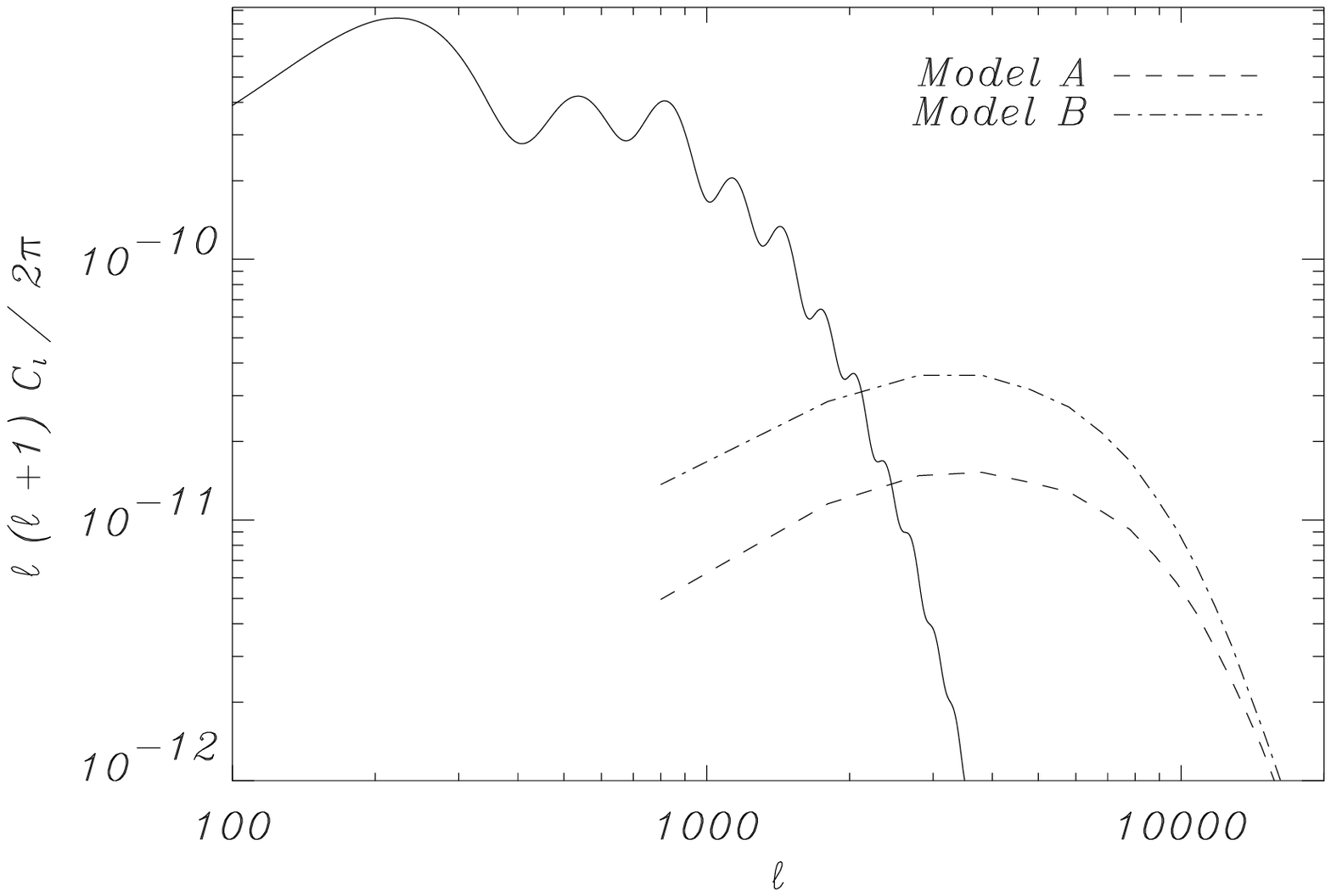}
\includegraphics[width=\columnwidth]{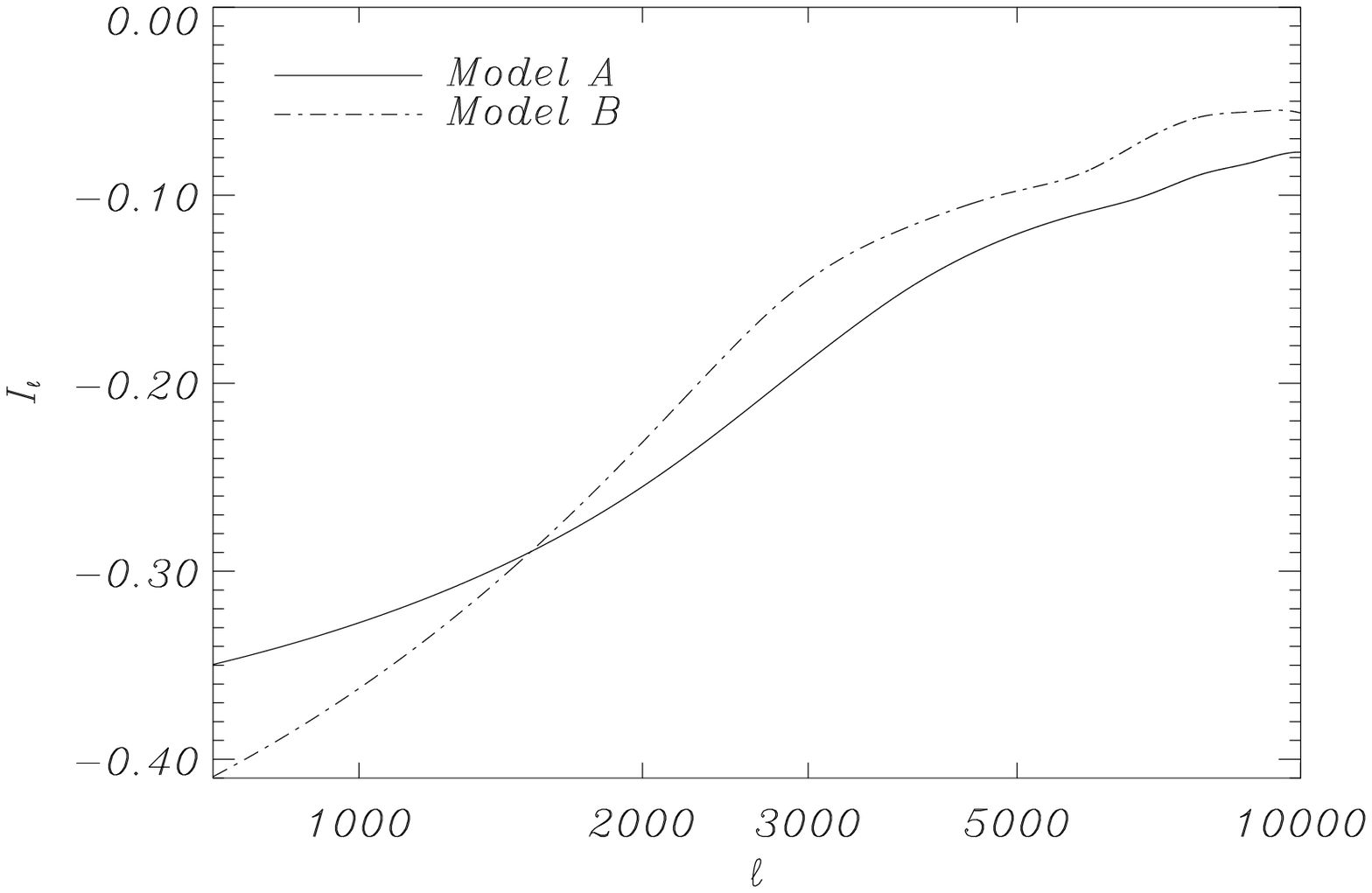}
\caption{Top: Angular power spectrum for the SZ effect derived
from Press-Schechter prescription, for two different models. These
models (A and B) correspond to the same cosmological model, same
normalisation ($\sigma_8=0.9$), but different scaling relations
for the temperature and the core radius with the mass and redshift
of the cluster (see details in text). The curves were obtained
averaging 15 simulations (1\degr side) for each case.  
Bottom: Same as before, but for the bispectrum. }
\label{compara_bp_Il}
\end{figure}

For definiteness, we will use the notation 'Model A' for the scaling
relations used in the main text (derived from fittings to
X-ray observations), and 'Model B' for the new
scaling relations considered in this appendix. 
Figure \ref{compara_bp_Il} shows the resulting power spectrum
($C_\ell$) and bispectrum ($I_\ell$) for these two models. We can see 
that the qualitative behaviour of the bispectrum is the same.

\begin{figure}
\includegraphics[width=\columnwidth]{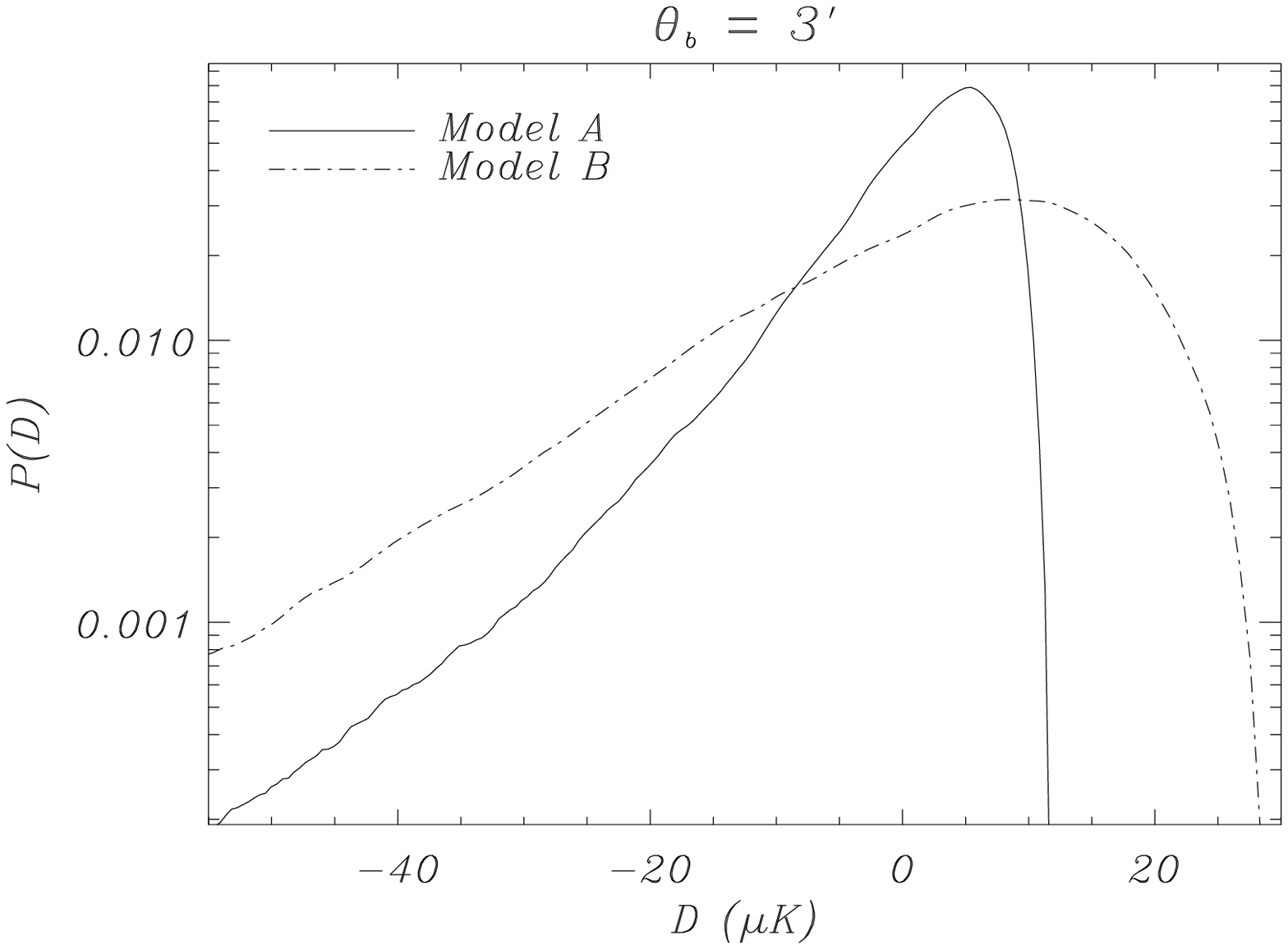}
\includegraphics[width=\columnwidth]{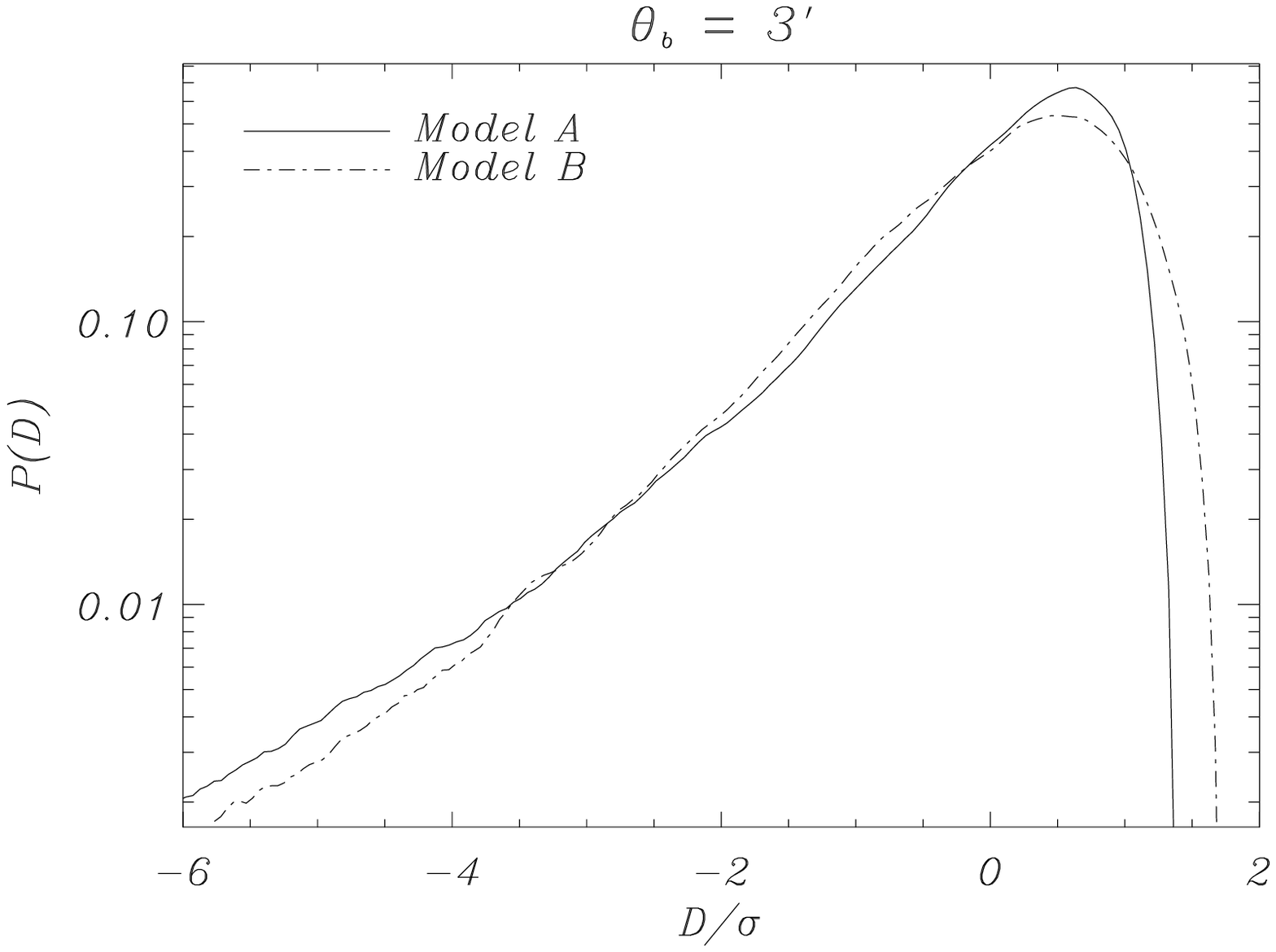}
\caption{Top: P(D) functions for the two models considered in
figure \ref{compara_bp_Il}, smoothing the maps with
a gaussian beam of size $\theta_b=3'$.
Bottom: Same figure, but rescaling the curves by their $\sigma$
(obtained as $\sigma^2 = \int D^2 P(D) dD$. }
\label{compara_pd_2scalings}
\end{figure}

Figure \ref{compara_pd_2scalings} shows the P(D) function for
these two models. Despite the different width of these curves (due
to the different total power in the spectrum, see figure
\ref{compara_bp_Il}), when rescaling them by their $\sigma$ 
($\sigma^2 = \int D^2 P(D) dD$), the shape of both curves
is very similar. 

Finally, we have explored how the shape of
the $P(D)$ is sensitive to the $\sigma_8$ value. 
We then obtain 15 simulations 
for each one of the following values of
$\sigma_8$: 0.6, 0.7, 0.8, 0.9, 1.0 and 1.1, keeping
the same cosmological model ($\Omega_m=0.3$, 
$\Omega_\Lambda = 0.7$ and $h=0.67$). For the scaling of the
temperature and core radius with mass and redshift, we
adopt the values of the 'Model B' described above.

\begin{figure}
\includegraphics[width=\columnwidth]{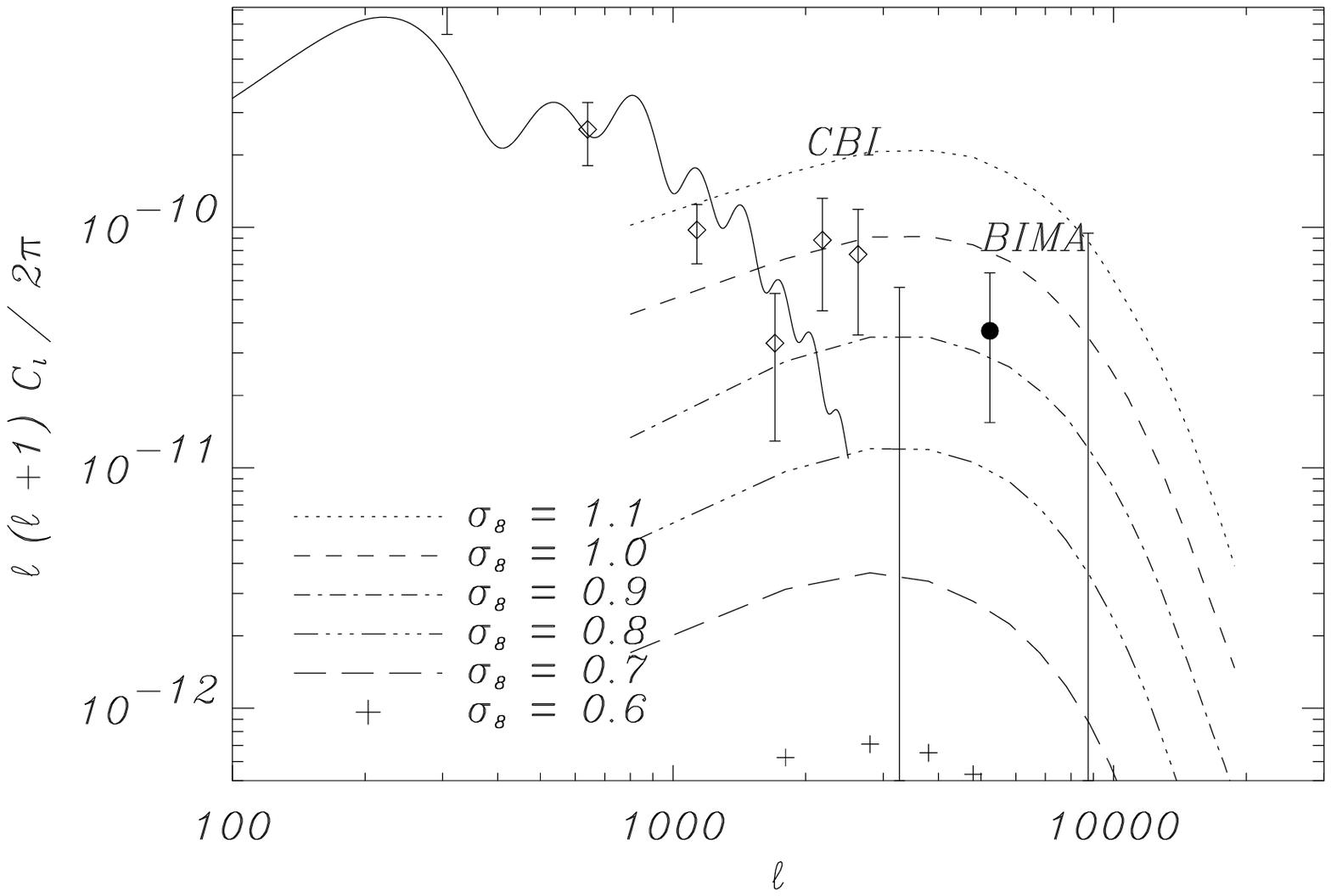}
\includegraphics[width=\columnwidth]{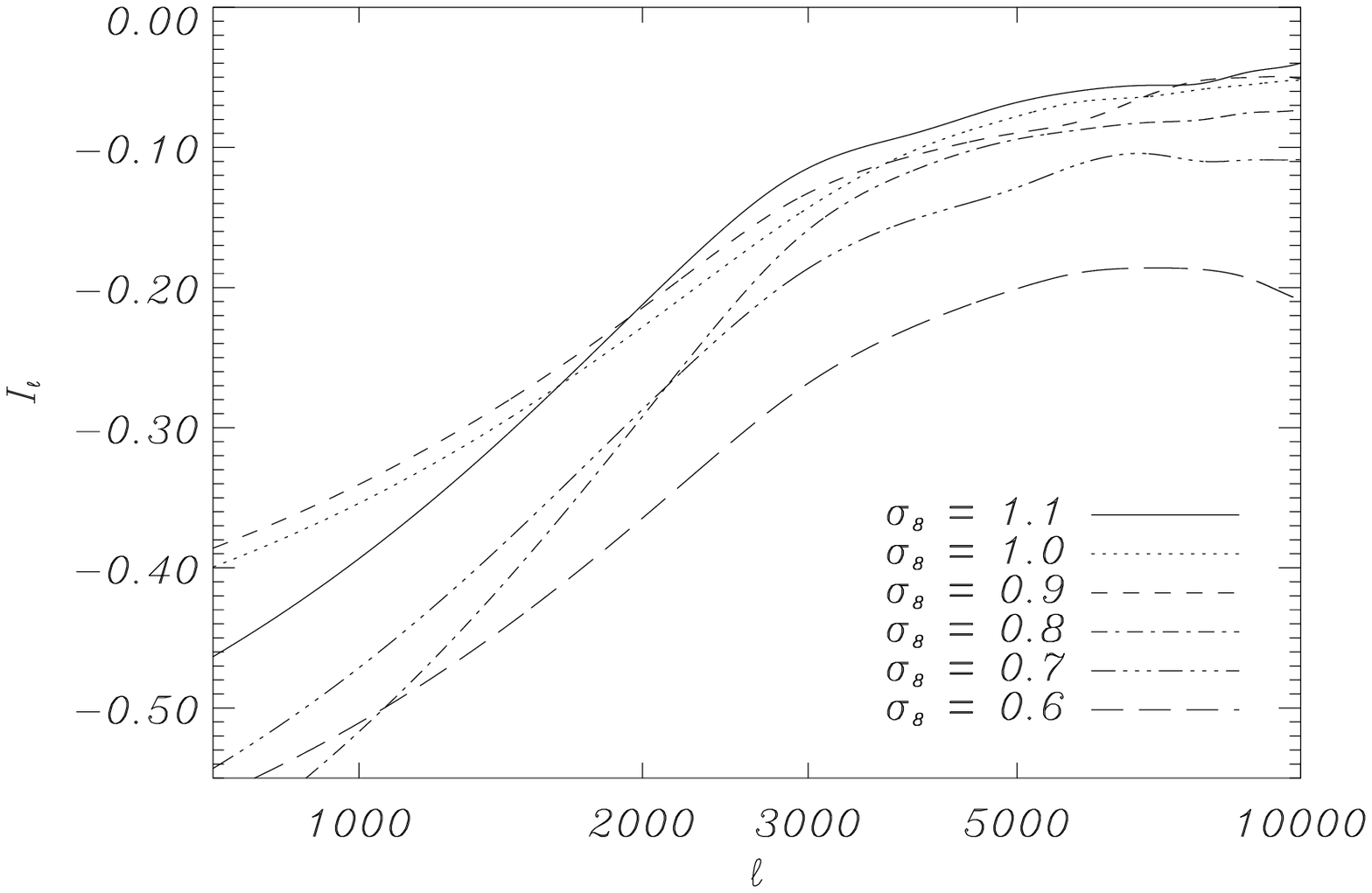}
\caption{Top: Power spectrum of the SZ effect for different
normalisations ($\sigma_8$ values). Each curve has been obtained from
averaging 15 realizations of 1\degr size (error bars are not shown). 
For comparison, it is also shown the corresponding primordial 
CMB power spectrum, and the current observational measurements from CBI and BIMA.
Bottom: Bispectrum of the SZ effect for the same values of
$\sigma_8$ considered above. }
\label{bp_sigma8}
\end{figure}

It is well-known that the SZ angular power spectrum scales as
$C_\ell \propto \sigma_8^{6--9}$ (see, e.g. \cite{komatsu99}). 
In our particular case, our power spectra derived from the simulations
shows an scaling $C_\ell \sim \sigma_8^9$. Figure \ref{bp_sigma8} 
shows the obtained power spectra and bispectrum for these values. 
The shapes of the bispectrum curves change slightly when changing the
value of $\sigma_8$. However, we can find an scaling for a given
multipole. Given that SZ power spectrum peaks 
around $\ell \sim 3000$, we have done the study for $\ell = 3000$, 
obtaining that the
bispectrum scales as $I_\ell \propto \sigma_8^{1.1}$.

\begin{figure}
\includegraphics[width=\columnwidth]{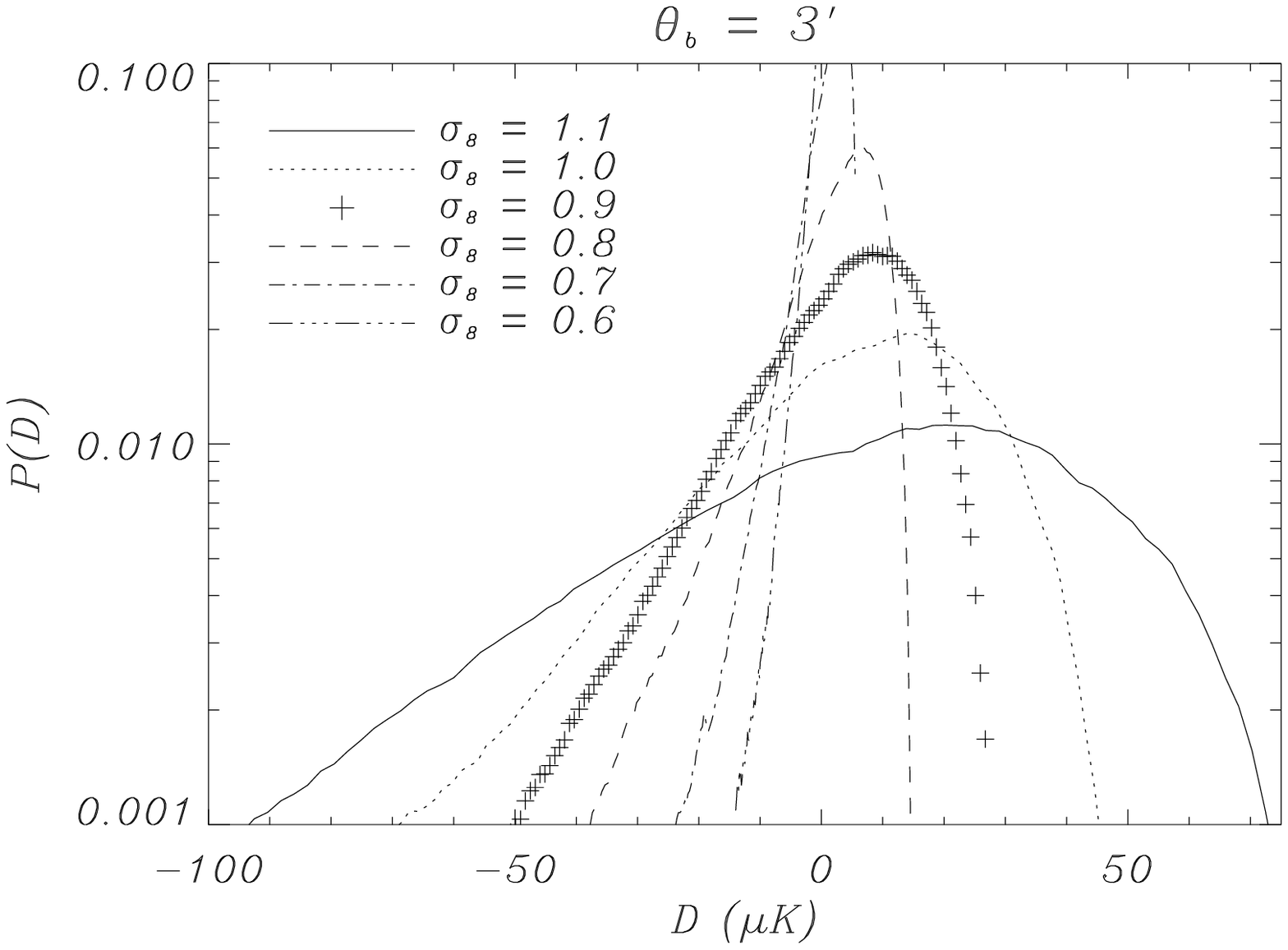}
\includegraphics[width=\columnwidth]{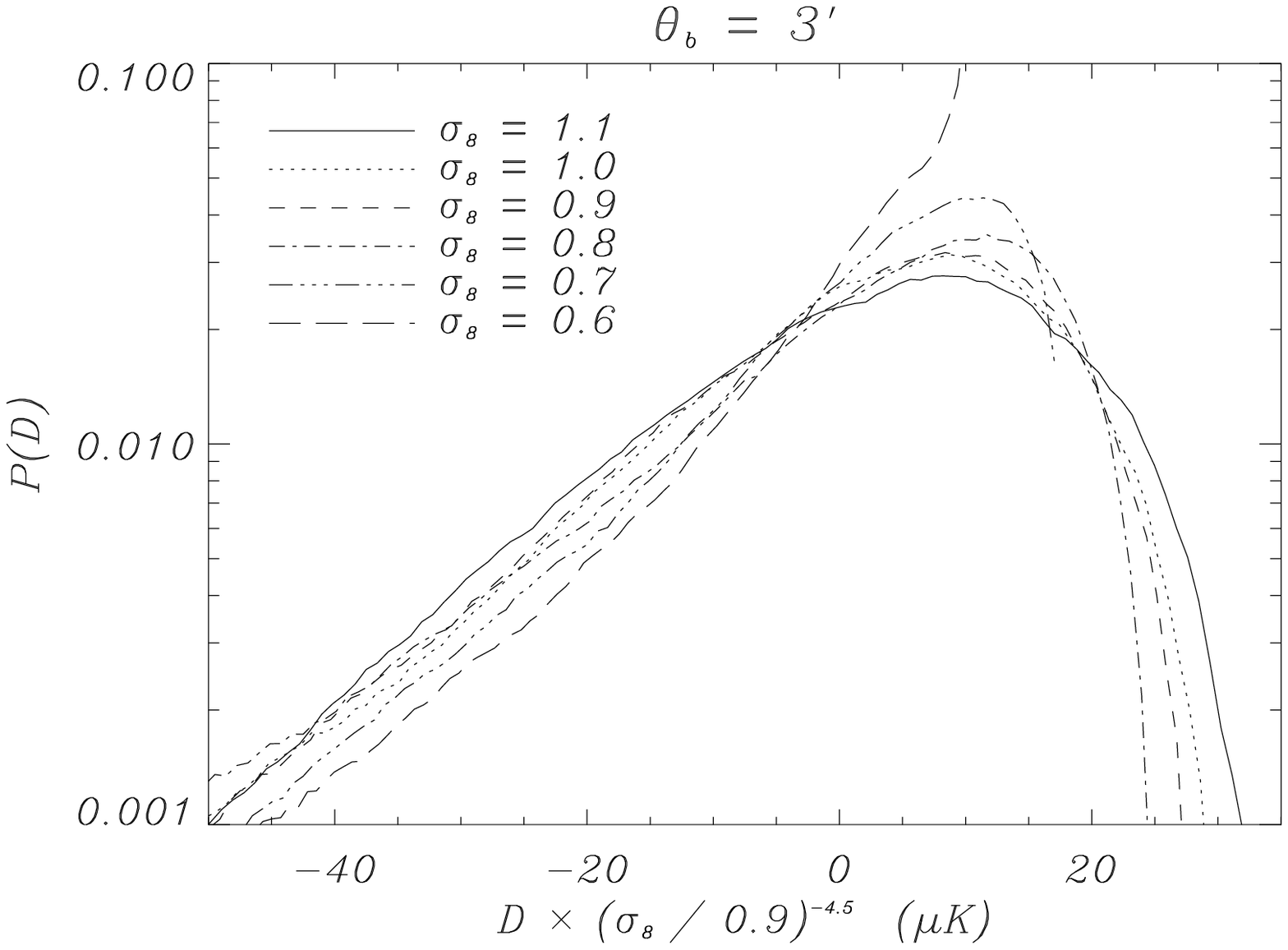}
\caption{Top: P(D) curves for the same realizations considered
in figure \ref{bp_sigma8}. The maps have been smoothed with a
$3\arcmin$ FWHM gaussian beam. 
Bottom: Same P(D) curves, but rescaled by $D \rightarrow D
\sigma_8^{-4.5}$. We can see that now the asymptotic regions are parallel.}
\label{pd_sigma8}
\end{figure}

The corresponding $P(D)$ functions are shown in 
figure \ref{pd_sigma8}, using a beam size of $\theta_b=3'$. 
Given that we are particularly interested in the
intermediate asymptotic region, we obtained the
scaling of the 'a'--factor in that region ($P(D) \sim e^{aD}$).
From our simulations, we have 
\begin{equation}
a \approx 0.055 \Bigg( \frac{\sigma_8}{0.9} \Bigg)^{4.5} 
\Bigg( \frac{\theta_b}{2\arcmin}\Bigg) ^{-0.2} \qquad [\mu K^{-1}]
\end{equation}
for the region $1' \le \theta_b \le 5'$ and $0.6 \le \sigma_8 \le 1.1$. 
This is precisely the scaling with $\sigma_8$ that 
we would expect at first order, given
that the shape of the $P(D)$ function is roughly proportional
to the $\sigma$ of the SZ map, and the latter is proportional
to $C_\ell^{1/2}$. 
For illustration, in the lower panel of 
figure \ref{pd_sigma8} we show the different $P(D)$ curves, 
rescaled by $D \rightarrow D \sigma_8^{-4.5}$, for the case $\theta_b=3'$. 
As we can see, their asymptotics become similar.

\begin{thebibliography}{99}
\bibitem[\protect\citeauthoryear{Atrio-Barandela \& M{\" u}cket}{1999}]{atrio99} 
Atrio-Barandela, F.~\& M{\" u}cket, J.~P.\ 1999, ApJ, 515, 465 

\bibitem[\protect\citeauthoryear{Barcons}{1992}]{barcons92} Barcons, X.\ 1992, ApJ, 396, 460 

\bibitem[\protect\citeauthoryear{Banday et al.}{1996}]{banday96} Banday,
A.~J., Gorski, K.~M., Bennett, C.~L., Hinshaw, G., Kogut, A., \& Smoot,
G.~F.\ 1996, ApJL, 468, L85 


\bibitem[\protect\citeauthoryear{Birkinshaw}{1999}]{birkinshaw99}
Birkinshaw, M.\ 1999, Phys. Rep., 310, 97 

\bibitem[\protect\citeauthoryear{Bond et al.}{2002}]{bond02} Bond,
J.~R. et al.,\ 2002, ApJ, submitted (astro-ph/0205386)

\bibitem[\protect\citeauthoryear{Bower}{1997}]{bower97} Bower, R.~G.\ 1997, MNRAS, 288, 355 

\bibitem[\protect\citeauthoryear{Carlstrom, Holder, \& Reese}{2002}]{carlstrom02} Carlstrom, 
J.~E., Holder, G.~P., \& Reese, E.~D.\ 2002, ARA\&A, 40, 643 

\bibitem[\protect\citeauthoryear{Cavaliere et al.}{1973}]{cavaliere73} 
Cavaliere, A., Friedland, A., Gursky, H., \& Spada, G.\ 1973, ApJ, 182, 405 

\bibitem[\protect\citeauthoryear{Cavaliere \&
Setti}{1976}]{cavaliere76} Cavaliere, A.~\& Setti, G.\ 1976, A\&A, 46, 81

\bibitem[\protect\citeauthoryear{Cay{\' o}n et al.}{2000}]{cayon00} Cay{\' o}n, L.~et 
al.\ 2000, MNRAS, 315, 757 

\bibitem[\protect\citeauthoryear{Cole \& Kaiser}{1988}]{cole88} Cole, S.~\& Kaiser, N.\ 
1988, MNRAS, 233, 637 

\bibitem[\protect\citeauthoryear{Condon}{1974}]{condon74} Condon, J.~J.\ 1974, ApJ, 188, 
279

\bibitem[\protect\citeauthoryear{Condon \& Dressel}{1978}]{condon78} Condon, J.~J.~\& 
Dressel, L.~L.\ 1978, ApJ, 222, 745 

\bibitem[\protect\citeauthoryear{Condon}{1984}]{condon84} Condon, J.~J.\ 1984, ApJ, 287, 461 

\bibitem[\protect\citeauthoryear{Cooray et al.}{1998}]{cooray98} Cooray, A.~R., Grego, 
L., Holzapfel, W.~L., Joy, M., \& Carlstrom, J.~E.\ 1998, AJ, 115, 1388 

\bibitem[\protect\citeauthoryear{Cooray}{2000}]{cooray00} Cooray, A.\ 2000,
Phys. Rev. D, 62, 103506 

\bibitem[\protect\citeauthoryear{Cooray}{2001}]{cooray01} Cooray, A.\ 2001, Phys. Rev. D, 64, 
63514 

\bibitem[\protect\citeauthoryear{Cooray \& Melchiorri}{2002}]{cooray02} Cooray, A.\&
Melchiorri, A.\ 2002, Phys. Rev. D, accepted (astro-ph/0204250)

\bibitem[\protect\citeauthoryear{Dawson et al.}{2002}]{dawson02} Dawson, K.~S, Holzapfel,
W.~L., Carlstrom, J.~E., LaRoque, S.~J., Miller, A., Nagai, D. \& Joy,
M.\ 2002, ApJ, submitted (astro-ph/0206012)

\bibitem[\protect\citeauthoryear{de Luca, Desert, \& Puget}{1995}]{deluca95} de Luca, A., 
Desert, F.~X., \& Puget, J.~L.\ 1995, A\&A, 300, 335 

\bibitem[\protect\citeauthoryear{Diego et al.}{2001}]{diego01} Diego, J.~M., 
Mart{\'{\i}}nez-Gonz{\' a}lez, E., Sanz, J.~L., Cay{\' o}n, L., \& Silk, 
J.\ 2001, MNRAS, 325, 1533 

\bibitem[\protect\citeauthoryear{Eke, Cole, \& Frenk}{1996}]{eke96} Eke, V.~R., Cole, 
S., \& Frenk, C.~S.\ 1996, MNRAS, 282, 263 

\bibitem[\protect\citeauthoryear{Ettori, De Grandi, \& Molendi}{2002}]{ettori02} Ettori, 
S., De Grandi, S., \& Molendi, S.\ 2002, A\&A, 391, 841 

\bibitem[\protect\citeauthoryear{Fabian}{1975}]{fabian75} Fabian, A.~C.\ 1975, MNRAS, 172, 149 

\bibitem[\protect\citeauthoryear{Ferreira, Magueijo, \&
Gorski}{1998}]{ferreira98} Ferreira, P.~G., Magueijo, J., \& Gorski,
K.~M.\ 1998, ApJ, 503, L1 

\bibitem[\protect\citeauthoryear{Fischer \& Lange}{1993}]{fischer93} Fischer, M.~L.~\& 
Lange, A.~E.\ 1993, ApJ, 419, 433 

\bibitem[\protect\citeauthoryear{Franceschini et
al.}{1989}]{franceschini89} Franceschini, A., Toffolatti, L., Danese, 
L., \& de Zotti, G.\ 1989, ApJ, 344, 35 

\bibitem[\protect\citeauthoryear{Fomalont et al.}{1988}]{fomalont88} Fomalont, E.~B., 
Kellermann, K.~I., Anderson, M.~C., Weistrop, D., Wall, J.~V., Windhorst, 
R.~A., \& Kristian, J.~A.\ 1988, AJ, 96, 1187 

\bibitem[\protect\citeauthoryear{Fomalont et al.}{2002}]{fomalont02} Fomalont, E.~B., 
Kellermann, K.~I., Partridge, R.~B., Windhorst, R.~A., \& Richards, E.~A.\ 
2002, AJ, 123, 2402 

\bibitem[\protect\citeauthoryear{Fomalont et al.}{1993}]{fomalont93} Fomalont, E.~B., Partridge, R.~B., 
Lowenthal, J.~D., \& Windhorst, R.~A.\ 1993, ApJ, 404, 8 

\bibitem[\protect\citeauthoryear{Haehnelt \&
Tegmark}{1996}]{tegmark96} Haehnelt, M.~G.~\& Tegmark, M.\ 1996, MNRAS, 279, 545 

\bibitem[\protect\citeauthoryear{Hewish}{1961}]{hewish61} Hewish, A.\
1961, MNRAS, 123, 167

\bibitem[\protect\citeauthoryear{Hobson \& Magueijo}{1996}]{hobson96}
Hobson, M.~P.~\& Magueijo, J.\ 1996, MNRAS, 283, 1133

\bibitem[\protect\citeauthoryear{Holder}{2002}]{holder02} Holder, G.~P.\ 2002, ApJL, submitted 
(astro-ph/0207633)

\bibitem[\protect\citeauthoryear{Holder}{2002b}]{holder02b} Holder,
G.~P.\ 2002b, ApJ, accepted  (astro-ph/0205467)

\bibitem[\protect\citeauthoryear{Jenkins et al.}{2001}]{jenkins01}
Jenkins, A., Frenk, C.~S., White, S.~D.~M., Colberg, J.~M., Cole, S.,
Evrard, A.~E., Couchman, H.~M.~P., \& Yoshida, N.\ 2001, MNRAS, 321, 372 

\bibitem[\protect\citeauthoryear{Kesden, Cooray \& Kamionkowski}{2002}]{kesden02}  Kesden, M.,
Cooray, A.~\& Kamionkowski, M.\ 2002, Phys. Rev. D, submitted (astro-ph/0208325)

\bibitem[\protect\citeauthoryear{Kneissl et al.}{2001}]{kneissl01}
Kneissl, R., Jones, M.~E., Saunders, R., Eke, V.~R., Lasenby, A.~N.,
Grainge, K., \& Cotter, G.\ 2001, MNRAS, 328, 783 

\bibitem[\protect\citeauthoryear{Komatsu \&
Kitayama}{1999}]{komatsu99} Komatsu, E.~\& Kitayama, T.\ 1999, ApJ, 526, L1 

\bibitem[\protect\citeauthoryear{Komatsu \& Seljak}{2002}]{komatsu02} Komatsu, E.~\& Seljak,U.\ 2002, MNRAS, submitted (astro-ph/0205468) 

\bibitem[\protect\citeauthoryear{Komatsu et al.}{2002}]{komatsu02b} Komatsu,
E., Wandelt, B.~D., Spergel, D.~N., Banday, A.~J., \& G{\' o}rski, K.~M.\
2002, ApJ, 566, 19 

\bibitem[\protect\citeauthoryear{Korolev, Sunyaev, \& Yakubsev}{1986}]{korolev86} Korolev, V.~A., Sunyaev, R.~A., \& Yakubsev, L.~A.\ 1986, Soviet 
Astronomy Letters, 12, 339 

\bibitem[\protect\citeauthoryear{Lin, Chiueh, \& Wu}{2002}]{lin02}  Lin, H., Chiueh, T.,~\& Wu,
X.\ 2002, astro-ph/0202174

\bibitem[\protect\citeauthoryear{Longair \& Sunyaev}{1969}]{longair69} 
Longair, M.~S., \& Sunyaev, R.~A.\ 1969, Nature, 223, 719

\bibitem[\protect\citeauthoryear{Luo}{1994}]{luo94} Luo, X.\ 1994, ApJ, 427, L71 

\bibitem[\protect\citeauthoryear{Mason et al.}{2002}]{mason02} Mason, B.~S. et al.,\ 2002,
ApJ, submitted (astro-ph/0205384)

\bibitem[\protect\citeauthoryear{Mohr, Mathiesen, \& Evrard}{1999}]{mohr99} Mohr, 
J.~J., Mathiesen, B., \& Evrard, A.~E.\ 1999, ApJ, 517, 627 

\bibitem[\protect\citeauthoryear{Molnar \& Birkinshaw}{2000}]{molnar00} Molnar, S.~M.~\& 
Birkinshaw, M.\ 2000, ApJ, 537, 542 

\bibitem[\protect\citeauthoryear{Ostriker \& Steinhardt}{1995}]{ostriker95}
Ostriker, J.~P.~\& Steinhardt, P.~J.\ 1995, Nature, 377, 600 

\bibitem[\protect\citeauthoryear{Padmanabhan}{1993}]{padmanabhan}
Padmanabhan, T.\ 1993, Structure formation in the universe. 
Cambridge, UK: Cambridge University Press

\bibitem[\protect\citeauthoryear{Peebles}{1993}]{peebles93} Peebles,
P.~J.~E.\ 1993, Principles of physical cosmology. Princeton Series in
Physics, Princeton, NJ: Princeton University Press. 

\bibitem[\protect\citeauthoryear{Press \& Schechter}{1974}]{press74}
Press, W.~H.~\& Schechter, P.\ 1974, ApJ, 187, 425 

\bibitem[\protect\citeauthoryear{Rowan-Robinson \& Fabian}{1974}]{rowan74} 
Rowan-Robinson, M.~\& Fabian, A.\ 1974, MNRAS, 167, 419 

\bibitem[\protect\citeauthoryear{Rubi{\~ n}o-Mart{\'\i}n, Atrio-Barandela, \& Hern{\'a}ndez-Monteagudo}{2000}]{rubino00} Rubi{\~ n}o-Mart{\'\i}n, 
J.~A., Atrio-Barandela, F., \& Hern{\' a}ndez-Monteagudo, C.\ 2000, ApJ, 
538, 53 

\bibitem[\protect\citeauthoryear{Santos et al.}{2002}]{santos02}
Santos, M.~G.~et al.\ 2002, Physical Review Letters, 88, 241302 

\bibitem[\protect\citeauthoryear{Sanz, Herranz, \& Mart{\'{\i}}nez-G{\' 
o}nzalez}{2001}]{sanz01} Sanz, J.~L., Herranz, D., \& 
Mart{\'{\i}}nez-G{\' o}nzalez, E.\ 2001, ApJ, 552, 484  

\bibitem[\protect\citeauthoryear{Sheth \& Tormen}{1999}]{sheth99}
Sheth, R.~K.~\& Tormen, G.\ 1999, MNRAS, 308, 119

\bibitem[\protect\citeauthoryear{Scheuer}{1957}]{scheuer57} Scheuer, P.~A.~G.\ 1957, 
Proc.Camb.Phil.Soc, 53, 764 

\bibitem[\protect\citeauthoryear{Scheuer}{1974}]{scheuer74} Scheuer, P.~A.~G.\ 1974, 
MNRAS, 166, 329 

\bibitem[\protect\citeauthoryear{Schulz \& White}{2002}]{schulz02}
Schulz, A.~E. ~\& White, M. \ 2002, ApJ submitted (astro-ph/0210667)

\bibitem[\protect\citeauthoryear{Seljak, Burwell, \&
Pen}{2001}]{seljak01} Seljak, U.~;, Burwell, J., \& Pen, U.\ 2001,
Phys.Rev.D, 63, 63001 

\bibitem[\protect\citeauthoryear{Springel, White, \&
Hernquist}{2001}]{springel01} Springel, V., White, M., \& Hernquist, L.\
2001, ApJ, 549, 681 [erratum: 2001, ApJ, 562, 1086]

\bibitem[\protect\citeauthoryear{Sunyaev}{1980}]{rashid80} Sunyaev, R.~A.\ 1980, Soviet Astronomy Letters, 6, 213 

\bibitem[\protect\citeauthoryear{Sunyaev \& Zeldovich}{1970}]{sunyaev70} Sunyaev, R.~A.~\& Zeldovich, Y.~B.\ 1970, Ap\&SS, 7, 3

\bibitem[\protect\citeauthoryear{Sunyaev \& Zeldovich}{1972}]{sunyaev72} Sunyaev, R.~A.~\& Zeldovich, Y.~B.\ 1972, Comments on Astrophysics, 4, 173

\bibitem[\protect\citeauthoryear{Sunyaev \& Zeldovich}{1980}]{sunyaev80} Sunyaev, R.~A.~\& Zeldovich, I.~B.\ 1980, ARA\&A, 18, 537 

\bibitem[\protect\citeauthoryear{Taylor et al.}{2002}]{taylor02}
Taylor, A.~C. et al.\ 2002, MNRAS, submitted (astro-ph/0205381)

\bibitem[\protect\citeauthoryear{Tegmark \& de
Oliveira-Costa}{1998}]{tegmark98} Tegmark, M.~\& de Oliveira-Costa,
A.\ 1998, ApJ, 500, L83 

\bibitem[\protect\citeauthoryear{Toffolatti et al.}{1998}]{toffolatti98} Toffolatti, L., 
Argueso Gomez, F., de Zotti, G., Mazzei, P., Franceschini, A., Danese, L., 
\& Burigana, C.\ 1998, MNRAS, 297, 117

\bibitem[Viana \& Liddle(1999)]{viana99} Viana, P.~T.~P.~\& 
Liddle, A.~R.\ 1999, MNRAS, 303, 535 

\bibitem[\protect\citeauthoryear{Zhang, Pen \& Wang}{2002}]{zhang02}
Zhang, P., Pen, U., \& Wang, B.\ 2002, ApJ, 577, 555 

\bibitem[\protect\citeauthoryear{Zeldovich \& Sunyaev}{1969}]{zeldovich69} Zeldovich, 
Y.~B.~\& Sunyaev, R.~A.\ 1969, Ap\&SS, 4, 301 

\end{thebibliography}
\end{document}